\documentclass[12pt]{JHEP}
\newcommand{\refeq}[1]{(\ref{#1})}
\newcommand{\intd}{\mathrm{d}}
\newcommand{\ex}{\mathrm{e}}
\newcommand {\non}{\nonumber}
\newcommand{\oneover}[1]{\frac{1}{#1}}
\usepackage{amsmath,epsfig}
\usepackage{amssymb,amsfonts}
\usepackage{latexsym}
\relax
\renewcommand{\theequation}{\arabic{section}.\arabic{equation}}
\def\hri#1#2{\href{http://arxiv.org/abs/#1}{[ArXiv:#1]#2}}
\def\hre#1#2{\href{http://arxiv.org/abs/#1/#2}{[ArXiv:#1/#2]}}

\def\be{\begin{equation}}
\def\ee{\end{equation}}
\def\bea{\begin{eqnarray}}
\def\eea{\end{eqnarray}}

\newcommand\fverb{\setbox\pippobox=\hbox\bgroup\verb}
\newcommand\fverbdo{\egroup\medskip\noindent%
                        \fbox{\unhbox\pippobox}\ }
\newcommand\fverbit{\egroup\item[\fbox{\unhbox\pippobox}]}

\newcommand{\bear}{\begin{eqnarray}}

\newcommand{\eear}{\end{eqnarray}}
\newcommand{\de}{\partial}
\newcommand{\<}{\langle}
\renewcommand{\>}{\rangle}
\newbox\pippobox

\def\ie{{\it i.e.~}}
\def\lab{\label}
\def\6{\partial}
\def\f{\Phi}
\def\a{\alpha}
\def\nn{\nonumber}

\def\half{\frac12}

\def\le{\left}
\def\ri{\right}
\def\cO{{\cal O}}

\def\pa{\partial}

\def\C0{{\bf C_0}}
\def\Y0{{\bf Y_0}}
\def\G0{{\bf G_0}}
\def\e{\epsilon}
\def\m{\mu}
\def\n{\nu}

\def\sp{\;\;\;,\;\;\;}

\def\p{\partial}

\def\z{\zeta}
\def\sq
\def\a{\alpha}
\def\b{\beta}
\def\l{\lambda}

\def\tr{{\rm Tr}}

\def\db{\delta b}

\def\eps{\epsilon}
\def\cG{{\cal G}}
\def\La{\Lambda}
\def\tT{\tilde{T}}
\def\cF{{\cal F}}

\title{Holography and Thermodynamics of 5D Dilaton-gravity}
\author{U. G{\"u}rsoy$^1$,
\href{http://hep.physics.uoc.gr/~kiritsis/}{E. Kiritsis}$^{2}$,
L. Mazzanti$^{3}$,
 F. Nitti$^{4}$\\
$^1$\href{http://www1.phys.uu.nl/wwwitf}{Institute for Theoretical Physics, Utrecht University;
Leuvenlaan 4, 3584 CE Utrecht, The Netherlands.}\\
~\\
$^2$\href{http://hep.physics.uoc.gr/}{Department of Physics, University of Crete
71003 Heraklion, Greece}\\
~\\
$^3$\href{http://cpht.polytechnique.fr/cpht/cordes/}{CPHT, Ecole Polytechnique, CNRS,
 91128, Palaiseau, France}\\
 (UMR du CNRS 7644).\\
~\\
$^4$\href{http://www.apc.univ-paris7.fr}{APC, Universit\'e Paris 7, \\ B\^atiment Condorcet, F-75205, Paris Cedex 13, France}}

\preprint{CPHT-RR088.1108\\
SPIN-08/57\\
ITP-UU-08/74}      

\abstract{The asymptotically-logarithmically-AdS black-hole solutions of 5D dilaton gravity with
a monotonic dilaton potential are analyzed in detail.
Such theories are holographically very close to pure Yang-Mills theory in four dimensions.
The existence and uniqueness of black-hole solutions is shown.
It is also shown that a Hawking-Page transition exists at finite temperature if and only
 if the potential corresponds to a confining theory. The physics of the transition matches in detail
 with that of deconfinement of the Yang-Mills theory. The high-temperature phase asymptotes to a free gluon gas at high temperature
 matching the expected behavior from asymptotic freedom. The thermal gluon condensate is calculated
 and shown to be crucial for the existence of a non-trivial deconfining transition.
 The condensate of the topological charge is shown to vanish in the deconfined phase.  }


\begin{document}

\def\g{\gamma}
\def\go{\g_{00}}
\def\gi{\g_{ii}}

\maketitle 

\section{Introduction and Summary}

In the past decade we have witnessed a rebirth of theoretical efforts to address the strong coupling problem in gauge theories.
The tool has  been the large-$N_c$ expansion but a
new twist has emerged, whereby the relevant dual string theories
live in an  appropriately curved higher dimensional space-time.

 The prototype example has been the AdS/CFT correspondence as exemplified by the
(well studied by now) duality of ${\cal N}=4$ super Yang-Mills theory and IIB string theory on $AdS_5\times S^5$\cite{malda,witten,gkp}.
Further studies focused on providing examples that are closer to real world QCD,
\cite{D4,mnks}. It is fair to say that we now have a good
holographic understanding of phenomena
like confinement, chiral symmetry and its breaking as well as several
related issues.  The finite temperature dynamics of gauge theories,
has a natural holographic counterpart in the thermodynamics of  black-holes
on the gravity side, and the thermal properties of various holographic constructions
have been widely studied, \cite{D4,bh1,herzog,weiz,Evans,erdm}, exhibiting
the holographic version of deconfinement and chiral restoration transitions.

The simplest top-down string theory model of QCD involves $D_4$ branes with supersymmetry breaking
boundary conditions for fermions \cite{D4},
as well as a flavor sector that involves pairs of $D_8-\overline D_8$ probe branes inserted in the bulk,
 \cite{sas}. The qualitative thermal properties of this model
closely mimic what we expect in QCD, \cite{weiz}.
Although such theories reproduced the qualitative features of IR QCD dynamics,
they contain Kaluza-Klein  modes, not expected in QCD,
with KK masses of the same order as the dynamical scale
of the gauge theory. Above this scale the theories deviate from QCD.
Therefore, although the qualitative features of the relevant phenomena are correct,
 a quantitative matching to real QCD is difficult.

Despite the hostile environment  of  non-critical theory,
several attempts have been made  to
understand holographic physics in lower dimensions
 in order to avoid the KK contamination,
based on two-derivative gravitational actions, \cite{ks}.
Indeed, large N QCD is expected to be described by a 5-dimensional theory.
The alternative problem
in non-critical theories is that curvatures are of string scale size and the truncation of the theory
to the zero mode sector is subtle and may be misleading.

 A different and more phenomenological bottom-up approach was developed and is now known as AdS/QCD.
The original idea described  in \cite{ps} was successfully applied to the meson
sector in \cite{adsqcd1}, and its thermodynamics was analyzed in \cite{herzog,BallonBayona}.
The bulk gravitational background consists of a slice of AdS$_5$, and a constant dilaton.
There is a UV and an IR cutoff. The confining IR physics is imposed by boundary
conditions at the IR boundary.
This approach, although crude,  has
been partly successful in studying meson
physics, despite the fact that the dynamics driving chiral symmetry breaking must be
imposed by hand via IR boundary conditions.
Its shortcomings however include a glueball spectrum that does not fit very well the lattice data,
the fact that magnetic quarks are confined instead of screened, and asymptotic Regge trajectories for
glueballs and mesons that are quadratic instead of linear.

A phenomenological fix of the last problem was suggested by
introducing a soft IR wall, \cite{soft}.
Although this fixes the asymptotic spectrum of mesons and meson
dynamics
 is in principle self-consistent, it does not allow a consistent treatment
of the glue sector both at zero and finite temperature.
In particular, neither dilaton nor metric  equations of motion are solved.
Therefore the ``on-shell'' action is not really on-shell. The entropy computed from the BH horizon does
not match the entropy calculated using standard thermodynamics from the
free energy computed from the action, etc.
Phenomenological metrics for the deconfined phase were also suggested, \cite{adr,keijo}
capturing some aspects of the expected thermodynamics.

The theoretical advances were paralleled by a very successful experimental effort at RHIC, \cite{rhic}.
The consensus on the existing data is that shortly after the collision a ball of quark-gluon plasma (QGP) forms that is at thermal equilibrium,
and  subsequently expands until its temperature falls below the QCD transition (or crossover) where it finally hadronizes.
Relativistic hydrodynamics describes very well the QGP  \cite{lr}, with a shear-viscosity to entropy density ratio close to that
of ${\cal N}=4$ SYM, \cite{pss}.
The QGP is at strong coupling, and it necessitates a treatment beyond perturbative QCD approaches, \cite{review}.
Moreover, although the shear viscosity from  ${\cal N}=4$ seems to be close to that ``measured'' by
 experiment, lattice data indicate that in the relevant
RHIC range  $1\leq {T\over T_c}\leq 3$ the QGP
seems not to be a fully  conformal fluid.
 Therefore the bulk viscosity is expected to play a role near the phase transition
\cite{kkt,m}.  The lattice techniques have been successfully used to study the
thermal behavior of QCD, however they are not easily extended to the computation
of hydrodynamic quantities. They can be used however, together with parametrizations of correlators in order to pin down parameters
\cite{m}. To date it seems that the holographic approach is  a promising one in this direction.

In the bottom-up holographic model of AdS/QCD, the bulk viscosity is zero as conformal invariance is essentially not broken
(the stress tensor is traceless).
In the soft-wall model, no reliable calculation can be done for glue correlators and therefore
transport coefficients are ill-defined. Similar
remarks hold for other phenomenologically interesting observables as the
drag force and the jet quenching parameter \cite{her,lrw,gub}.

In order to go beyond the inadequacies of existing  bottom-up holographic models,
input has been put together both from string theory and QCD
in order to craft an improved holographic QCD model, \cite{ihqcd}.
It is a five-dimensional Einstein dilaton system, with an appropriately chosen dilaton potential.

The vacuum solution involves an asymptotically logarithmically AdS solution near the boundary.
The bulk  field $\l$, dual to the 't Hooft coupling,
is vanishing logarithmically near the boundary in order to match the expected QCD behavior.
This implies that the potential must have a regular Taylor expansion as $\l\to 0$, and that $\l=0$ is not an extremum.
This is unlike almost all asymptotically AdS solutions discussed so far in the literature.
In particular the canonically normalized scalar (the dilaton) is diverging at the boundary $r\to 0$ as $\phi\sim -\log(-\log r)$.
The coefficients of the UV Taylor expansion of the potential are in one-to-one correspondence with the holographic $\beta$-function.

In the IR, the potential must have an appropriate behavior so that the theory is confined, has a mass gap and a discrete spectrum.
This selects a narrow range of asymptotics that roughly obey
\be
V(\l)\sim \l^{2Q}\sp \l\to \infty~~~~~{\rm  with}~~~~~ {2\over 3}\leq Q< {4\over 3}.
\label{1}\ee
The vacuum solution always ends in a naked singularity in the bulk. Demanding that this is a ``good''
singularity in the classification of Gubser
\cite{gub2} implies  $Q<4/3$. However, here  we use a narrower notion of what we mean
by ``good'' singularity: we accept only singularities that are {\em repulsive} to
physical fluctuations, i.e. such that no extra boundary conditions are needed there.
 This requirement further restricts $Q<{2\sqrt{2}\over 3}$ in (\ref{1}).
Simple interpolations between the UV and IR asymptotics reproduce very well the low-lying glueball spectrum as well as the perturbative running of the
't Hooft coupling \cite{ihqcd}.

Five-dimensional Einstein dilaton systems with a monotonic dilaton potential  (no extrema) provide an interesting class of gravitational theories
that display diverse behaviors as a function of the IR asymptotics of the potential.
Potentials with asymptotics growing slower than (\ref{1}) with $Q=2/3$ do not exhibit confinement.
Potentials  with asymptotics growing faster  than (\ref{1}) with $Q=2\sqrt{2}/3$ do  exhibit confinement but the naked singularity
is too strong and extra boundary conditions are needed at the singularity in order to study the spectrum of fluctuations.

In this paper we will analyze the existence and structure of black-hole solutions,
and their thermodynamics,  in the
class of gravitational theories described above.
We will take the horizon to be a flat three-dimensional torus,
but our techniques extend to the case where the horizon is a three-sphere.
Our preliminary results in this direction  have been published in \cite{ft}.
The thermodynamics of similar systems has also been analyzed in \cite{gub3}.
Our aim is to eventually describe the finite-temperature structure of a holographic model closely resembling
pure large-$N_c$ Yang Mills theory. The latter is widely analyzed using
lattice techniques (see e.g. \cite{karsch,teperlucini}), which indicate  that the theory
exhibits a first order   confinement-deconfinement
phase  transition at a non-zero critical temperature of the order of the strong coupling scale $\Lambda$.
While one of the main motivations for having a realistic holographic description
of  finite-temperature QCD is the computation of transport coefficients and
other quantities relevant for heavy ion collision experiments,  in this paper we will only be concerned
with equilibrium thermodynamics, as this is already a daunting task.
We leave the hydrodynamics for the very near future \cite{GKMicN}.

As  we show in this paper,  the correspondence between the 5D
Einstein-dilaton  setup
advocated  in \cite{ihqcd} and 4D large $N_c$ pure Yang-Mills extends to the
finite temperature regime in  a remarkable way.  One of the main results of this work is {\em the proof of the
the existence, in the case of confining theories, of a first order Hawking-Page phase
transition between the thermal gas and black-hole solutions.}
Moreover, the 5D  black-holes in  confining  theories  provide a realistic holographic
description of  the thermodynamics of the  deconfined phase of 4D,  large $N_c$ pure Yang-Mills,
that emerges from lattice studies.

The black-holes that we discuss obey the same UV asymptotics as the zero-temperature solution,
namely they are asymptotically-logarithmically  $AdS$ with a logarithmically vanishing dilaton. Close to
the $AdS$ boundary the metric  is that of  a 5D AdS-Schwarzschild black-hole in Poincar\'e
coordinates (i.e. with flat horizon), up to logarithmic corrections.  Although asymptotically
$AdS$ black  holes in Einstein-dilaton theories have received considerable attention
(see e.g. \cite{EDBH}),
these examples were always associated with the existence of an exactly $AdS$ solution with
constant dilaton, corresponding to an extremum of the dilaton potential. In contrast,
the 5D black-holes we discuss here are of a new type, since the dilaton potential
in our model is always monotonic. The $AdS$ point is at  infinity in field space,
 therefore the models we discuss {\it do not have a  pure $AdS$ solution}.

In this paper  we derive
a series of general results about the thermodynamics of the 5D system,
that do not depend on the specific form of
the potential but only on its small $\l$  and large $\l$ asymptotics. All these results point
to the remarkable similarity between the thermodynamics of 5D models in the confining class,
and the thermodynamics of 4D large $N_c$ Yang-Mills. While the detailed
 quantitative comparison between a specific model and the lattice results for thermal Yang-Mills
 will appear elsewhere \cite{GKMN3},
these general  results are the main focus of this paper.
Below we briefly summarize them.  \\

\subsubsection*{\em Parameters of the solutions}

The parameters of the action are the 5D Planck mass $M_p$, and the various parameters that determine the shape of the
dilaton potential
$V(\l)$. In particular,  the value  $V(0)$  sets the $AdS$ length scale $\ell$.
In addition, every  black hole solution  is characterized by  the five integration constants of the 5th
order dilaton-gravity system of equations.

We show how to identify these parameters in the dual gauge theory and eventually how to fix them.
\begin{itemize}
\item We keep the form  of the potential  generic,  except that it should be  a monotonic function of
 $\l$ and it should obey the small and large $\l$ asymptotics we mentioned above. A specific potential
can be fixed by requiring that the zero-$T$ spectrum agrees with the lattice data, as was done in \cite{ihqcd}.
\item  The specific  value of the  $AdS$ length $\ell$ is irrelevant for any physical
observable, and sets the overall units of the 5D solution.
\item The 5D Planck scale (in $AdS$ units)
is fixed by matching the free field asymptotics of the QCD free energy
in the high-temperature limit. This universally fixes $M_p\ell = 1/(45\pi^2)$.
The physical Planck mass, that determines the strength of gravitational interactions
(and of  the interactions between glueballs in the dual theory)  includes an extra factor of $N_c^{2/3}$,
that guarantees the suppression of quantum corrections in the large $N_c$ limit in our setup.

\item Among the 5 integration constants of the equations of motion, only two are physical and independent:
the value of the dilaton at the horizon,  and an overall scale  $\Lambda$, related to the dilaton
asymptotics near the UV  boundary.
The former determines the black hole temperature and entropy; the latter is also present in
the  vacuum solution, and it is dual to $\Lambda_{QCD}$. Fixing the
UV asymptotics is equivalent to selecting a specific value for $\Lambda$.
\end{itemize}
Since our model should be thought of as coming from a non-critical string theory,
an additional parameter is the string scale $\ell_s$. This is invisible in the
 5D Einstein-dilaton  setup we assume throughout this paper,
and we will not discuss
it any further. It can be fixed by
comparing the  effective string tension
(calculated from the linear part of the static quark potential)
to  the lattice data, as was done in  \cite{ihqcd}.

\subsubsection*{\em Existence and uniqueness of the black-hole solutions}

The $T=0$ solution defines a vacuum background.
Once we specify the UV asymptotics of a black-hole solution to be  the same as for the vacuum background,
 there is  only one additional parameter that characterizes the solution,
that we take to be
 the value of the dilaton at the horizon, $\l_h$. For any {\em monotonic} dilaton potential $V(\l)$
we show that,  for each $\l_h$, ranging between zero and
infinity, there exists a unique black-hole solution, with  a  temperature $T$, entropy $S$ (horizon area)
and free energy ${\cal F}$  functions of $\l_h$ only. Thus, the  thermodynamics is encoded in the
functions $T(\l_h)$, $S(\l_h)$ and ${\cal F}(\l_h)$, namely the temperature, entropy and free energy.
Moreover, we show that the limit $\l_h\to \infty$ of the black-hole solution coincides with the
{\em unique} zero-temperature solution that, for a given potential, displays a ``good'' (i.e.
repulsive)  singularity.

{ Although our models allow an infinite number of black-hole solutions
(each with a different value of $\Lambda$) with the  same mass, this does not imply that
these black-holes admit scalar hair. The reason is that each different value of $\Lambda$ corresponds to
a different asymptotic for the metric and dilaton. In other words, $\Lambda$ plays
the role of an extra ``charge'' that can be measured at infinity. Moreover,
due to the absence of extrema in the dilaton potential,  there is no
  pure $AdS$-Schwarzschild solution  with a constant dilaton.
Thus, our black  hole solutions satisfy the no-hair theorems for
asymptotically $AdS$  gravity coupled to scalars (see \cite{Hertog} for a recent discussion). }\\

\subsubsection*{\em Deconfinement phase transition}

We show  that any 5D theory,  whose zero-temperature solution is dual
to a UV-free and  IR-confining 4D theory,  also exhibits a Hawking-Page type of phase transition,
dual to a deconfining phase transition in 4D. { The transition is always first order, except in the
special case when the IR behavior  of the  vacuum solution  is at the borderline between
confining and non-confining geometries: in this case the (string frame) solution
is asymptotically a flat metric with a linear dilaton, and the phase transition
is second order.  }
Conversely, non-confining theories do not exhibit a phase transition: black-holes
dominate the thermal ensemble for any non-zero temperature $T$.

The phase structure of black-holes in confining theories is similar to what is
found in the original Hawking-Page situation \cite{HP}, namely:
 \begin{itemize}
\item Black hole solutions  exist only above a certain temperature $T_{min}$;
\item Generically, for any $T>T_{min}$ there exist (at least) two different black-hole solutions with the same
$T$ and the same UV asymptotics (large and small BHs);
\item Above a certain critical temperature $T_c>T_{min}$, it is
(one of the) large black-holes that dominate the thermal ensemble, while for $0<T<T_c$ it is
the ``thermal gas'' solution (with the same metric and dilaton as the zero-temperature solution)
that dominates.  On the other hand, the small black-holes never dominate the ensemble.
\item { Typically, the big black-holes have positive specific heat, and are thermodynamically stable,
whereas the small black-holes have negative specific heat and are unstable. There
may be however exceptions to this rule, where in a  limited  range of $\l_h$ the small black-holes
are also stable.}
\item { In the borderline case (asymptotically linear dilaton), there is only  a single black-hole for $T>T_{min}$, and
 the second order transition occurs exactly at $T_{min}$. }
\end{itemize}
There is one important difference with the Hawking-Page case, however. There, the black-holes are
 global $AdS$-Schwarzschild with horizon topology $S^3$, and the theory
is dual to a conformal field theory on the $S^1\times S^3$ \cite{D4};
Here  instead we are dealing with
black-holes whose horizon has topology $T^3$, and the phase transition
is associated with dynamical confinement instead of the non-trivial topology of space. \\

\subsubsection*{\em Similarities of the black-hole phase  and the deconfined phase in Yang-Mills}

The thermodynamics of  black-holes in 5D duals of confining theories
shares many features with the deconfined phase of 4D Yang-Mills at large $N_c$.
\begin{itemize}
\item The appropriately regularized free energy $ {\cal F}/N_c^2 $ acts as an order parameter for the phase transition.
This is similar to the case of ${\cal N}=4$ SYM on the sphere, and as expected in pure YM in flat space;
\item Another order parameter is the Polyakov loop which vanishes in the confined phase and
becomes non-trivial above the deconfinement transition. This is paralleled by the dual gravity
computation. A string worldsheet that encircles the Euclidean time direction and extends in the
radial direction has infinite action in the confining  geometry, hence the vev of the loop vanishes.
On the other hand  it becomes  finite in the black-hole geometry yielding a finite value for the associated vev\cite{D4}.
\item The latent heat per unit volume is of order  $N_c^2 T_c^4$;
\item At very high temperature the thermodynamic quantities behave like in  a conformal theory, although
the approach to conformality is  logarithmically slow.
With a suitable choice of the relation between the 5D Planck scale and the $AdS$ length,
in the limit $T\to \infty$ we find a free gas,  as appropriate for a gravity dual of pure Yang-Mills
and unlike strongly coupled theories like ${\cal N}=4$ sYM .
\item The speed of sound  is small near the phase transition, and it
approaches the conformal value $c_s^2\to 1/3$ at high temperature\footnote{Reproducing this
behavior was the main motivation of \cite{gub3}, and it emerges quite naturally in our setup}.
\end{itemize}

\subsubsection*{\em The topological vacuum density}

The vacuum expectation value of the topological  density, $\<\tr F \wedge F\>$ can be computed
by including, on the gravity side,  a 5D axion, dual to the YM theta parameter \cite{ihqcd}.
We show that in the black-hole phase (deconfinement) the
profile of the axion is necessarily trivial, unlike the low temperature phase (confinement).
This causes  the vev of the topological density to vanish.
It is in agreement with the large $N_c$ expectations and with the lattice calculations
in finite temperature Yang-Mills \cite{ltheta}\\

\subsubsection*{\em Explicit calculation of  the free energy  and the role of the gluon condensate}

We compute explicitly the free energy of the black-hole solutions, relative
to the vacuum,  as the difference between the on-shell  actions.
A crucial role in this computation and in  the dynamics of the phase transition
is played by (the holographic dual
of) the thermal vev of $\tr F^2$. This quantity appears in the near-boundary
expansion of the difference between the black-hole metric scale factor $b(r)$ and its
zero-temperature analogue, $b_o(r)$. If $r$ denotes the conformal coordinate of both metrics,
the $AdS$ boundary is at $r=0$. We show that, once the UV asymptotics are fixed to be the
same for all the solutions, then as $r\to 0$:
\be\label{dbintro}
b(r) - b_o(r) = b_o(r)\left({\cal G} {r^4\over \ell^3} + {\rm subleading} \right).
\ee
The quantity ${\cal G}$ is proportional to $\<\tr F^2\>_T-\<\tr F^2\>_o$.  It is
a function of temperature, and it provides
a measure of   the deviation from conformality. It plays a crucial role
for the existence of the phase transition.

Indeed,  an explicit calculation shows that the free energy difference between a black-hole and
the vacuum solution is given schematically (omitting 3-space volume and other numerical factors) by:
\be
{\cal F} \sim {\cal G} - {T S\over 4}.
\ee
The second term is negative definite, and it is the only one present in a conformal field
theory. The gluon condensate is therefore crucial for the existence of a phase transition.

One important point is that the free energy written above
 receives contributions from the UV boundary alone: to compute ${\cal F}$
it is sufficient to know the metric close to the $AdS$ boundary. This is important for at least two
reasons:
\begin{enumerate}
\item In the context of $AdS$/CFT  all the information  about observable quantities must be encoded in
near-boundary data. No explicit contributions are obtained from the IR regime. The IR influences only
indirectly by fixing normalizable modes near the boundary via regularity conditions.

\item The specific dynamics of the IR singularity is irrelevant for the computation of the thermodynamics.
By contrast, in other studies  claiming the existence of deconfining phase transitions in
simple 5D models \cite{herzog,Evans},  it is the IR boundary or singularity that gives
the required positive contribution to the free energy.
\end{enumerate}

\subsubsection*{\em Matching the $\beta$-function to the trace anomaly}

We provide  a non-trivial check of the gauge/gravity correspondence applied
to the 5D theories of \cite{ihqcd}: the matching of the (flat space)
 conformal anomaly, encoded in the YM  $\beta$-function,  to lowest order in a
small $\l$ expansion.\footnote{This is the initial and simplest step in a rigorous program for the renormalization
of asymptotically logarithmically AdS theories, that will appear shortly, \cite{gupa}.}

From the free energy, we can compute the trace of the thermal stress tensor in the
deconfined phase, which turns out to be proportional to the gluon condensate:
\be
\<T^\mu_\mu\> ~~\sim~~ {\cal G}.
\ee
On the other hand, in 4D Yang-Mills theory the same quantity $\<T^\mu_\mu\>$ obeys the
dilatation Ward identity:
\be
\<T^\mu_\mu\> = {\beta(\l)\over 4\l^2} \<\tr F^2\>.
\ee
By a holographic computation we can find the relation between $\<\tr F^2\>$ and ${\cal G}$ (the latter
being defined by eq. (\ref{dbintro})), to lowest
order in the $\l\to 0$ limit, and can show that the two expressions for $T^{\m}_{\m}$
coincide precisely.\\

Part of the analysis in this paper is performed with the help of some new technical tools that
we believe are interesting by themselves:
\begin{itemize}
\item The {\em thermal generalization of the superpotential:} { This is  widely used in the
zero-temperature counterpart (see e.g. \cite{PET,superpotential}). A superpotential $W$ allows to recast Einstein's equations for the dilaton and scale factor
in the first order form, and to decouple them  from
the equations governing the evolution of the thermal function appearing in the metric;}
\item The {\em scalar variables:} this is a pair of functions of $\l$,  $X$ and $Y$, that are invariant under radial diffeomorphisms.
They  satisfy a coupled system of first-order differential equations. These  functions
 encode all the information
about the UV and IR properties of the full solution. From them one can easily derive
all the thermodynamic observables and relations.
\end{itemize}

The paper is organized as follows. In Section \ref{sec2} we review the setup,  the vacuum solutions,
and the results about confinement found in \cite{ihqcd}. We analyze the
possible types of singularity  and we give a more exhaustive
 analysis of this issue as compared to \cite{ihqcd}.

In Section 3 we describe the black-hole solutions their existence and uniqueness properties and
define the relevant thermodynamic quantities. We then compute the free energy difference
between the black-hole and vacuum solutions, as a function of entropy, temperature, and the
value of the gluon condensate.

In Section \ref{holotrace} we  show that, to lowest order in $\l$ as $\l\to 0$, the trace anomaly computed from the
equation of state matches the holographic computation of the the vev of $\tr F^2$.

In Section \ref{phasetrans} we prove that confinement at zero temperature is in one-to-one correspondence
with the presence of a phase transition at a finite temperature $T_c$.  We find
the explicit form of the   black-hole solutions in the two opposite regimes when the black-hole
horizon is very close to or very far from the UV boundary. We then show that in confining theories
there is always a finite, minimum black-hole temperature, whereas in non-confining theories black-holes exist with arbitrarily small temperatures.
This fact, together with the first law of black-hole
thermodynamics, is used to prove the main statement of this section in the  particular
case when there are only two black-hole solutions for each temperature. The proof in the most
general case is left to Appendix G.

In Section 6 we study the dynamics of the 5D axion, dual to the Yang-Mills vacuum angle,
 showing that above the critical temperature the axion profile is necessarily constant, and
the topological density has zero vev.

In Section 7 we define the {\em scalar variables}, and show that their use helps in
computing all the thermodynamic quantities, as well as the UV and IR asymptotic properties
of the solutions. In particular, in Subsection 7.2  we compute the near-boundary expansion
of the black-hole metric components and dilaton profile, with respect to the vacuum solution.
Section 8 contains a brief outlook.

Most technical details are left to the Appendices. In Appendix A we give details about Einstein's equations
in various frames, and the relation between fluctuations in different frames. In Appendix B we revisit
the case of a constant potential and derive $AdS$ and dilaton flow solutions and the corresponding
black-holes. In Appendix C we give the details of  the computation of the black-hole on-shell action
and ADM mass. In  Appendix D we provide the high-$T$ asymptotics of the gluon condensate. Appendix E is devoted
to the discussion of the general solution to the zero-temperature superpotential equation, and the classification
of zero-temperature singularities. In Appendix F we introduce the finite temperature generalization of the
superpotential, and use it to solve explicitly the black-hole equations for large $\l_h$. In Appendix G we
give the general proof  of the statement in Section 5, in the case where more than 2 black-holes exist
for certain temperature ranges. In Appendix H we provide some details of the computations with
scalar variables. In particular in Subsection H.4 we show that a black-hole
solution with a regular horizon  always connects with the UV boundary, thus providing
the proof of the existence of black-hole solutions for arbitrary $\l_h$. In Appendix I we give the
high-$T$ asymptotics of various quantities. Finally in Appendix J we show some interesting
analytical solutions of the system.

\section{Review of vacuum solutions}\label{sec2}

The holographic duals of large $N_c$ Yang Mills theory proposed in \cite{ihqcd} are
based on five-dimensional Einstein-dilaton gravity with a dilaton potential. The basic fields
for the pure gauge sector  are the
5D metric $g_{\mu\nu}$ (dual to the 4D stress tensor) and a scalar field $\Phi$ (dual
to $Tr F^2$). The action for these fields is taken to be\footnote{See Appendix A for our
sign conventions.}:
\begin{equation}
   {\cal S}_5=-M^3_pN_c^2\int d^5x\sqrt{g}
\left[R-{4\over 3}(\partial\Phi)^2+V(\Phi) \right]+2M^3_pN_c^2\int_{\partial M}d^4x \sqrt{h}~K.
 \label{a1}\end{equation}
Here, $M_p$ is the  five-dimensional Planck scale and $N_c$ is the number of colors.
The last term is the Gibbons-Hawking term, with $K$ being the extrinsic curvature
of the boundary. The effective five-dimensional Newton constant
is $G_5 = 1/(16\pi M_p^3 N_c^2)$, and it is small in the large-$N_c$ limit.

The vacuum solutions are of the form
\be\label{sol1}
ds^2 = b(r)^2 \left(dr^2 + \eta_{ij} dx^idx^j\right), \qquad \Phi = \Phi(r),
\ee
where the metric is written in conformally-flat coordinates.

The radial coordinate $r$ corresponds to the 4D RG scale.
In the holographic dictionary, we identify the 4D energy scale $E$ with the metric
scale factor,  $E=E_0 b(r)$, up to an arbitrary energy unit $E_0$. Also, we identify
 $\l\equiv e^\Phi$ with the  running 't Hooft  coupling $\l_t\equiv N_cg_{YM}^2$,
up to an {\it a priori}
unknown multiplicative factor, $\l = \kappa \l_t$.
 All  physical observables are independent of the parameter $\kappa$, as explained in Appendix \ref{l-frame}\footnote{More precisely, this statement applies to quantities that can be computed within the Einstein frame. Quantities that involve
the string frame metric may depend on $\kappa$ in a nontrivial way. In this paper we will not be concerned with any
such quantity, and we leave a more detailed discussion of this issue for an upcoming work \cite{GKMN3}.}.

With the above identifications, one can give a  holographic definition of the $\beta$-function
of the system in terms of the background solution:
\be
\beta(\l) =  {d\lambda\over d\log E} = \l {\dot{\Phi}\over \dot{A}},  \qquad A(r) \equiv \log b(r).
\label{beta1}\ee
Above and throughout this paper a dot stands for a derivative with respect to the radial (conformal) coordinate $r$.

With the ansatz (\ref{sol1}), Einstein's  equations are
\begin{equation}
6{\dot{b}^2\over b^2}+3{\ddot{b}\over b}=b^2 V,
\qquad
6{\dot{b}^2\over b^2}-3{\ddot{b}\over b}={4\over 3}\dot{\Phi}^2,
\label{eqs1}\end{equation}
The dilaton field equation is not
an independent equation, but it follows from (\ref{eqs1}).

It is sometimes useful to work with  the  {\it
domain wall coordinates}, in which the metric reads:
\be\label{ucoord}
ds^2 = du^2 + e^{2A(u)} \eta_{ij} dx^idx^j, \qquad  dr  = e^{-A(u)} du, \qquad b(u) = e^{A(u)}.
\ee
 In this frame Einstein's equations take the
form:
\be\label{eqs2}
3A'' + 12A'^2 =  V(\Phi), \qquad A'' = -{4\over 9} \Phi'^2.
\ee
where a prime denotes a derivative w.r.t. $u$.
In particular, the second equation implies that the scale factor
of an asymptotically $AdS_5$ spacetime (for which $A \sim -u/\ell$ as $u\to -\infty$) is monotonically decreasing, and it therefore provides a consistent
definition of the holographic energy scale.

 Einstein's equations can be put in first order form by defining a {\it superpotential} $W(\Phi)$, i.e.  one solution to the equation:
\be\label{super1}
 V(\Phi) =
-{4\over 3}\left({d W\over d\Phi}\right)^2 + {64\over 27} W^2.
\ee
With this definition, Einstein's equations (\ref{eqs2}) become:
\be
\Phi'(u) = {dW \over d\Phi} , \quad A'(u) =   -{4\over 9} W (\Phi).
\label{eqs1st}\ee

The system of eqs. (\ref{super1}-\ref{eqs1st}) has three integration
constants, one of which is an artifact due to reparametrization invariance\footnote{This can be seen
by choosing $\Phi$ as a coordinate: then the first equation in (\ref{eqs1st}) becomes vacuous, and only two first order equations remain.}.
The  solution is completely specified by a choice of  $W(\Phi)$,
up to an integration constant that consists in a simultaneous rescaling
of $b(r)$ and $r$, and only affects the overall scale of the system.
In other words, all nontrivial physics is encoded in $W(\Phi)$.

The general solution of eq. (\ref{super1}) is discussed  in detail in Appendix \ref{superapp}.
As we will  discuss at the end of this section, for any  $V(\Phi)$,
there is a {\em single choice} of $W(\Phi)$ that satisfies
some reasonable physical conditions. We are thus left with a one-parameter
family of solutions, distinguished only by a choice of scale. This
parallels the situation in the gauge theory.

Another useful reformulation of the Einstein's equations is in terms of the logarithmic derivative of $W(\Phi)$, which
is directly related to the $\beta$-function:
\be\label{X}
X(\Phi) = -{3\over 4}{d\log W \over d\Phi} = {\beta(\l)\over 3\l}
\ee
The complete solution of the system is encoded in this function. It is determined from the potential by solving a
first-order equation:
\be\label{X0eq}
\frac{dX}{d\f} = - \frac43(1-X^2)\le(1+\frac{3}{8X}\frac{d\log V}{d\f}\ri).
\ee
Once $X$ is known the scale factor and the dilaton are obtained from it by solving the first order equations
(\ref{Ap}) and (\ref{Fp}). Thus, this formulation reduces the Einstein equations to three first order equations.
In Section \ref{covvar}, we shall present a natural generalization of this formulation to the black-hole solutions.

The precise relation between $X(\f)$ and $W(\f)$ is given by,
\be\lab{relXW} W(\f) = \frac{9}{4\ell}
e^{-\frac43\int_{-\infty}^{\f}X(t) dt},  \ee 
where $\ell$ is the asymptotic $AdS$ length (see next subsection).  
This relation holds
both for the zero T theory and at finite T. In the latter case,
the functions $X$ and $W$ are replaced by their finite T
counterparts, as explained later.

The small-$\l$ and large-$\l$ asymptotic of $W(\l)$ (or $X(\l)$) determine the solution in the UV
and  the IR of the geometry, corresponding to the large- and small-$b$ regions,
respectively.

\subsection{UV asymptotics}

In the UV,  asymptotic freedom with logarithmically running coupling
 requires the background to be asymptotically Anti-de Sitter. The perturbative $\beta$-function,
$\beta \sim -b_0 \l^2 - b_1 \l^3 + \ldots$,
requires an expansion of $X$ in the form:
\begin{equation}\label{x0}
    X(\l) = -\frac{b_0}{3}\l - \frac{b_1}{3}\l^2 + \cO(\l^3)
\end{equation} where $b_0$ and $b_1$ are the $\b$-function coefficients.
Using (\ref{X}) one finds the expansion of the superpotential as,
\be\label{superexp}
W(\l) = {9\over 4 \ell} \left(1  + w_0\l + w_1 \l^2 +\ldots\right),
\ee
which implies  a potential of the form
\be\lab{UVpot}
V(\l) = {12\over \ell^2} (1 + v_0\l + v_1 \l^2 + \ldots).
\ee
Here $\ell$ is the $AdS$ length, and the dimensionless parameters $w_i, v_i$ are fixed in
terms of  the $\beta$-function coefficients. In particular the small-$\l$ expansion parameters $w_i$
of the superpotential are {\em universal}, and do not depend on the particular choice of solution
of eq. (\ref{super1}): as shown in Appendix \ref{superapp},  different solutions of eq. (\ref{super1})
differ by subleading non-analytic terms.
For a general potential (\ref{UVpot}), the $\b$-functions coefficients and the parameters of the potential
are related as follows:
\be\label{x0x1} b_0 =  \frac98 v_0=\frac94 w_0, \qquad b_1 = \frac94 v_1 - \frac{207}{256}v_0^2=\frac92 w_1 - \frac94 w_0^2.
\ee
Let us here also define the ratio,
\be\lab{b}
b = \frac{b_1}{b_0^2},
\ee
which will prove useful in what follows.
Note that $b$ is invariant under the rescaling  $\l\to\kappa\l$.

The UV region corresponds to $r\to 0$
in conformal coordinates, and the asymptotic
solution is given by:
\bea
&& b(r) = {\ell\over r} \left[1 +{4 \over 9} {1\over \log r \Lambda} - {4\over 9}~b~
{\log (-\log r\Lambda )\over \log^2 r\Lambda} +\ldots 
\right],\label{bUV} \\
 &&b_0 \l(r) =  -{1\over \log r \Lambda} + b~{\log(- \log r\Lambda )\over \log^2 r\Lambda} +\ldots
\label{lUV}\eea

The scale $\Lambda$ appearing in the expansion is the only physical integration constant, and
it is the holographic manifestation of the strong coupling scale in QCD perturbation theory. The UV
boundary conditions for the metric and dilaton are completely specified by the choice of this scale.
In practice $\Lambda$ is determined by a combination of
the initial conditions  of $\l$ and $A$, given at a point $r_0$ close to the boundary, as,
\be\lab{LQCD-2}
\Lambda~\ell = \exp\left[A(\l_0)-\frac{1}{b_0\l_0}\right](b_0\l_0)^{-b} + \cdots.
\ee
The ellipsis refer to contributions that vanish as one takes the cut-off away,
$\l_0\to 0$. The coefficients $b_0$ and $b$ are defined in (\ref{x0x1}) and (\ref{b}).
The derivation of (\ref{LQCD-2}) follows from Appendix \ref{XYeq} and the eqs. (\ref{bUV}-\ref{lUV}) above.

\subsection{IR asymptotics}

The IR properties such as confinement of the electric color charges (signaled
by an area law for the Wilson loop)  and the features
of the glueball spectrum are determined by the behavior of $W(\l)$  (or $X(\l)$) for large $\l$. In
particular, the Wilson loop  follows an area law if and only if  $W(\l)$ grows as $\l^{2/3}$ or
faster. The same condition ensures a mass gap in the spectrum. In other words
one has the criterion:

\begin{center}
{\it Confinement $\Leftrightarrow$ $W(\l) \geq O(\l^{2/3})$ as $\l \to \infty$.} \end{center}

The form of the IR geometry depends on the details of the asymptotics. The Einstein-frame scale
factor is guaranteed to decrease monotonically from the UV to the IR, and eventually
the spacetime terminates in a singularity at some $r=r_0$. We classify the singularity into
{\em good} and {\em bad} according to the following criterion \cite{ihqcd}: \\

{\em A good singularity is screened, i.e.  it is repulsive to physical modes.}\\

\noindent On the other hand, {\em bad} singularities are such that finite energy modes can probe arbitrarily
deep into the region close to the singularity. Typically this means that one needs to specify extra boundary
conditions  at the singularity, i.e. the information provided with  the classical action is not
enough to compute physical quantities. For good singularities, all (physical) boundary conditions
must be imposed in the  UV region. Therefore we believe
 that only ``good'' singularities have a meaningful holographic interpretation.

The most interesting  geometries
are those with the singularity at $r_0=\infty$, and with the asymptotics:

\be
b(r) \sim e^{-\left({r\over L}\right)^\a} , \qquad \l(r) \sim e^{3/2\left({r\over L}\right)^\a}
\left({r\over L}\right)^{{3\over 4}(\a-1)}, \qquad r\to \infty
\label{IR}\ee
Here, the length scale $L$ is set by the same integration
constant that fixes $\Lambda$ in eq. (\ref{bUV}).

In such solutions the curvature of the string-frame metric vanishes in the extreme IR.
These solutions occur
when  $W(\l)$ and $X(\l)$ behave for large $\l$ as:

 \be
W(\l) \sim \l^{2/3} (\log \l)^{{\a-1\over 2\a}}, \qquad X(\l) \sim-{1\over 2}  -
 {3\over 8}{\a-1\over \a} {1\over \log \l} +\ldots , \quad \l\to \infty,
\label{WX}\ee
which in turn requires the  potential to grow as

\be
V(\l) \sim \l^{4/3} (\log \l)^{{\a-1\over \a}}, \qquad \l \to \infty,
\label{potconf}\ee
{\em These solutions are confining iff $\a\geq1$}\footnote{Solutions such that $b(r)$  decays as a
power-law as $r\to\infty$ are  not confining.}.

The parameter $\a$ determines  the
 asymptotic spectrum of
normalizable fluctuations around the solution, which corresponds to the spectrum
of composite
states ({\it glueballs}) of the gauge theory, with masses that scale as:
\be
m_n \sim n^{(\a-1)/\a}.
\ee
For a linear glueball spectrum, $m_n^2\sim n$, one should choose $\a=2$.

The borderline confining case,  $\a =1$ has interesting properties:
the asymptotic geometry in the string frame reduces to flat space with
a linear dilaton. The spectrum has a mass gap and it is discrete up to a certain
energy level, above which it becomes continuous.
We will see that this case also has special thermodynamic properties.

Solutions with a singularity at a {\em finite value} $r_0$ of the conformal coordinate
are also confining, and correspond to $W(\l)$ growing as $\l^Q$ with $Q>2/3$.
Close to the singularity $r=r_0$ the scale factor vanishes as
\be
b(r) \sim (r_0-r)^\delta, \qquad Q = {2\over 3}\sqrt{1+\delta^{-1}}.
\ee
 The glueball spectrum has quadratic growth,
$m_n^2 \sim n^2$, as in the hard wall models.

The case $\delta<1$ ($Q> 2\sqrt{2}/3$) should
be discarded, since  in this case the singularity is a {\em bad} one
according to the our classification, i.e.
it is not screened from the physical fluctuations \cite{ihqcd}.

 An example of this type
(with $\delta=1/3$)  is
the ``dilaton flow'' solution of 5D Einstein-dilaton gravity
with a negative cosmological constant, discussed in \cite{Kehagias,DF}, which was argued
to be dual to an $SO(6)$ invariant mass deformation  ${\cal N}=4$ SYM.
Although this description can be adequate in the UV, calculation  of any
physical quantity   requires extra  knowledge about the
details of the singularity, which is not available in the classical gravity
approximation.

\subsection{The superpotential vs. the potential}

The action (\ref{a1}) is defined in terms of $V(\l)$, not $W(\l)$ or $X(\l)$. Therefore  it is important to know
what other large-$\l$ asymptotics for $W$ and $X(\l)$ can occur for a given  $V(\l)$.
 This problem is analyzed in Appendix \ref{superapp} (for $W$) and in Appendix \ref{fixedXY} (for $X$), where the
form of the general solution of eqs. (\ref{super1}) and (\ref{X0eq})
 is discussed in detail.  Essentially, for any given monotonic $V(\l)$,
a solution with a good infrared singularity, if it exists, is  unique.

In the UV region, $\l\to 0$, all solutions to eq. (\ref{super1})
 have the same expansion, given by  eq. (\ref{superexp}) with the same
coefficients $w_i$. In other words, in the UV  all solutions to Einstein's equations
 flow to the same log-corrected $AdS$ (eqs (\ref{bUV}-\ref{lUV})).

In the IR, the situation is more complicated.
 We  consider a potential $V(\l)$
defined over the whole range $0< \l<\infty$, and such that for large enough $\l$
it is well approximated by the form:
\be\label{asregion0}
V(\l) \simeq V_{\infty} \, \l^{2Q} (\log \l)^P
\ee
 for some real $P$ and $Q$\footnote{{Although we parametrize the IR asymptotic in this particular form,
all our discussion
also applies to any potential that has  intermediate IR growth between two values of $Q$ or $P$
that share the same behavior, for example $V(\l)\sim \l^Q (\log \l)^P (\log \log \l)^R\ldots $.
However if $Q$ or $P$ take values that mark the boundary between two different
qualitative behaviors of the solution (e.g. $Q=4/3$, see below),
extra work  is needed to understand
the intermediate asymptotics.}}. We will assume $V(\l)$ is  a positive,
 monotonic function,  to avoid the presence of conformal
fixed points at finite $\l$. Thus, we take $V_{\infty}> 0$ and $Q\geq 0$.

All the interesting physics is found for $Q\leq 4/3$.
As shown in Appendix \ref{superapp}, if $Q\leq 4/3$,  there exist three classes of solutions to the
superpotential equation:
\begin{enumerate}
\item {\bf Special:} A {\em single} solution such that $W(\l) \sim \sqrt{V(\l)}$ for $\l\to \infty$.
\item {\bf Generic:} A continuous family  with leading asymptotics
\be \label{bad}
W (\l) \simeq C  \l^{4/3} \qquad \l \to \infty
\ee
where $C$ is an arbitrary constant.
\item {\bf Bouncing:}  A continuous family which never
reaches the asymptotic large-$\l$ region: the variable  $\l$ attains
a maximum value $\l_*$, then decreases again to zero towards a region
where
\be
W \simeq \tilde{C}\l^{-4/3}, \qquad \l \to 0,
\ee
\end{enumerate}

On the other hand, if $Q>4/3$
only the bouncing  solution exists\footnote{This includes the case when the potential grows faster than
(\ref{asregion0}) for any $Q$, e.g. if $V(\l)\sim e^{c\l}$.}, and the dilaton never reaches infinity.

The special solution has asymptotic behavior:
 \be\label{W0as}
W_{o} (\l) \simeq W_{\infty}\l^Q\,(\log \l)^{P/2} 
 \qquad \l \to \infty, \qquad  W_{\infty} = \sqrt{27 V_\infty\over4(16-9Q^2)}.
\ee
It presents  a  good IR singularity for $Q< 2\sqrt{2}/3$, and
it is confining for $Q>2/3$, or $Q=2/3$ and $P>0$.
The generic and  bouncing solutions, on the other hand, always have bad singularities.

As we shall discuss in Section 5,
the special solution is also the only  one  that can be obtained in the zero-mass limit
of black-hole solutions of the same bulk theory.  This gives another characterization
of the special solution, and singles it out as the only physically
sensible choice \cite{gub2}.

Thus, given a potential $V(\l)$   with asymptotics  (\ref{potconf}),
there is a {\em single} solution with the large-$\l$ behavior (\ref{WX}) corresponding to a ``good'' singularity.
All other solutions
have  {\em bad} singularities in the IR, and  cannot be lifted to black-holes. Requiring
the absence of bad singularities  is what ultimately fixes the integration constant
of the $W$ equation in the zero-temperature system, or equivalently, the integration constant of (\ref{X0eq}).

As shown in Appendix \ref{SuperappUV},  changing this integration constant
adds a perturbation in the metric and dilaton that goes
 as $r^4$ close to the boundary.  Thus, this corresponds to changing the expectation value of
 the corresponding dimension $4$ operator, i.e. $\tr F^2$. In other words, the integration constant
in the superpotential controls the value of   $\< \tr F^2\> $ in the gauge theory and there is a unique
value such that no bad singularities appear.
\footnote{This situation has an analogue in
the case of  constant  potential, $V(\l) = 12/\ell^2$: also in this  case there is a single ``good''
solution,  $AdS_5$ spacetime  with constant dilaton.
Even in this
theory  it is well known \cite{Kehagias,DF}
that there is a continuous family of  solutions with a running dilaton, all of which have
bad singularities in the interior.
This is presented  in detail both at zero and finite temperature in appendix \ref{ads}.}.

To summarize: the physically
interesting situation when a good solution exists  {\em and} corresponds to confined color, is
the case $2/3\leq Q \leq 2\sqrt{2}/3$.

\section{Finite-temperature  solutions  and thermodynamics}

We now consider the dilaton-gravity system described in the previous section with a good potential
according to the aforementioned criteria and study it
at  finite temperature $T$. As usual, this can be implemented
by going to Euclidean signature and compactifying the Euclidean
time (that for simplicity of notation will be still called $t$) on
a circle with period $\beta= 1/T$. This breaks the Poincar\'e invariance
of the vacuum  to spatial rotations, and allows for a larger
class of solutions. According to the $AdS$/CFT prescription, the partition
sum is constructed by considering all solutions with fixed UV boundary
conditions\footnote{Later in this section we will be more specific about what we mean
by ``fixed UV boundary conditions.''}.
From now on we will introduce a subscript ``$o$''
for the quantities related to the zero-temperature solution.

The thermal solutions  are of two types:
\begin{enumerate}
\item {\bf Thermal gas solution:} this is the same  as (\ref{sol1}),
\be\label{thermal}
ds^2 = b^2_o(r)\left(dr^2 - dt^2 +  dx_mdx^m\right), \qquad \Phi = \Phi_o(r),
\ee
except for the identification $t\sim t + i\beta$.  It corresponds to  a gas of
thermal  excitations above the same vacuum described by the original solution,
from which  it inherits all the non-perturbative features (confinement,
spectrum,  values of condensates, etc. )

\item {\bf Black hole solutions:} they are of the form
\be
ds^2=b(r)^2\left[{dr^2\over f(r)}-f(r)dt^2+dx_mdx^m\right], \qquad \Phi = \Phi(r)
 \label{a7}\ee
and are characterized by the presence of a horizon $r_h$ where $f(r_h)=0$. This
implies that such a solution, if it exists, corresponds to a non-confined
phase, since the confining string tension is proportional
to ${\rm Min}_r(\sqrt{g_{xx}(r)g_{tt}(r)}) =0$  \cite{sonnenschein}. In the Euclidean
version, deconfinement is signaled by a non-zero value of the Polyakov loop,
as discussed in \cite{D4}.
Since we want to study the
theory on $S^1 \times R^3$ we consider black-holes with flat horizon topology.

Notice that in general
the functions $b(r)$
and $\Phi(r)$ appearing in eq. (\ref{a7}) are different from
their zero-temperature counterparts, and have also a nontrivial temperature
dependence.

\end{enumerate}

In the rest of this section we will discuss the features of the black-hole
solutions to the general system (\ref{a1}), and the thermodynamics
in the canonical ensemble.

\subsection{5D Einstein-dilaton Black holes}

We require that the solution has the same UV asymptotics as the one at zero temperature:
an $AdS$ boundary at $r=0$ where $b(r) \sim \ell/r$ and $e^\Phi$
vanishes logarithmically; we have to impose $f(0)=1$, so that the black-hole solution (\ref{a7})
coincides with the zero-temperature and  thermal gas solutions, (\ref{thermal}) in the UV limit $r\to 0$.

A black-hole solution with a regular horizon is characterized by the existence
of a surface  $r=r_h$, where the dilaton and scale factor are regular, and
\be
f(r_h) = 0 , \quad\dot{f}(r_h) < 0.
\ee
The Euclidean version of the solution is defined only for $0<r<r_h$.
The  horizon $r=r_h$ is a regular surface if Euclidean time is identified
as $\tau \to \tau + 4\pi/|\dot{f}(r_h)|$. This determines
the temperature of the solution as:
\be \label{temperature}
T = {|\dot{f}(r_h)|\over 4\pi}.
\ee

The independent field  equations are:
\begin{equation}
6{\dot{b}^2\over b^2}-3{\ddot{b}\over b}={4\over 3}\dot{\Phi}^2 \sp {\ddot{f}\over \dot{f}}+3{\dot{b}\over b}=0,
\label{a11}\end{equation}
\begin{equation}
6{\dot{b}^2\over b^2}+3{\ddot{b}\over b}+3{\dot{b}\over b}{\dot{f}\over f}={b^2\over f} V.
\label{a12}
\end{equation}
Integrating once the second equation of (\ref{a11}), we obtain:
\be\label{a17b}
\dot{f} =- {C\over b^3},
\ee
for some integration constant $C$. This  shows that $\dot{f}$ cannot change
sign. For a black-hole, $f(r)$  has to decrease from $f=1$ at the boundary to
$f=0$ at the horizon, therefore  $C>0$.

The general solution for $f$ is
\be
f(r)=1 - C\int^{r}_{0}{dr'\over b(r')^3},
 \label{a18}\ee
where we have chosen the second integration constant so that $f(0)=1$\footnote{Recall that
$b(r) \sim r^{-1}$ as $r\to 0$, so the second term in eq. (\ref{a18}) vanishes
at the boundary.}. Setting $C=0$ and $f(r)=1$ we recover the zero-temperature Einstein's equations.

The quantity $C$ is related to
the horizon location as:
\be
C = {1 \over  \int_{0}^{r_h}{dr'\over b(r')^3}}
 \label{C}\ee

Note that $b(r)$ is regular in the whole region of integration.
We can compute the temperature by eq. (\ref{temperature}):
\be
\beta={1\over T}={4\pi\over |\dot{f}(r_h)|}=4\pi  b^3(r_h)\int_{0}^{r_h}{du\over b(u)^3} = {4\pi  b^3(r_h)\over C} .
 \label{T}\ee

The horizon area is given by
\be
{\cal A}(r_h) =   b^3(r_h)  V_3,
\label{area1} \ee
where $V_3$ is the volume of 3-space,
and it is related to the entropy as usual by $ S = {\cal A}/4G_5$.

In the particular case  $V(\Phi)= 12/\ell^2$, $\dot{\Phi}=0$,
 we have the $AdS$-Schwarzschild solution
in Poincar\'e coordinates,
\be\label{Adssol}
b(r) = {\ell\over r}, \qquad f(r) = 1 - \left(r\over r_h\right)^4, \qquad T = {1\over \pi r_h}, \qquad {\cal A} = \left({\ell\over r_h}\right)^3V_3.
\ee
Notice that in this case the scale factor is temperature-independent. This is not true
in general: for $V$ depending non-trivially on $\Phi$, different  $f(r)$ will result in
different $b(r)$.

Near the AdS boundary (UV), this difference can be made more precise:

\begin{itemize}
\item
As shown in Appendix \ref{superfinTUV}, near the AdS boundary
$b(r)$ and $\l(r)$ have $AdS$ asymptotics, an  expansion in inverse logs of the same form as
 $b_o(r)$ and $\l_o(r)$,  eqs. (\ref{bUV}-\ref{lUV}), specified by an integration constant $\Lambda$.
In particular, $\l(r)\to 0 $ in the UV for all the solutions.
Fixing the UV boundary conditions therefore, means specifying the scale $\Lambda$ appearing
in this expansion.  \\

{\em Here, and from now on, by stating that two solutions obey the {\em ``same UV boundary conditions''}, we require that {\em ``the
scale $\Lambda$ appearing in the expansion in the perturbative UV log  must be the same''}}\footnote{As it should
be clear, it is not enough to specify that the metric be asymptotically $AdS$ and that $\l(r)$ asymptotes
some fixed value as $r\to 0$, since for all solutions $\l(0)=0$.}. \\

\item  Assuming for $b(r)$ and $\l(r)$ the {\em same} value of the integration constant $\Lambda$, as for
$b_o(r)$ and $\l_o(r)$, then:

\bea
&& b(r) = b_o(r)\left[1 + \,{\cal G}\, {r^4\over \ell^3} +\ldots\right],  \qquad r \to 0,
\label{b-bo} \\
&& \l(r) = \l_o(r)\left[1 + \,{45\over 8}{\cal G}\,{r^4\log \Lambda r \over \ell^3} +\ldots\right],  \qquad r \to 0\label{l-lo}\\
&& f(r) = 1 -{C\over 4} {r^4\over \ell^3} + \ldots \qquad r\to 0, \label{fUV}
\eea

where $C$ is defined in (\ref{a17b}), and ${\cal G}$  is a temperature-dependent constant with the
dimensions of energy.
Eq. (\ref{fUV}) is obtained  from the expression
(\ref{a18}) and the fact that $b(r)\to \ell r^{-1}$ as $r\to 0$;
eqs. (\ref{b-bo}-\ref{l-lo})   will be derived explicitly in Section \ref{soluv}.
\end{itemize}

According to the standard rules of the correspondence, the quantity ${\cal G}(T)$
 is interpreted (up to a multiplicative constant, to be determined later)
as the difference between the vev's  of the
corresponding dimension-four operator in the black-hole and
in the vacuum solution. Since we have assumed that $\Phi$ couples to $Tr F^2$ as $\int e^{-\Phi}Tr F^2$,  the appropriate
 operator is the gluon condensate $ \l^{-1} Tr F^2$. The precise relation between
 ${\cal G}$ and $\< Tr F^2\>$ will be obtained in Section \ref{holotrace}.

\paragraph{Integration constants} The quantities $C$ and ${\cal G}$ are related to two of the five
independent integration constant of the system of Einstein's
equations. For $C$, this is clear from its definition, eq.
(\ref{a17b}): it determines the temperature of the black-hole.
${\cal G}$  can be regarded as the integration constant for the
thermal generalization of the superpotential equation, given  in
Appendix \ref{superfiniteT}. In the zero-temperature case  it was
fixed
to single out the special solution, with the
``good''  IR behavior;
  in the black-hole it is fixed by the
requirement of regularity of the horizon\footnote{{In our analysis we cannot determine uniquely what is
the value of the gluon condensate in a given background, but only the differences between $\tr F^2$
in two backgrounds with the same asymptotics. In order to compute the v.e.v. of the gluon condensate
unambiguously in a given background one would need to perform the full procedure of holographic renormalization,
which for the asymptotics we are considering is not yet fully developed but will be available soon, \cite{gupa}.}}.
Two more integration
constants are fixed by setting $f(0)=1$, and by choosing the scale
$\Lambda$ in the UV perturbative expansion, i.e. by requiring that
the solution has the same boundary behavior as at $T=0$. The last
integration constant, as in the $T=0$, is unphysical and is due to
reparametrization invariance. It can be eliminated by rewriting the
solution using $\l$ as a coordinate. As we show in Appendix
\ref{counting}, a better way of counting integration constants is
by giving ``initial values'' directly at the horizon. This results
in the following statement:\\

\noindent
{\em For any positive and monotonic potential  $V(\l)$ that grows no faster
\footnote{This restriction  is necessary to ensure that (vacuum and black-hole) solutions
 that extend to arbitrarily large values of $\l$ exist.} than $\l^{4/3}$ as $\l\to \infty$,
 and for any value $\l_h$,  there exists  one and only one black-hole such that:
\begin{enumerate}
\item $\l \to \l_h$ at the horizon
\item it has the same UV asymptotics as the zero-temperature solution (\ref{bUV}-\ref{lUV})
\end{enumerate}
}

The existence, for each $\l_h$,  of a black-hole solution with regular
horizon that extends all the way to the UV $AdS$ boundary, is shown in Appendix \ref{conUVIR}
using the method of scalar variables. Uniqueness, on the other hand, follows
from the discussion in Appendix \ref{counting}.

Thus,  the value of the dilaton at the horizon,  $\l_h$, is the most
natural candidate to classify the black-holes, and all the thermodynamic quantities like
e.g. temperature and entropy 
which are single  valued functions of $\l_h$ (the same does not necessarily hold if
one writes them as a function of the horizon position, i.e. $T(r_h)$ is not necessarily
single valued. We have found numerically examples of this behavior).

\subsection{Thermodynamics}

In this section we compute the free energy differences  of  various
solutions at a given temperature. This will
allow us to compute  all other thermodynamic quantities. The details  of the relevant calculations
 can be found in Appendices \ref{onshell} and  \ref{free-en}.

The free energy at fixed temperature $\b^{-1}$ of a given solution
is given by:
\be \label{th0}
\beta {\cal F} = {\cal S}^{\e},
\ee
where  ${\cal S}^{\e}$ is the regularized Euclidean action evaluated on the solution. The action needs
to be regularized, due to the divergences near the $AdS$ boundary. To achieve this,
we take 3-d space  to be a torus with finite  volume $V_3$, and  cut-off the radial direction in the UV up to
a minimum radius $r=\e >0$, so  all the integrals are limited below by $\e$.
Free-energy differences will be finite (and proportional to $V_3$) as $\e \to 0$ since
 the large-volume divergences do not depend on the detailed solutions but only on the asymptotics.

As shown in Appendix \ref{onshell}, the regularized  Einstein action (\ref{a1})
 evaluated on a black-hole solution  is:

\be\label{actionbh}
{\cal S}^{\e}={\cal S}^{\e}_{E}+{\cal S}^{\e}_{GH}+ {\cal S}^\e_{count}=2\beta\,\sigma\,
\left[3b^2(\e)f(\e)\dot{b}(\e)+{1\over 2}\dot{f}(\e)b^3(\e)\right] + {\cal S}^\e_{count},
\ee
where we have defined:

\be\label{sigma}
 \sigma\equiv M^3_p N_c^2 V_3.
\ee
The counterterm action $ {\cal S}^\e_{count}$ is required to make the above expression finite in the
limit $\e\to 0$ \cite{skenderis}. As we can see, eq. (\ref{actionbh}) depends solely on the metric evaluated  at the UV
cutoff: the contribution to the bulk Einstein action coming from the horizon region vanishes.

Instead of explicitly calculating ${\cal S}^\e_{count}$, we can define
the free energy by subtracting a given reference background, following the prescription of \cite{HH}.
We should take as reference
the thermal gas background with the same temperature and the same $\Lambda$ as the black-hole,
obtained by setting $f(r)=1$, and replacing $b(r)$ with $b_o(r)$. However, this is correct for the
unregularized action that extends all the way to $r=0$. When we deal with regularized geometries,
we must make sure that \cite{HH}
\begin{itemize}
\item[i)] the intrinsic  geometry of the 4-dimensional boundary be the same
for the two solutions,
\item[ii)] the boundary values of scalar field, $\l$ and $\l_o$ are the same.
\end{itemize}

To satisfy  i),   we must demand that
the proper lengths of the time circles of the solutions  (\ref{thermal})
and (\ref{a7}) and the proper volume of 3-space  be the same at $r=\epsilon$.
Denoting
by $\tilde{\beta}$ and $\tilde{V}_3$ the period of the time coordinate and the volume of 3-space
in the thermal gas case,  this requirements imply:
\be\label{diffbeta}
\tilde{\beta}\,b_o(r) \Big|_{cut-off}=\beta ~b(r)\sqrt{f(r)}\Big|_{cut-off}, \qquad \tilde{V}_3\,b_o^3(r)\Big|_{cut-off}=V_3 ~b^3(r)\Big|_{cut-off}.
\ee

The condition ii) means that we must require $\l(\e)=\l_o(\e)$. This actually implies that the two backgrounds
are characterized by  {\em different} values $\Lambda$ and $\tilde{\Lambda}$ of the scale
that defines the UV boundary conditions. This makes
the calculation of differences such as $b(\e)-b_o(\e)$ quite complicated, since one cannot use directly
the UV expansion (\ref{b-bo}), which relies on the two scales being equal. However, since we are dealing
with only one scalar field, we can  equivalently keep $\tilde{\Lambda}=\Lambda$ and set the cut-off of the
background at a {\em different} location $r=\tilde{\e}$,
such that\footnote{This strategy will not work in the case of multiple bulk  scalar fields.}
\be\label{diffl}
\l(\e) = \l_o(\tilde{\e}).
\ee
Since the cut-off coordinates  do not coincide in the two solutions, in the  conditions
(\ref{diffbeta}) one now has to evaluate each side of the equality at the appropriate
value of the coordinate $r$.
 Now we can
use eq. (\ref{l-lo}) to determine the needed shift in the boundary positions:
\be\label{deltaeps}
\tilde{\e} - \e =- {45\over 8}{{\cal G}\over \ell^3} {\e^4\over \dot{\l}_o(\e)} = - {45\over 8}{{\cal G}\over \ell^3} \e^5 (\log \e \Lambda)^2.
\ee

The regularized action for the thermal gas background, in the case that the IR singularity is of the good type,  is
given by:
\be\label{actionth}
\tilde{\cal S}^{\tilde{\e}} = 2\tilde{\beta}\, \tilde{\sigma} \,\left[3b_o^2(\tilde{\e})\dot{b}_o (\tilde{\e})\right] +   {\cal S}^{\tilde{\e}}_{count}, \qquad \tilde{\sigma} =  M^3_p N_c^2 \tilde{V}_3.
\ee
As for the black-hole,  eq. (\ref{actionth}) only receives contributions from the metric evaluated at the UV cut-off.
Indeed, evaluating the Einstein term on shell in general gives an extra negative term localized in the IR,
of the form ${\cal S}_{IR} = 2\tilde{\beta} \tilde{\sigma}b_o^2(r_0) \dot{b_o}(r_0) \leq 0$, where $r_0$ is
the position of the singularity in the vacuum background. {\em This contribution vanishes exactly, if $r_0$
is a good singularity}. This is good news, since this means that  the details of the physics of the
singular region are irrelevant for the calculation of the free energy (as for other physical quantities \cite{ihqcd}),
and only the UV data matter. Similarly a Gibbons-Hawking term at the singularity (which
in principle could be there) also does not give any new contribution, since on shell it
is also proportional to $b_o^2(r_0) \dot{b_o}(r_0)$ and it vanishes for good singularities.\footnote{
For a recent  example of a similar study,
  where the free energy does receive contributions from the deep IR, see \cite{Evans}.}

Therefore the   free energy (difference)  ${\cal F}$ is:
\be\label{diff} \beta {\cal F} =
\lim_{\e\to 0} ({\cal S}_{\e}-\tilde{\cal S}_{\tilde{\e}}).
\ee
In the difference above, it is {\em guaranteed} that the contribution
coming from the counterterms exactly cancel even at finite
$\epsilon$, since these terms are build out of invariants of the
induced boundary geometry and the boundary values of the scalar
field\footnote{{This cancelation may not hold in the case
of the counterterms that diverge as $\log \epsilon$, i.e. those that give rise to
the conformal  anomaly. However, the free energy difference due to
these counterterms, if any, is of the order $\epsilon^4 \log \epsilon$. Therefore
these terms do not result in any finite contribution to $\Delta {\cal F}$
}}. Therefore this subtraction prescription makes it
unnecessary to know the explicit form of $S_{count}$.

Using the results (\ref{actionbh}) and (\ref{actionth})
in  (\ref{diff}), together with the relations (\ref{diffbeta}),
we have:

\be\label{free}
{\cal F}=  \sigma \, \lim_{\e\to 0} \,\left\{ 6 b^2(\e) \sqrt{f(\e)} \left[\dot{b}(\e) \sqrt{f(\e)}  - {b^2(\e)\over b^2_o(\tilde{\e})}\dot{b}_o(\tilde{\e})\right] + \dot{f}(\e) b^3(\e)\right\}
\ee

Using  the UV expansions (\ref{fUV}-\ref{b-bo}) and the relation (\ref{deltaeps}) n eq (\ref{free}), and taking the limit $\e\to 0$,
we obtain  the final  result for the free energy,

\be\label{free-2}
{\cal F}=\sigma \left( 15\,{\cal G}(T) - {C\over 4}\right) =   15 \,\sigma \,{\cal G}(T) - {1\over 4} TS,
\ee

where the entropy $S$ is given by  the area of the horizon:
\be\lab{entropy}
S={{\cal A}\over 4G_5}=4\pi \sigma b^3(r_h) = \sigma \frac{C}{T}.
\ee
In the second equality we have used eq. (\ref{sigma}) and $G_5 = 1/(16\pi M_p^3 N_c^2)$, and in the third one
eq. (\ref{T}).

The black-hole energy $E$ can be obtained either by the thermodynamic formula
$E = {\cal F}+TS  = - \de_\b(\b {\cal F}) $, or by computing the ADM mass,
and consistency requires that the two computations give the same result. This is indeed the case: as shown
in Appendix \ref{ADM}, the black-hole mass is given by

\be\label{bhenergy}
E = \sigma \left(15 \, {\cal G} + {3\over 4}\,C\right)  = {\cal F} + TS.
\ee

The presence of the gluon condensate term in eq. (\ref{free-2}) is the source of  the breaking of
conformal invariance, since in a conformal theory the relation
${\cal F} = - T S/4$ is exactly satisfied. As we will see in Section \ref{holotrace}, the holographic
computation of the conformal anomaly in flat space matches the field theory result.

Notice that second term in eq. (\ref{free-2}) is negative for any $T$.
It is the  presence of a non-trivial gluon condensate that may  allow
a change in sign of the free energy, corresponding to a first order
phase transition. While the calculation  of ${\cal G}(T)$, in general,
 is only possible numerically, in Section \ref{phasetrans} we will give
a general argument to determine whether or not a given Einstein-dilaton
theory exhibits a phase transition at some critical temperature.

Finally, we note that the  Gibbons-Hawking term contributes importantly to this expression. A simple calculation shows that
the  Einstein term in the action contributes as $\cF_E = -5\sigma{\cG} - T~S/4$ whereas the Gibbons-Hawking contribution
is $\cF_{GH} = +20\sigma{\cG}$. Note that the Einstein term itself is negative-definite, therefore the GH term is crucial
for the existence of a phase transition. { This is {\it unlike} the usual  Hawking-Page phase transition
in global $AdS$, where the change in sign of the  free energy is due solely to the Einstein term. }

\section{The holographic conformal anomaly (in flat 4-space)}\label{holotrace}

The expectation value of the gluon condensate plays an important role in the thermodynamics
of the system we are investigating. As we will see,  it can be related to
the thermal version of the anomalous trace of the  stress tensor. In this section we will  show
that  a holographic calculation of the trace anomaly in flat space
to lowest order in $\l$ matches the
four-dimensional result as advocated in \cite{ihqcd}. This is a non-trivial
check of the validity of the gauge/gravity duality in our setup, in particular of
our identification of the holographic $\beta$-function.

In four-dimensional Yang-Mills theory, breaking of scale invariance is expressed by the operator
equation:
\be\label{traceop}
T^\mu_\mu =  {\beta(\l_t)\over 4\l_t^2} \tr F^2,
\ee
where $\l_t$ is the 4D  't Hooft coupling.

Defining the pressure $p$  and energy density $\rho$,
\be
p = -{{\cal F} \over V_3}, \qquad \rho = {{\cal F} + TS \over V_3},
\ee
the trace  of the thermodynamic stress tensor can be obtained immediately from  eq. (\ref{free-2}):
\be\label{tracean}
\langle T^\mu_\mu\rangle _R  = \rho - 3 p = 60 M^3_p N_c^2 {\cal G}(T),
\ee
where we have used the definition of $\sigma$, eq. (\ref{sigma}).
The left hand side of (\ref{tracean}) is the trace of the renormalized  thermal
stress tensor, $\<T^\mu_\mu\>_R =\<T^\mu_\mu\>-\<T^\mu_\mu\>_o $, and it is proportional to ${\cal G}\sim \<\tr F^2\> $,
in qualitative agreement with (\ref{traceop}).

To check the detailed agreement between  eqs. (\ref{traceop}) and (\ref{tracean}),
 we need to derive the precise
relation between  ${\cal G}$ and $\<\tr F^2\>$. In what follows we work to lowest order in $\l$.
To compute $\<\tr F^2\>$ holographically,  we use the prescription
established in \cite{KlebanovWitten}:
we recall that
for any fluctuation of a bulk scalar with a canonically normalized
kinetic term,
\be\lab{cannor}
S_{fluc} = \half\ell^{-3}\int\sqrt{g}d^5x (\6\chi)^2,
\ee
and coupling to a boundary operator $\cO(x)$ as
\be
S_{coup} = \int d^4 x \chi \cO,
\ee
one can read off the vev of the operator $O(x)$  from the  UV asymptotics of $\chi(x,r)$:
if the UV expansion is of the form
\be\lab{UVfluc}
\chi \approx r^{\Delta_-} \chi_0(x) + r^{\Delta_+} \chi_1(x),
\ee
with $\Delta_+$ being the canonical dimension of the dual operator, and $\Delta_-=d-\Delta_+$ for a $d$-dimensional
gauge theory,  then the vev of the operator is given by the formula:
\be\lab{VEV}
\langle \cO(x)\rangle = \le(2\Delta_+ - d\ri) \chi_1(x) .
\ee

Now, we apply this prescription  to the dilaton fluctuation. In our setup the dilaton $\Phi$  is related to the
't Hooft coupling $\l_t$ by $e^\Phi \equiv \l = \kappa \l_t$\footnote{
We will keep the multiplicative constant $\kappa$ unspecified, as it will drop out of the calculation,
i.e. matching of the anomalies  does not depend on the value of $\kappa$.}.
The coupling of the dilaton to the YM field strength is given by
\be\lab{YMac1}
S_{coup} = -\int d^4x\, {1\over4 \l_t} \tr~F^2 = -{\kappa\over 4}\int  d^4x\,  e^{-\Phi}\,\tr~F^2  .
\ee
Thus, the dilaton fluctuation couples as,
\be\lab{YMac2}
\delta S_{coup} = {\kappa\over 4}\int  d^4x\, \delta\f \, e^{-\f}\,\tr~F^2.
\ee

From the bulk action (\ref{a1})  we learn that $\delta\f$  and the canonically normalized fluctuation
 (\ref{cannor}) are  related by\footnote{In doing this computation one should actually use a gauge-invariant
  fluctuation. This can be defined as $\delta\Phi_{G.I.} = \delta\Phi - \dot{\Phi}/\dot{A} \delta \psi$,
where $\psi$ is the part of the metric fluctuation
that couples to $T^\mu_\mu$, and is proportional to $\delta A$. However, close to the UV boundary,
$\delta \psi \sim r^4$, $\delta \Phi \sim r^4\log r$ and $\dot{\Phi}/\dot{A} \sim (\log r)^{-1}$,
therefore $\delta\Phi_{G.I.} \to \delta\Phi$ as $r\to 0$.}
\be\lab{cannor2}
\chi = \le(\frac83 (M_p~\ell)^3~N_c^2 \ri)^{\half}\delta\f.
\ee
Using   the relation $e^{\f} = \kappa\l_t$ we find the coupling
\be\lab{opr}
\int d^4 x \, \chi \,\cO =\int  d^4 x \, \chi\left[ {\kappa\over4 \l}  \le(\frac83 (M_p\ell)^3 N_c^2 \ri)^{-\half} \tr~F^2\right].
\ee

The particular dilaton fluctuation  we are interested  is the difference $\delta \Phi = \Phi-\Phi_o$: this allows to compute the
difference between the thermal and vacuum values of  $\langle \tr~F^2(0) \rangle$. Notice that this difference is
well defined, contrary to e.g. $\langle \tr~F^2(0) \rangle_o$, which suffers from UV ambiguities. Moreover, it is a purely {\em  normalizable}
fluctuation close to the boundary (i.e. it only contains terms like the  second one in (\ref{UVfluc}), since we assumed that the  black-hole
and vacuum backgrounds obey the same UV boundary conditions.

The fluctuation $\delta \Phi$ is  obtained from eq. (\ref{l-lo}).
To leading order in powers of  $\l$ (or equivalently in inverse powers
of $\log (r\Lambda)$):
\be\label{deltaphi}
\delta \Phi \simeq  {45\over 8} \,\cG\,{r^4\over \ell^3}  \log (r \Lambda)  ,
\ee

Recalling  the leading dilaton asymptotics $(b_0 \l)^{-1}  = -\log(r~\Lambda)$, and using (\ref{cannor2}) and  (\ref{deltaphi}) together with
(\ref{VEV}) and  (\ref{opr})  yields the  relation between $\cG$ and $\tr~F^2$:
 \be\lab{conds}
\langle \tr~F^2 \rangle_T - \langle \tr~F^2 \rangle_o  = -\frac{240}{\kappa b_0} M_p^3 N_c^2\,\, \cG .
\ee

Eq. (\ref{conds}) is the holographic computation of the vev of  the  r.h.s. of  the trace identity (\ref{traceop}), to leading order
in the expansion in $\l$. Using the expansion $\beta(\l) = - b_0 \l^2+\ldots$ together with the relation $\l = \kappa \l_t$, we obtain from (\ref{conds}):
\be\lab{anomaly}
{\beta(\l_t)\over 4\l_t^2} \langle \tr F^2\rangle_R = 60 M_p^3 N_c^2\,\, \cG .
\ee
which exactly matches the l.h.s, eq. (\ref{tracean}). This is a nontrivial consistency check of our setup.

Notice that the matching of the conformal anomaly  is independent
of both $\kappa$ and $b_0$, i.e.  it is insensitive to the relative normalization of the dilaton field
$\l$ with respect to the true 4D Yang-Mills coupling, $\l_t$.

\section{Thermal phase transitions}\label{phasetrans}

In this section we will derive the building blocks necessary to obtain
the main theoretical result of this paper:
{\em
 confining backgrounds exhibit a first order phase transition}, whose features
precisely mimic those
of the large-$N_c$ Yang-Mill deconfinement
 transition. We also show the converse:
 non-confining theories do not  have a phase transition at finite $T$.

The primary information we will need to extract is the dependence of the
temperature on the horizon position $r_h$, or better on the value of the
dilaton at the horizon, $\l_h$. Although this is difficult to obtain
for generic $\l_h$, it is nevertheless  possible to determine
the asymptotic form of
$T(\l_h)$ for $\l_h$ very small and very large
(in a sense that will be defined below), corresponding to very large and very small horizon area, respectively.

\subsection{Horizon in the UV region}\label{UVasymps}

As we have  discussed in Section 3,  the black-hole metric approaches  asymptotically the
$AdS$-Schwarzschild  solution for small $r$, with $b(r)\simeq \ell/r$ and $f(r)\simeq 1 - (C/4) r^4/\ell^3$,
and with a logarighmically running dilaton, $\l(r) \simeq (-b_0\log r \Lambda)^{-1}$.

For a large enough value of the constant $C$,  $f$ vanishes at a small enough $r$
such that this approximation   is valid all the way to the horizon.
In this case, $f(r) $ vanishes at $r_h^4 \simeq 4 \ell^3 C^{-1}$. This approximation
gets better for smaller values of $r_h$ (i.e. horizon closer to the
$AdS$ boundary).
 Then, the temperature and entropy  are  given approximately by the
$AdS$ formulas (see eqs. (\ref{Adssol}) and  (\ref{entropy})) :
\be \label{Tads}
 T  \simeq {1\over \pi r_h},  \qquad S = 4\pi \sigma \left({\ell\over r_h}\right)^3,
\ee
with $\sigma$ defined  by eq. (\ref{sigma}).
These expressions can be converted into functions of $\l_h$, by using  the UV asymptotics of $\l(r)$,
(\ref{lUV}) evaluated at the horizon\footnote{This is justified in the limit of small
 $r_h$ (small $\l_h$), since as we have discussed in Section 3,
 the deviations from the zero-temperature solution are $O(r^4)$ for small $r$, so eq. (\ref{lUV})
is a good approximation all the way to the horizon.}:

\be\label{rhlh}
r_h \simeq  {1\over \Lambda} \l_h^{-b}e^{-{1\over b_0\l_h}}
\ee

\noindent
In particular, $T(\l_h)$ is a decreasing function near the boundary.

The relation (\ref{Tads}) is corrected by the logarithmic running in the UV.
Neglecting non-perturbative contributions, the scale factor is given by eq. (\ref{bUV}),
from which we can obtain the logarithmic corrections to the
thermal function $f(r)$ through eq. (\ref{a18}) and (\ref{C}).
Through (\ref{T}) we obtain the corrected relation between $T$ and $r_h$ (see Appendix \ref{free-en-2} for
the details):
\be\label{Trh}
T =  {1\over \pi r_h} \left[1  - {4\over 9} {1\over (\log r_h \Lambda)^2 } +\ldots\right].
\ee

The results above allow us to compute the temperature dependence
of the term ${\cal G}(T) $  in the free energy coming from the gluon
condensate. From eq. (\ref{free-2})
we can write\footnote{  It is enough to write the relation below as an indefinite integral,
since adding a  constant does not affect  the result for the high temperature
asymptotics (\ref{CT}). A more
precise relation, with integration limits, will be  given in Section \ref{intrep}}:
\be
15\sigma\,{\cal G} = {\cal F} + {1\over 4} T S = -\int S dT + {1\over 4} T S,
\ee
where  $S = 4\pi \sigma  b^3(r_h)$. For small $r_h$, we can use the
UV asymptotic form for the scale factor, eq. (\ref{bUV}), and the relation
(\ref{Trh}), to obtain the asymptotic form of ${\cal G}$.  The calculation
is carried out explicitly in Appendix \ref{free-en-2},  and the result is:
\be\label{CT}
{\cal G} \to  {\pi^4\over 45}  {\ell^3 \,T^4 \over (\log {\pi T\over \Lambda})^2},\quad \qquad T\to \infty.
\ee

This equation has important consequences. It implies that the gluon
condensate contribution to the free energy is subleading at large $T$,
by a factor $(\log T)^{-2}$, with respect to the term $TS/4 \sim T^4$. Therefore,
for very large BH's we can write:
\be\label{fe-highT}
{\cal F} \to -{TS\over 4} \simeq -(\pi^4 \sigma  \ell^3) \, T^4 , \qquad T\to \infty.
\ee
On the other hand, at  very high temperatures,  pure $SU(N_c)$
Yang-Mills theory behaves as a free
gas with $\sim N_c^2$ degrees of freedom, and its free energy  is
approximated  by the
Stefan-Boltzmann formula\footnote{ This is {\it unlike} the case of ${\cal N}=4$ super YM.
There the theory in the UV is strongly coupled and a non-trivial strong coupling calculation is needed
to establish the correct coefficient, \cite{klebanov}.}
\be\label{sb}
{{\cal F}_{YM}\over V_3} \simeq -{\pi^2\over 45}N_c^2 \, T^4, \qquad T\to \infty.
\ee
Comparing the last two equations allows us to fix $\sigma = V_3N_c^2 /(45 \pi^2)$. From
the definition of $\sigma$, eq. (\ref{sigma}),   this is equivalent to
fixing the 5D Planck scale in $AdS$ units, in  a model independent way:
\be\lab{Planck}
(M_p \ell)^3  = {1\over 45 \pi^2}.
\ee

\subsection{Horizon in the IR region}\label{IRasymps}

Now, we are going to answer the question: what is the behavior of the function $T(r_h)$,
for large $r_h$? We will find that the answer depends exclusively on whether or
not the corresponding zero-temperature solution (more specifically, the {\em special} solution
in the classification of Section 2) is confining. In the non-confining case, the black-hole
temperature decreases to zero for large $r_h$; on the contrary,
for confining theories, $T(r_h)$ asymptotically grows with $r_h$.
 This distinction has  dramatic
 consequences for the thermodynamics of the model: as we will show in the next section,
it implies that confinement is in one-to-one correspondence with the
existence of a phase transition at a finite $T_c$.

As a byproduct of our analysis, we will find that when the
horizon is deep in the IR ( i.e. the black-hole mass is very small), the black-hole
geometry is  well approximated by the geometry of the zero-temperature special
solution (provided it exists).
 More precisely: the special solution is the only zero-T geometry
that can be lifted to a black-hole of arbitrarily small mass.

To find the large $r_h$ behavior of the black-hole temperature, we must solve eqs.
 (\ref{a11}--\ref{a12}) for $r$ and $r_h$ ``close to the singularity'' of the
zero-temperature solution.
 This question is somewhat ambiguous, since in different zero-$T$
backgrounds the singularity can be at a finite or infinite value
of the conformal coordinates, and it is unclear how exactly to identify the
``asymptoti'' region.

A coordinate-independent resolution of this ambiguity
consists in solving the equations using $\Phi$ as a radial coordinate,
and identifying the asymptotic regions according to the value of 't Hooft coupling  $\l\equiv e^\Phi$.
Indeed,  in the zero-temperature background this quantity
covers the whole range from $0$ in the UV to $+\infty$ in the deep IR,
no matter what is the position of the IR singularity in conformal coordinates. So it makes sense
to define  the black-holes whose horizon is in the deep IR (with respect to the zero-temperature
background) as those where $\l$ attains a large value at the horizon.
In any given black-hole solution $\l(r)$ is a monotonically increasing function,
and $\l_h\equiv \l(r_h)$ is the maximum value it can attain.
More precisely,
we consider ``large''  $\l$  as the region $\l\gg\l_0 $
such that the potential $V(\l)$ is
well approximated by its asymptotic form:
\be\label{asregion}
V(\l) \simeq V_{\infty} \, \l^{2Q} (\log \l)^P, \qquad \l\gg\l_0,
\ee
for some real $P$, and $Q\geq 0$. The actual value $\l_0$ of course depends on the specific  form of $V(\l)$.
Recall that confining theories correspond
to $Q>2/3$ or $Q=2/3$, $P>0$, and to make sure that there exists a special solution with  a good (i.e.
repulsive) singularity we must further restrict $Q<2\sqrt{2}/3$. The zero-temperature superpotential
for such solution obeys the asymptotics:
\be\label{Wasregion}
W_o(\l) \simeq W_{\infty}  \, \l^{Q} (\log \l)^{P/2}, \qquad \l\gg\l_0
\ee

As shown in Appendix \ref{counting},
the horizon  value  $\l_h$ uniquely determines the temperature, once the UV asymptotics
are kept fixed.   In the next two subsections we will determine
the behavior of   $T(\l_h)$  for  $\l_h$ in the asymptotic region, $\l_h\gg \l_0$.
An efficient way to attack this problem is to generalize
the superpotential technique described in Section 2
to the finite-temperature case. This will allow us to give a solution of
the system, for large $\l_h$,   in the entire large-$\l$ region. Another method that uses
diffeomorphism invariant variables is described in Section \ref{covvar}.

\subsubsection{Solution in the large-$\l$ region} \label{superIR}

We will now investigate the features of the black-hole solutions,
in the case when the horizon is deep in the IR.
Consider the situation where  the horizon
value of the 't Hooft coupling  $\l_h$,  is large enough to
be  deep in the IR asymptotic region, $\l_h\gg\l_0$, defined in the previous section.

In this case it is possible to find analytically a good approximation to the solution
 in the region
$\l_0 \ll \l \leq \l_h$, and to match it to the zero-temperature solution in the part of the
asymptotic region far from the horizon,  $\l_0 \ll \l \ll \l_h$.
This will fix all the integration constants, and give the temperature
unambiguously as a function of $\l_h$ for $\l_h\gg \l_0$.

An explicit (approximate) solution for large $\l_h$ is found by defining a {\em thermal superpotential},
i.e. a function $W(\Phi)$, that generalizes the zero-temperature superpotential $W_o(\Phi)$
to the black-hole background, such that the scale factor and dilaton equations (in domain wall coordinates)
reduce to the form:
\be\label{eq1sttherm}
A'(u)  = -\frac49 W, \qquad \Phi'(u) = {dW \over d\Phi}
\ee
The details of this  formalism are  developed in Appendix \ref{superfiniteT}. As shown there, the superpotential
$W(\Phi)$
and the logarithm $g$ of the thermal function $f$  appearing in the metric satisfy a pair of coupled differential equations in the
variable $\Phi$,  which take the form:

\bea
&&\left(\de_\Phi g + {\de^2_\Phi g\over \de_\Phi g}\right)\de_\Phi W + \de^2_\Phi W = {16\over 9} W, \qquad g \equiv \log f \label{supg2}\\
&&-{4\over 3}W(\de_\Phi W) (\de_\Phi g )- {4\over 3}\left(\de_\Phi W\right)^2 + {64\over 27}W^2 =
e^{-g} V(\Phi). 
\label{supW2}
\eea

In these variables the system of Einstein's equations splits into two separate decoupled systems, and
solving eq. (\ref{supg2}-\ref{supW2}) determines the metric and dilaton through eqs. (\ref{eq1sttherm}) solely
from the knowledge of $W(\Phi)$, as in the zero-temperature case. Here, though, the superpotential cannot
be taken as an independent function, but it depends on temperature through the coupling to $g(\Phi)$ in eq. (\ref{supW2})

As shown in Appendix \ref{appIRas}, the  solution of the system  (\ref{supg2}-\ref{supW2}) in the whole
region where (\ref{asregion}) is valid   has the form:
\be
 f\left(\Phi\right)\simeq 1- \left(\Phi\over \Phi_h\right)^R e^{-K\left(\Phi_h-\Phi\right)},
 \qquad   W(\Phi) \simeq W_{\infty}  \, \l^{Q} (\log \l)^{P/2} 
\ ,\ \qquad  \Phi_0\ll \Phi<\Phi_h, \label{solfr4}
\ee
where the constants $K$ and $R$ are given by:
\be
K \equiv \left({16\over 9Q} -Q\right); \qquad R \equiv  -{P\over 2} \left(1+{16\over 9Q^2}\right). \label{solfr3}
\ee
More precisely, the approximate equality signs in  eq.  (\ref{solfr4})
stand for dropping terms of $O(1/\Phi)$ in $W$ and $\log[1-f]$.

Notice that the asymptotic solution for the superpotential has automatically the same asymptotics
of the {\em special} solution $W_o(\l)$, eq. (\ref{Wasregion}): the presence of a regular horizon
at large $\l_h$ has lifted the degeneracy present in the zero-temperature superpotential equation, selecting
a single solution, which happens to be the unique one with the good singularity! We will come back to this
point later in this Section, and in Section 7.

The approximation $W \simeq W_o$  is valid in the whole asymptotic region. Thus, the metric
and dilaton have the same asymptotics as in the $T=0$
case, all the way to the horizon.
Moreover, $f(\Phi)$ is very close to unity already in the asymptotic region,
specifically where   $\Phi_0\ll \Phi\ll \Phi_h$.
Thus, for all smaller values of $\Phi$,
including those outside the asymptotic region, $f(\Phi)\simeq 1$ and the superpotential
equation reduces to the zero-temperature one, with special solution  $W_o(\Phi)$. This implies
that $W \simeq W_o$ is a good approximation in the {\em whole range } of $\Phi$,
not only asymptotically. The equations (\ref{eq1sttherm}) for the metric and dilaton
therefore reduce to the zero-temperature equations, whose solutions are classified
by the value of the integration constant $\Lambda$.
 This implies
that we can fix the remaining integration constant of the system (the scale $\Lambda$) to the
same value as in the $T=0$ background when we integrate (\ref{eq1sttherm}). With this choice,
all the  integration constants in the asymptotic solution are fixed. Thus, for large $\l_h$ the
complete thermal solution for the scale factor and dilaton is {\em everywhere} well approximated
by the zero-temperature form:
\be
A (r) \simeq A_o (r), \qquad  \Phi(r) \simeq \Phi_o(r).
\ee
From eq. (\ref{solfr4}), expressed in $r$-coordinates (by setting $f(r)\equiv f(\Phi(r))$),
 we observe that, at any fixed $r$, taking the limit $\Phi_h\to +\infty$ brings the metric to match the
vacuum form with $f(r)=1$. Therefore the  limit $\l_h \to \infty$ corresponds to the (point-wise) limit
of the black-hole going to the zero-temperature metric.

In the particular case of power-law asymptotic and
singularity at $r=\infty$,
corresponding to $Q=2/3$, $P=(\a-1)/\a$ (see \cite{ihqcd}),  the solution for large $r$ is :
\bea
 &&A (r) \sim - \left({r\over L}\right)^\a, \label{solAr2}\\
&& \Phi(r) \simeq {3\over 2} \left({r\over L}\right)^\a + {3\over 4}(\a-1)\log  \left({r\over L}\right), \label{solPhir2}\\
&&f(r)\simeq 1 - \left({r\over r_h}\right)^{1-\a}
 \exp\left[-\left({r_h\over L}\right)^\a +   \left({r\over L}\right)^\a\right]. \label{solfr6}
\eea
where $L$ is the same length scale  appearing in the zero-temperature solutions (\ref{IR}).
The relation between temperature and horizon position for these black-holes is:
\be \label{tempsmall2}
T  = {|\dot{f}(r_h)|\over 4\pi} =  {1\over 4\pi} {r_h^{\a-1}\over L^{\a}}\left[1 + O\left(\left({L\over r_h}\right)^\a\right)\right],
\ee
Notice that, for $\a>1$, i.e. when the theory is confining,  the temperature {\em increases} as a function
of $r_h$. In the next section we will see that this behavior is characteristic of general  confining
asymptotics. \\

\noindent
{\bf A no-go theorem}\\
As discussed in detail in Appendix \ref{superapp}, the asymptotic behavior (\ref{asregion}) does
not fix uniquely the asymptotics of the zero-temperature solution, but rather allows various
possibilities summarized in Appendix \ref{general}.
An important conclusion that can be derived  from the discussion above is that  black-holes
 which probe the asymptotic large-$\l$ region {\em necessarily match the ``special''
zero-temperature solution} with asymptotic (\ref{W0as}), as $\l_h\to \infty$. All other solutions cannot
be lifted to black-holes with arbitrarily small mass (i.e. whose metric is arbitrarily
 close to the zero-temperature metric) . For an alternative derivation of the same results, see Section
 \ref{matchcov}. We thus have the following: \\

\noindent
{\bf No-go theorem:} {\em The only vacuum solutions of Einstein-dilaton gravity, with a potential
satisfying (\ref{asregion}), that can be continuously lifted to a regular black-hole with
arbitrarily small mass,
are the ones stemming from a superpotential with the ``special''  asymptotics (\ref{W0as}).
 }\\

In particular, this implies that the two continuous families of zero-$T$ solutions with ``generic''
and ``bouncing'' superpotentials (all of which have bad singularities)
cannot be promoted to black-holes with an arbitrarily small mass (i.e. such that the metric looks
like that of those zero-$T$ solutions almost everywhere).

The  zero-$T$ superpotentials with bad singularities  were
already discarded as pathological in \cite{ihqcd}, since they exhibit
singularities which are not screened from physical fluctuations, and do not have a well
defined eigenvalue problem. If we subscribe to the criteria laid out in \cite{gub2},
the no-go theorem above gives another reason to discard these kinds of solutions: in \cite{gub2}
a  singular background was considered acceptable if the singularity can be ``hidden'' behind
a regular black-hole horizon by  an infinitesimally small deformation of the metric\footnote{The
physical reason behind this criterion was that, if the singular background has a holographic
interpretation as the zero-temperature vacuum of a 4D theory, then it should be possible
to obtain it continuously as the $T\to 0$ limit of a thermal state, described holographically
by a black-hole. This reasoning assumes that a small black-hole,  which is infinitesimally close
to the vacuum background, has also an infinitesimally small temperature. In the
following section we will see that this is not necessarily true: there are
cases (e.g. in the presence of a deconfining phase transition at finite $T$) when  the $M\to 0$
limit of the black-hole corresponds to a {\em high temperature} limit. Thus, in the most interesting cases of confining backgrounds, the criterion of \cite{gub2} loses part of
its physical motivation.}.\\

\subsubsection{Big black-holes and small black-holes}

Now, let us determine the temperature of the asymptotic
solution obtained, for general $P$ and $Q$ in (\ref{asregion}). It is easiest to do
so as a function of $\Phi_h$. We have (see eq. (\ref{temperature}))
\be\label{T2}
4 \pi T = \left|{df(r)\over dr}\right|_{r_h} =  \left|{df(\Phi)\over d\Phi}\right| {d\Phi\over d u} {du \over dr}=
\left[   \left|{df(\Phi)\over d\Phi}\right| {d W\over d\Phi} e^{A(\Phi)}\right]_{\Phi_h}.
\ee
The quantities $d f/d\Phi$ and $d W/d\Phi$ are obtained from eq. (\ref{solfr4}).
On the other hand, $A(\Phi)$ can be obtained
by combining the  two  eqs. (\ref{eq1sttherm}),
\be
{d\Phi\over dA} =  -{9\over 4}{ d \log W\over d\Phi} \simeq -{9\over 4}\left[Q + {P\over2}{1\over\Phi}+\ldots\right],
\ee
which upon integration which gives:
\be\label{APhi}
A(\Phi) \simeq -{4\over 9 Q } \Phi + {2 P\over 9 Q} \log \Phi.
\ee
 Inserting these  expressions in eq. (\ref{T2}) we obtain:
\be\label{Tphih}
  T (\Phi_h) \simeq  {K Q\over 4\pi} e^{{(Q^2 - 4/9)\over Q}\Phi_h} \Phi_h^{P/2 + (2P/9Q)} \qquad \Phi_h\gg \Phi_0
\ee

From  equation (\ref{Tphih})  we deduce that the thermodynamic behavior of  black-hole solutions  is very different according to the IR confining properties
of the zero-T solution:
\begin{enumerate}
\item if {\bf $Q >2/3$, or $Q=2/3$ and $P>0$ (i.e. the  $T=0$ theory confines):}  \\
the temperature {\it increases} with $\Phi_h$, so very small black-holes
(large $\l_h$)
corresponds to  high temperature solutions.
\item if {\bf $ Q<2/3$,  or $Q=2/3$ and $P<0$ (the $T=0$ theory does not confine):}\\
The temperature decreases with $\Phi_h$, and the limit $\Phi_h \to \infty$
corresponds to $T\to 0$.
\end{enumerate}
Since, in all cases, $T$  {\it decreases} with $\Phi_h$ in the UV regime,
it follows that:\\

\noindent
{\em In confining theories, $T(\Phi_h)$ must have a minimum value $T_{min}$ below which there is
no black-hole solution. In this region  the only remaining solution is the thermal
gas. In particular,  $T_{min}$ cannot be zero: this would imply the existence of a point where $\dot{f} = 0$, but from eq.
 (\ref{a17b}) we observe that if $\dot{f}$ vanishes at some point where $b$ is
 regular, it must vanish  everywhere.}\\

\begin{figure}[h!]
 \begin{center}
\includegraphics[scale=0.9]{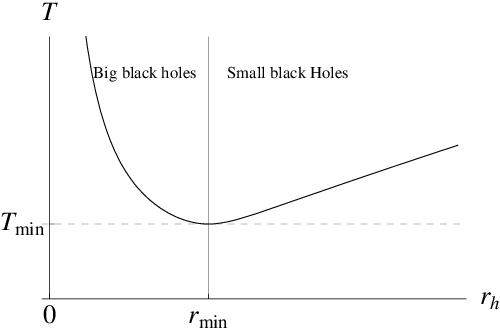}\\
(a)\\
\vspace{1cm}
\includegraphics[scale=0.9]{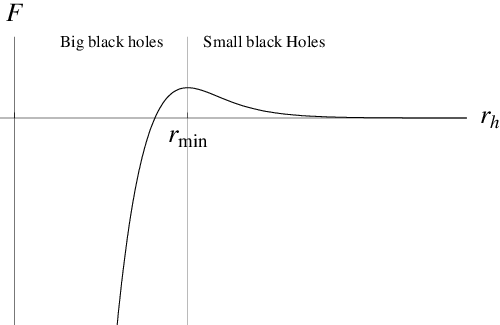}\\
(b)
 \end{center}
 \caption[]{Typical plots of the black-hole temperature (a) and free energy (b)
 as a function of the  horizon position $r_h$, in a confining background.
The temperature features a minimum at $r_{min}$ ,
that separates the large black-hole  from the small black-hole  branches.
}
\label{TFrh-BS}
\end{figure}

What we have shown implies that  confining theories admit (at least) two branches of black-holes
with the same temperature, starting at $T=T_{min}$ .  For $T<T_{min}$ there are no black-hole solutions, whereas  for high enough temperature there
are exactly two black-holes:
\begin{enumerate}
\item
 the {\em big black-hole} has its horizon closer to
the $AdS$ boundary, satisfying the relation (\ref{Tads}),
\be\label{Tads2}
r_h^{big} \simeq {1\over \pi T}.
\ee
\item   the {\em small black-hole} has
its horizon deep in the interior. With the asymptotics (\ref{solfr6}), the horizon position is related to the
temperature by (inverting (\ref{tempsmall2}):
\be\label{tempsmall}
r_h^{small} \simeq L \left({4\pi\over \a}\, TL\right)^{1/(\a-1)}, \qquad \a>1
\ee
\end{enumerate}

The typical situation (with two branches of solutions) is depicted in
figure \ref{TFrh-BS}.

{Generically, the big black-holes are thermodynamically stable whereas the small black-holes are  unstable.
Using $c_v = T dS/dT$ and (\ref{entropy}), we
obtain, \be\lab{dtdrh} \frac{dT}{d\l_h} =
\frac{3ST}{c_V}\frac{dA}{d\l_h}. \ee
If  $dA(\l_h)/d\l_h<0$, the condition for
thermodynamic stability i.e. $c_V>0$ coincides with $dT/d\l_h<0$, which is true for
big black  holes;  on the other hand for the
small black-holes $dT/d\l_h>0$, hence if $A(\l_h)$ is monotonic,
they have negative specific heat and are unstable. Although typically  $dA(\l_h)/d\l_h<0$ is
satisfied, since generally increasing $\l_h$ means going deeper in the IR, it is
hard to tell whether this is true for all values of $\l_h$: recall that there is  a ``hidden'' dependence  on $\l_h$ in $A(\l)$, due to the fact that the form of the scale factor is temperature-dependent.
 The condition  $dA(\l_h)/d\l_h<0$ is certainly obeyed
asymptotically both for small  $\l_h$ and for large $\l_h$,
 where all black-hole metrics  reduce to  the
zero temperature solution: for small $\l_h$,  $A(\l_h) \sim (b_0 \l_h)^{-1}$,  and for large $\l_h$,
$A(\l_h) \sim -2/3 \log \l_h$. However one might have some range  of $\l_h$ where $dA(\l_h)/d\l_h>0$,
thus the small black-holes are stable, or  the big ones unstable. We have found some numerical
evidence of this behavior, where  the violation of monotonicity of $A(\l_h)$ is always in a very
limited range of $\l_h$, and only in the small black-hole branch. Since the big black-holes are the thermodynamically dominant solutions above $T_c$, we don't expect them to
become unstable for $T>T_c$, since the dilaton potentials we use can always be written
in terms of a superpotential:  this guarantees the validity of the Positive Energy Theorem \cite{PET},
which in turn guarantees stability of the vacuum. }

\begin{figure}[h]
 \begin{center}
\includegraphics[scale=1.2]{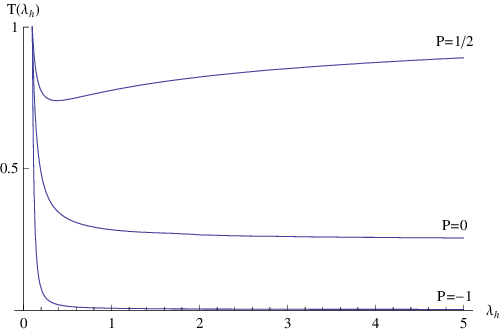}\\
(a)\\
\vspace{1cm}
\includegraphics[scale=1.2]{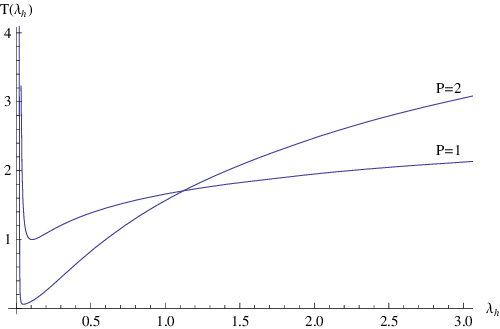}\\
(b)
 \end{center}
 \caption[]{The temperature as a function of the horizon value of $\l$ in the model specified
by the potential (\ref{explicitV}), for $Q=2/3$ and various values of $P$. The
other coefficients are fixed to $V_1=10$, $V_2=100$. The confining models
($P>0$) feature a minimum temperature at finite $\l_h$; in the non-confining model ($P=-1$) the
$T(\l_h)$ monotonically decreases to zero; In the borderline case ($P=0$) $T(\l_h)$ decreases
monotonically to a finite value as $\l\to \infty$.}
\label{Tnumerics}
\end{figure}

 We now illustrate the general behavior just described with  a few simple concrete examples, for
which we solved numerically Einstein's equations, starting from an explicit form of the dilaton
potential. In the case of  the explicit potential that was considered in \cite{ihqcd},
the numerical results for the thermodynamics were discussed in \cite{ft}, and they
indeed agree with the general results discussed above. Here we present the result
for a simple class of potentials, which have the IR asymptotics of the form (\ref{asregion})
for general $P$ and $Q$:
\be\label{explicitV}
{V(\l)\over 12} =1  +  \l + V_1\l^{2Q}\left[\log \left(1 + V_2 \l^2 \right)\right]^P
\ee
where we have set the $AdS$ scale to unity. Also, we have set   the coefficient of the
 linear term to one. This can be done by rescaling $\l$ and redefining the coefficients
$V_1$, $V_2$.

The numerical results for  the function $T(\l_h)$ for $Q=2/3$, fixed $V_i$,  and various values of $P$  are shown
in figure \ref{Tnumerics}. They confirm our general analysis: confining potentials ($P\geq 0$)
feature a minimum black-hole temperature, whereas in non-confining ones ($P<0$) the temperature
can be arbitrarily low.

\begin{figure}[h]
 \begin{center}
\includegraphics[scale=1.2]{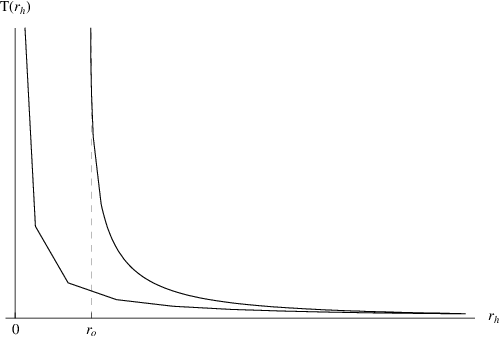}\\
\end{center}
\caption[]{Temperature as a function of  (a) $r_h$ 
in the model (\ref{explicitV}) for $Q=2/3$ and  $P=2$. The temperature diverges for $\l_h\to \infty$,
for which  $r_h\to r_0$. It is a single-valued function of $\l_h$, but not of $r_h$.}
 \label{finiter0}
\end{figure}

{ As discussed earlier, the temperature is guaranteed to be a single-valued
function of $\l_h$, but the same is not necessarily true if we consider  $T$ as
a function of the horizon position $r_h$. A rather striking example of this fact
is shown in figure \ref{finiter0}: it displays the plot of the curve $T(r_h)$
in a case ($Q=2/3$, $P=2$) where the singularity of the  vacuum solution
is at a finite value $r_0$ of the conformal coordinate. For comparison, the curve $T(\l_h)$ in
the same model, shown in figure \ref{Tnumerics} (b), is single-valued.
One unexpected feature is that in this case there are black-holes whose horizon is well beyond
the position of the zero-temperature singularity at $r=r_0$. Nevertheless,  as the temperature
increases along the small black-hole branch, we have $r_h\to r_0$ as expected. }

\subsubsection{An integral representation for the free energy}\lab{intrep}

As described above, in many cases there are more than one black-hole solutions, some of them having negative specific
heat. This is reflected in the fact that $\l_h$ (or $r_h$) as a function of $T$ is generically
multi-valued.  The formula for the
free energy that we derive here will encompass
 all of the different branches under one
integral representation. This form will be used to
prove the proposition in the next subsection. It is also a very convenient form for
numerical evaluation of the free energy, in the process of determining the thermodynamics of the
system numerically.
Here we present the discussion for the case of two branches of solutions,
 one small and one big black-hole for simplicity of the presentation.
However the final result is easily generalized to multi-black-hole cases.

Let us denote the free energies of the small and the big black-hole by $\cF_S$ and $\cF_B$ respectively.

Integrating the first law, $S=-d\cF/dT$ for the big BH, one obtains,
\be\lab{fB}
\cF_B(T) = \cF_{min} - \int_{T_{min}}^T S_B dT, \qquad T>T_{min}
\ee
where $\cF_{min} = \cF(T_{min})$.

\begin{figure}[h]
 \begin{center}
\includegraphics[scale=0.5]{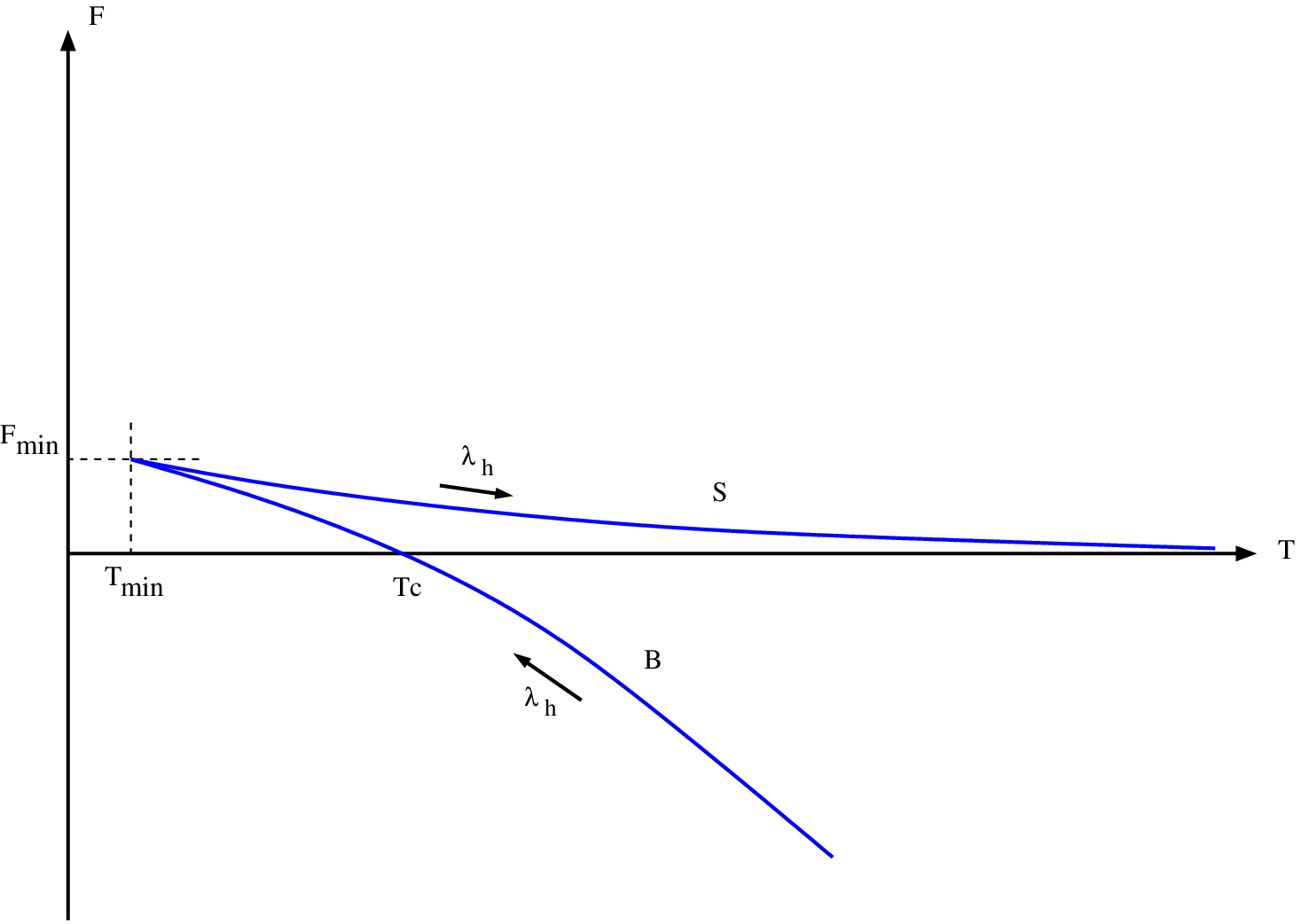}
 \end{center}
 \caption[]{Black hole free energy}
\label{FTsingle}
\end{figure}

In order to determine the integration constant in (\ref{fB}) one can
make use of the same formula but {\em on the small BH branch}. This is clearly suggested
from fig. \ref{FTsingle} where we depict a generic form for the function $\cF(T)$ in case of two-branches.
First, note that $\cF_{min}$ is
the same on both branches (see figure \ref{FTsingle}):
\be\lab{fmin} \cF_B(T_{min}) =  \cF_S(T_{min}) = \cF_{min}.
\ee
Thus one has:
\be\lab{fS}
\cF_S(T) = \cF_{min} - \int_{T_{min}}^T S_S dT, \qquad T>T_{min}
\ee
The small black-hole free energy vanishes in the limit of zero
black-hole size (where the metric coincides with the zero-temperature
background), which for confining backgrounds is the $T\to \infty$ limit.
This allows to write $\cF_{min}$ as:
 \be\lab{fmin2}
\cF_{min} =  - \int^{T_{min}}_{+\infty} S_S\,\, dT,
\ee

Combining (\ref{fB}) with (\ref{fmin2}) one obtains an integral representation
for the free energy that only depends on the area of the horizon,
and is valid on both  branches.
It can be put in a simpler form in the $\l_h$ variable. Using (\ref{entropy}) one obtains,
\be\lab{flh}
\cF_B(\l_h) = -4\pi \sigma \int_{\infty}^{\l_h} b^3(\tilde \l_h) \frac{dT}{d\tilde \l_h} d\tilde \l_h , \qquad \l_h<\l_{min},
\ee
where $\l_{min}$ is the horizon position of the minimum temperature black-hole.
Note that the two branches are combined in the integral as both $b$ and $T$ are single-valued as functions
of $\l_h$, but are not as functions of $T$.   In appendix \ref{intrep2} we present  further useful formulas regarding this
integral representation. We remark that RHS of (\ref{flh}) is finite everywhere except at $\l_h=0$
where $\cF_B\to -\infty$.  Finiteness near the singularity
$\l_h=\infty$  follows from the fact that, for all of the confining cases,  $b^3$ vanishes exponentially
faster than $dT(\l_h)/d\l_h$. 

We note the remarkable fact that the free energy is completely determined by the knowledge of area of
(the small and the big) black-hole horizons. This means that the entire thermodynamic properties of the
dual field theory is encoded in the horizon areas, as a function of $T$. We stress that the area of the big BH horizon only
is not sufficient; this misses the integration constant (\ref{fmin}). It is therefore not sufficient to
 determine $T_c$ for instance. An alternative way to say this is as follows: as we showed in
eq. (\ref{free-2}) there are two contributions to the free energy:
the entropy and the condensate. Here we learn that, {\em we need both branches to disentangle the two contributions.}


\subsection{Confinement and phase transitions}\label{conftrans}

In this section we show the one-to-one connection between color confinement (in the vacuum background)
and the presence of a deconfining transition:

\paragraph{Proposition:} \begin{itemize} \item[i.] {\em There exists a
confinement-deconfinement phase transition at finite $T$, if and only
if the zero-$T$ theory confines.} \item[ii.] {\em This transition is
of the {\bf first order} for {\bf all} of the confining geometries,
with a single exception:}
\item[iii.] {\em In the limit
confining geometry $A_o(r)\to - r/L$ (as $r\to \infty$), the phase
transition is of the {\bf second order} and happens at $T =
3/(4\pi L)$.}
\item[iv.] {\em All of the non-confining geometries at zero-$T$ are always in the
black-hole  phase at finite $T$. They exhibit a second order phase
transition at $T=0^+$.}
\end{itemize}

We outline our demonstration in  the coordinate system where $\l$ is chosen as the radial variable.
Being diffeomorphism invariant, our arguments apply to all of the confining zero-$T$ geometries that are
described in Section \ref{sec2}\footnote{One can state the part iii of the
theorem in a diffeomorphism-invariant way by stating that the borderline geometry is
defined as $\lim_{\lambda\to\infty} (X+\half)\log\l = 0$ \cite{ihqcd}.}.

\begin{figure}
 \begin{center}
 \leavevmode \epsfxsize=12cm \epsffile{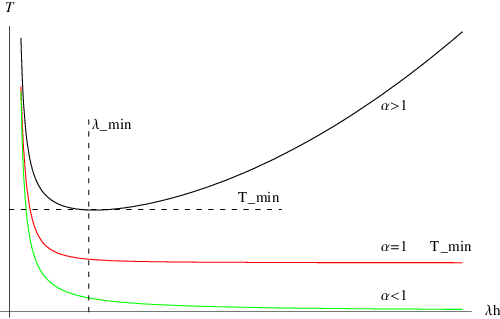}
 \end{center}
 \caption[]{Temperature as a function of $\l_h$ for the infinite r geometries of the
 type $A\to r^{\a}$. Black holes exist only above $T_{min}$
 whose precise value depend on the particular zero-$T$ geometry.}
 \label{trhillus}\end{figure}

We first consider the geometries that confine the color charge at zero $T$.
In Section (\ref{IRasymps}) we
have shown that there exists an extremum of $T(\l_h)$ if and only if the zero $T$ theory confines.
Thus, for confining potentials there exists such points $\l_{min}$ that satisfy,
\be\lab{tmin}
\frac{dT}{d \l_h} \bigg|_{\l_{min}} = 0 \qquad where \qquad T(\l_{min}) \equiv T_{min} > 0.
\ee
In figure \ref{trhillus}  a schematic plot of $T(\l_h)$ for the typical examples
of confining, non-confining and the borderline cases is shown.

\begin{figure}
 \begin{center}
 \leavevmode \epsfxsize=12cm \epsffile{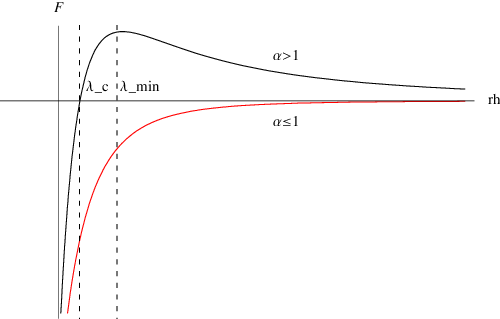}
 \end{center}
 \caption[]{Free energy density as a function of $\l_h$ for the infinite r geometries of the
 type $A\to r^{\a}$.}
 \label{frhillus}\end{figure}

Now, it follows from the positivity of the entropy and the first law of thermodynamics $S=-d\cF/dT$, that
the extrema of  $T(\l_h)$ coincide with the extrema of $\cF(\l_h)$. To see this we write
$-S=d\cF/dT = (d\cF/d\l_h)/(dT/d\l_h)$ and observe that in order for $S>0$ at all $\l_h$, the extrema of
$T(\l_h)$ should coincide with the extrema of $\cF(\l_h)$.
Hence, for the confining
potentials there exist at least one extremum of $\cF(\l_h)$. See figure \ref{frhillus} for a schematic plot
of $\cF(\l_h)$.

For simplicity, here
we shall assume that there exists a single $\l_{min}$ in the confining geometries, and
no extremum in the non-confining ones.
  We shall comment on the multi-extrema cases in the next subsection
 and carry out a detailed analysis  in Appendix \ref{multiX}.

\begin{figure}
 \begin{center}
 \leavevmode \epsfxsize=12cm \epsffile{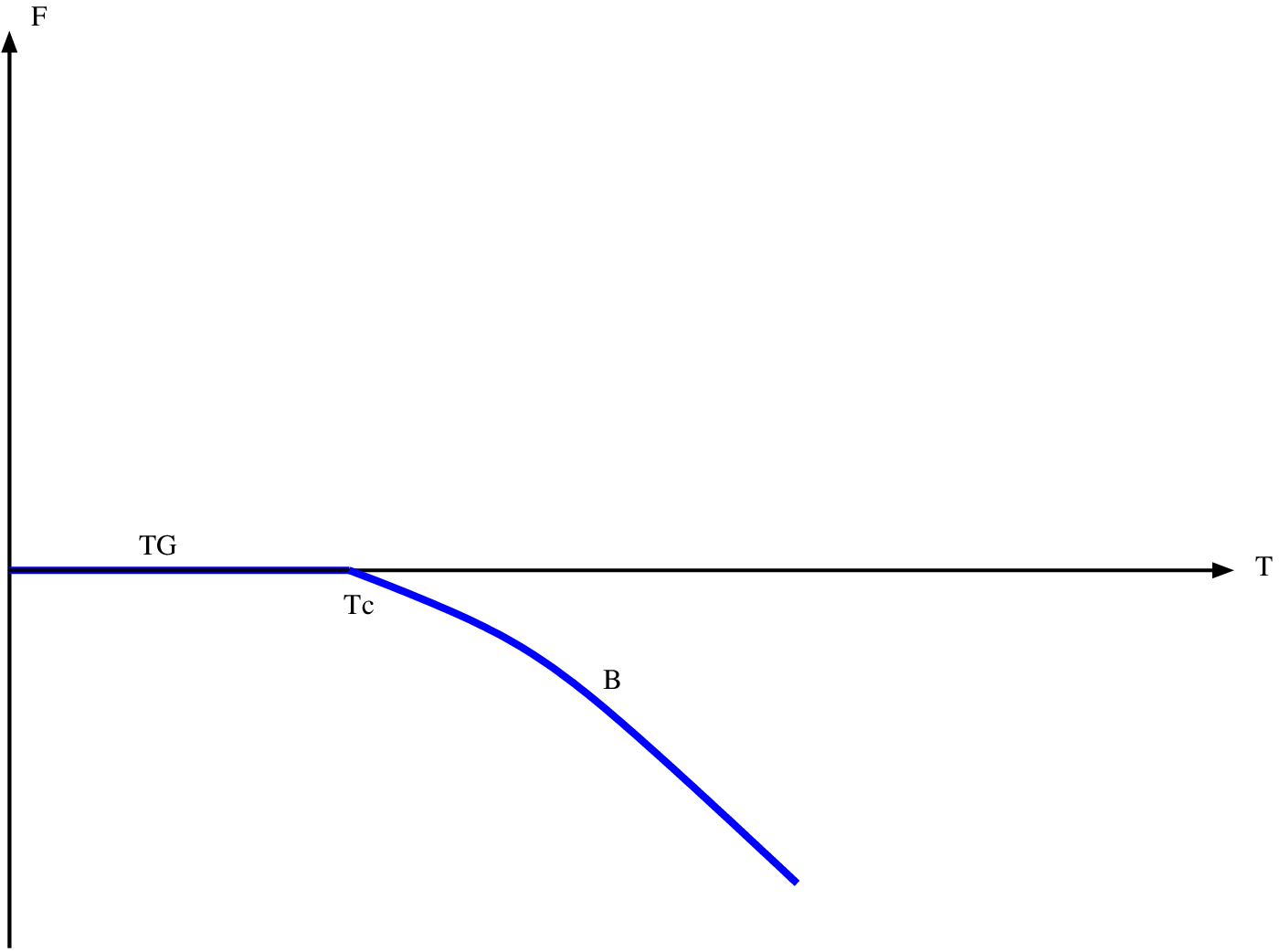}
 \end{center}
 \caption[]{A schematic plot of the free energy in the typical case of a confining geometry. "B" and "TG" denote the big black-hole and the thermal gas respectively. At $T_c$ there is a first order transition.
 This plot corresponds to evaluating $F(T)$ of fig. \ref{FTsingle} on the minimum energy configuration. }
 \label{FTmin}\end{figure}

Let us first reproduce here the integral representation
derived in the previous subsection:
\be\lab{Flh}
\cF(\l_h) = -4\pi M^3 V_3\int_{\infty}^{\l_h} b^3(\tilde \l_h) \frac{dT}{d\tilde \l_h} d\tilde \l_h.
\ee
The integrand is positive definite for $\l_h>\l_{min}$ and negative definite for $\l_h<\l_{min}$.
Therefore, evaluating it at  $\l_{min}$,  one has,
\be\lab{frmin}
   \cF(\l_{min})>0.
\ee
On the other hand, when one evaluates (\ref{Flh}) on the boundary, one finds
\be\lab{fboun}
   \lim_{\l_h\to 0 } \cF(\l_h)=-\infty.
\ee
This follows from the UV asymptotics described in Section \ref{UVasymps}.
Therefore there must  exist a point $\l_c$ where it vanishes:
\be\lab{fc}
   \cF(\l_c)= 0,
\ee
see figure fig, \ref{frhillus}.
This proves that {\em there exists a phase transition
if the zero-$T$ theory confines}. The transition temperature $T_c$ is greater than $T_{min}$ because $\l_c<\l_{min}$ and $dT/d\l_h<0$
at $\l_c$.
This result confirms our intuitive picture that, as the temperature
is increased, first the small and big BHs form at a temperature $T_{min}$, where the minimum energy configuration is still
 the thermal gas, and as $T$ is kept increasing, the big black-hole takes over the
thermal gas phase at a higher temperature $T_c$.
The true free energy of the system, \ie the function $F(T)$ {\em evaluated  on the minimum energy configuration},  is shown schematically in fig. \ref{FTmin}.

On the other hand, for the non-confining theories (see the case $\alpha<1$ in figs.
\ref{trhillus} and \ref{frhillus}), $T(\l_h)$ is always monotonically decreasing.
This is because, in the UV  $T\propto \exp[1/b_0\l_h]$. Hence it is monotonically decreasing in the UV and there exists no extremum
of this function. Therefore its derivative cannot change sign. From (\ref{Flh}), it follows that $\cF(\l_h)<0$ for all $\l_h$ and that
there  is {\em no phase transition for the non-confining potentials} at any finite-$T$.

We  have so far proven part i. of the proposition.
 Parts ii. and iii. are proven as follows. The order of
the phase transition is determined by the latent heat:
\be\lab{latent}
L_h = E(\l_c) = S(\l_c) T(\l_c).
\ee
It follows from (\ref{entropy}) that, $S(\l_c)$ is non-zero unless
$\l_c$ coincides with the singularity. Therefore $L_h>0$ and
the phase transition is {\em first order} for the standard confining geometries. For example, this is
the case for $\alpha>1$ in  infinite geometries, (see figs. \ref{trhillus} and \ref{frhillus}).

On the other hand, in the borderline case $\alpha=1$, $T_c=T_{min}$ and, at $\l_c$
the entropy vanishes, because $\l_c$ coincides with the singularity and $b(\l_c)\to 0$ there.
Thus, in this case the transition is of {\em second order}.  It would be interesting to find other
examples of second order phase transitions in the Einstein-dilaton system.

The last part of the proposition follows simply from the fact that,
for all the non-confining geometries, $\cF(\l_h)<0$ for all $\l_h$ and it vanishes only at the singularity.
At this point both $T$ and $S$ vanish as the area of the horizon vanishes. Thus there is a trivial
second order phase transition at $T=0^+$ and {\em the system is always in the black-hole phase} for
any finite-$T$.

\subsubsection{Geometries with multiple extrema}

We demonstrated our proposition under a single assumption, that the function $T(\l_h)$ has at most one single local extremum.
 We did not find any counter-example to this assumption in our numerical studies, and it is a logical possibility that, with the given assumptions for
$V(\l)$ (that it is a positive and monotonically increasing
function of $\l$ that limits to a constant at $\l=0$ and diverges
exponentially as $\l\to\infty$), multiple extrema cases never
occur. However, we can not rule out these possibilities by
analytic arguments, therefore they should be considered in order
to complete our demonstration. Moreover, as we discuss below, they
bear interesting possibilities for new types of phase transitions.

In general,  there may exist theories which admit more than one small and
one big black-hole. In these cases, the functions $\cF(\l_h)$ and
$T(\l_h)$ are complicated and admit many extrema. As a result $F(T)$, evaluated on the entire
set of solutions (not only the lowest energy solution)
may be a complicated multi-valued function with many cusps and crossings, see
fig. \ref{FTmulti} in Appendix G for an example.

{ The proof of part i. of the proposition extends without changes to the general
case with multiple extrema. To prove  points ii. iii. and iv. in full
generality,  however,  we must make
an additional assumption on the behavior of the entropy as a function of $\l_h$:
we must assume that a generic black-hole in a given branch has larger entropy
than a generic black-hole in the next (with larger $\l_h$) branch. This is a weak
version of monotonicity
 of $S(\l_h)$, and it is  a sufficient condition
for the full proposition to hold, although it might not be necessary\footnote{Strict monotonicity is too strong a condition, since
we have found numerical counterexamples where this is not satisfied.}.}

The details of the general case are presented in Appendix \ref{multiX}.
The upshot of the analysis in \ref{multiX} is that, regardless how complicated
the system is, $F(T)$ {\em evaluated on the minimum energy configuration} always have a
similar form to fig. \ref{FTmin} as in the single extremum case. Since, in the infinite volume limit,
 only the lowest energy configuration is
relevant for the thermodynamics, our demonstration in the previous subsection directly carries over to
cases of multiple extrema. The precise statement is that
{\em regardless how complicated the function $T(\l_h)$ is, there exists a confinement-deconfinement phase transition
if and only if the corresponding zero-$T$ theory confines.}

\begin{figure}[h]
 \begin{center}
\includegraphics[scale=0.5]{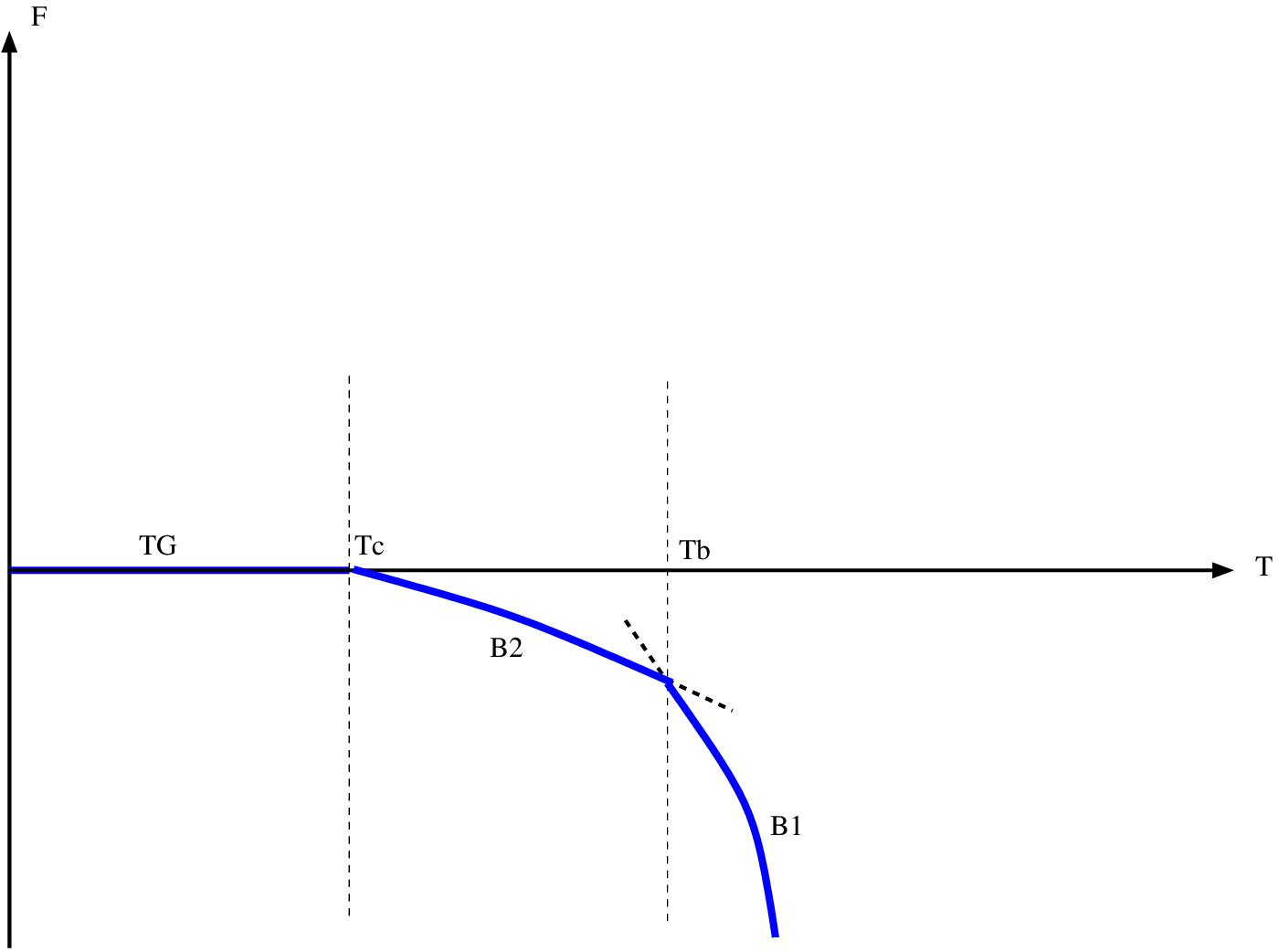}
 \end{center}
 \caption[]{The function $F(T)$ for a multiple extrema case, evaluated on the minimum
 energy configuration. Here $k=1$, i.e. there are two big-black-hole geometries
 denoted by $B_1$ and $B_2$ with a first-order transition between them at $T_b$. $TG$
 denotes the thermal gas geometry, that takes over below $T_c$.   }
\label{FTmultimin}\end{figure}

However, our analysis in Appendix \ref{multiX} shows another
interesting possibility: in the multiple extrema cases, there may
exist {\em first order phase transitions between different
deconfined vacua}. The temperature of these transitions are {\em
always higher than the deconfinement temperature $T_c$}. In
general, there may be an arbitrary number $k$, of such transitions
with $T_k>T_{k-1}>\cdots T_1>T_c$. In the dual geometric  picture,
these transitions occur between {\em different big black-hole
geometries}. For a number $k$ of such transitions, the function
$T(\l_h)$ should possess $k$ local minima and $k-1$ local maxima.
This means that, there should be $k$ different pairs of small and
big black-holes. This is a necessary condition but it is not
sufficient. The sufficient condition follows from the particular
shape of $F(T)$ that leads to such transitions between different
big black-hole branches. An example is discussed in Appendix \ref{multiX}.
We present the free energy for the minimum energy configuration,
for the case $k=1$ in fig. \ref{FTmultimin}. The fact the these
transitions are always first order follows from discontinuity in
$F'(T)$ at $T_l$, as shown in  fig. \ref{FTmultimin} and Appendix
\ref{multiX}.

It was argued that there could be a series  of phase transitions
in large-$N_c$ gauge theory
 that would correspond to a partial, step-by-step breaking of the center $Z_N$ of the gauge group,  \cite{teperlucini}.
  At large $N_c$  there is room for an arbitrary number of  such steps.
The order parameter corresponding to the l-th such transition
would be $\langle\tr~{\cal P}^l \rangle$, namely the l-th power of
the Polyakov loop. It is plausible that such phase transitions
may be in the same universality class as the ones described above.

Let us remark however, that, neither in
 the lattice studies of \cite{teperlucini} nor in our numerical investigations,
 one has encountered such transitions.
 They may exist as an exotic possibility in our set-up.

\subsection{Similarities with the Yang Mills deconfinement transition}

We have found that backgrounds which exhibit confinement, also exhibit a deconfinement
phase transition at some  finite temperature $T_c$, above which the black-hole phase dominates.
The qualitative features of the phase transition and of the thermodynamics of the deconfined
phase are remarkably similar to those found in four dimensional pure Yang Mills theory at large $N_c$.

Below we list some of the model-independent features of the gravity phase
transition
that match the gauge theory side. We will analyze the
quantitative agreement in concrete  models in a separate work \cite{GKMN3}.

\begin{itemize}

\item It is a confinement-deconfinement transition. In particular in the high temperature phase
the confining string tension vanishes.

\item The Polyakov loop is the order parameter for the confinement-deconfinement transition of $SU(N_c)$ YM theory.
The vev of the Polyakov loop $\langle {\cal P} \rangle$ is zero in the confined phase and it acquires a non-zero expectation
value above $T_c$. Here we see the analogous behavior in the dual geometric picture\cite{D4}: Holographically,
the Polyakov loop is
described by a classical string embedding that wraps the Euclidean time direction. In the thermal gas solution the time
circle is non-shrinkable, hence the action of the string is infinite, giving a zero vev for the Polyakov loop.
On the other hand, in the black-hole solution, the time circle shrinks to zero size at the horizon and one obtains
a non-trivial vev.

\item Both in large-$N_c$ Yang-Mills and in the gravity  theory
  the phase transition is first order, with
      a latent heat that scales as $N_c^2$. On the gravity side the latent heat
       per unit volume is given by
      (see  eq. (\ref{bhenergy}))
\be
L_h =  N_c^2 M_p^3\left( 15 {\cal G}(T_c)    + 3 \pi T_c b^3(r_c)\right) = N_c^2{4 \over 3} \ell^{-3} {\cal G}(T_c) =
N_c^2{4\pi \over 45} \ell^{-3} T_c b^3(T_c).
\ee
\item As in large-$N_c$ Yang-Mills, the quantity ${\cal F}/N_c^2$ serves as an order parameter: it is
of order one in the deconfined phase, and zero in the confined phase
\item The high temperature behavior is the same  of a free gluon gas, up to logarithmic corrections. This behavior
 can be seen in the temperature dependence of the pressure $p=-{\cal F}/V_3$, energy density $\rho$, and entropy density $s$ :
from eq. (\ref{fe-highT}) and standard thermodynamic relations, and with the choice (\ref{Planck})  for the Planck scale,
we have:
\be
 \left.\begin{array}{l} {\rho(T)\over T^4} \\ \\ {3p(T)\over T^4} \\ \\ {3\over 4}{s(T)\over T^3} \end{array}\right\} \to {\pi^2\over 15} N_c^2  \qquad T\to \infty.
\ee
 \begin{figure}[h!]
 \begin{center}
\includegraphics[scale=1.0]{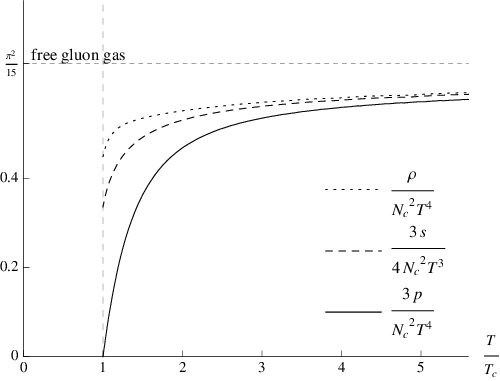}
 \end{center}
 \caption[]{The energy density $\rho$, entropy density $s$ and pressure $p$ in the black-hole phase for $T_c<T<5T_c$.}
 \label{esp}
\end{figure}
In Figure \ref{esp} we present   an  explicit example of the behavior of these quantities in the deconfined phase, up to a
temperature of $5T_c$, derived from our model with a potential of the form (\ref{explicitV}) with $Q=2/3$ and $P=1/2$.

The deviation from conformality is expressed by  the trace of the thermal stress tensor:
\be     \label{e3p}
{\rho - 3p\over T^4} \to  \left\{\begin{array}{ll} L_h/T_c^4 & \quad T\to T_c \\ &\\ {4\pi^2  N_c^2\over 135 }
 (\log T/T_c)^{-2} & \quad T\to \infty,\end{array}\right.
\ee
as can be derived by combining  eqs. (\ref{tracean}) and (\ref{CT}). A concrete example of the temperature dependence of
this quantity in the black-hole phase is shown in Figure (\ref{e-3pfig}).

\begin{figure}[h!]
 \begin{center}
\includegraphics[scale=1.0]{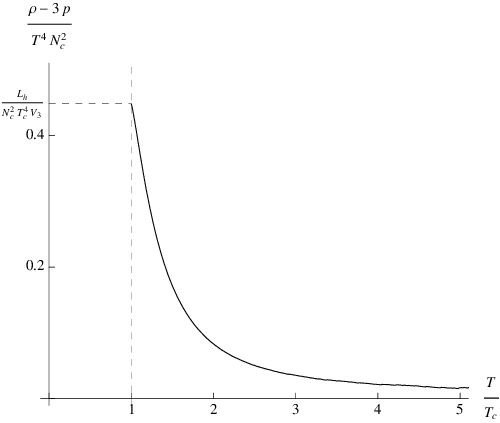}
 \end{center}
 \caption[]{Temperature dependence of the trace of the stress tensor in the black-hole phase for $T_c<T<5T_c$.}
 \label{e-3pfig}
\end{figure}

\begin{figure}[h!]
 \begin{center}
\includegraphics[scale=1.0]{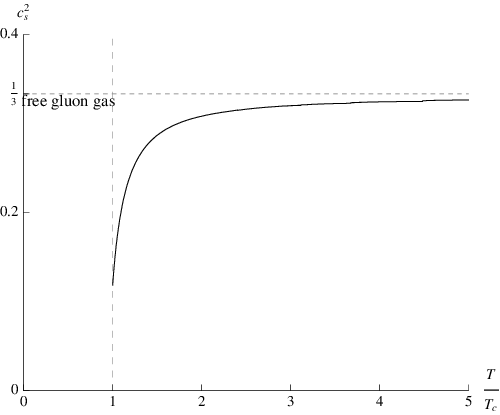}
 \end{center}
 \caption[]{The speed of sound  in the black-hole phase for $T_c<T<5T_c$.}
 \label{sound}
\end{figure}

\item The speed of sound is given by $c_s^2 \equiv  (\de p /\de \rho)_S = S/C_v $, where $C_v = \de E/\de T$
is the specific heat.   As expected in the deconfined phase of pure Yang-Mills in 4D, this quantity
approaches from below the conformal value,  $c_s^2 \to 1/3$ as $T\to \infty$. This can be seen
using the high-temperature expansion of the thermodynamic quantities derived in Appendix \ref{highT}.
A concrete example in our setup is shown in Figure \ref{sound}.

\end{itemize}

\section{The axion background at finite temperature.}

 The effect of a non-trivial vacuum angle in Yang-Mills is
captured by including in the bulk a five-dimensional axion.
The axion $a$ is dual to the instanton density $Tr[F\wedge F]$. In
particular its UV boundary value is the UV value of the QCD
$\theta$-angle. Moreover, its profile $a(r)$ in the vacuum
 solution may  be interpreted as the ``running''
$\theta$-angle in analogy with the dilaton, that we interpret  as the running coupling constant.
This was explained and justified in \cite{ihqcd}.

The axion action is suppressed by ${\cal O}(1/N_c^2)$ with respect to the action
for the other fields,  (\ref{a1}) :
\be
S_{\rm axion} = {M_p^3\over 2}\int d^5x \sqrt{-g} ~Z(\l)~\left(\de_\mu a\right)^2\lab{d12a}
\ee
where $Z(\l)$ has the following asymptotic expansions, \cite{ihqcd},
\be
\lim_{\l\to 0}Z(\l)=Z_a\left[1+{\cal O}(\l)\right]\sp \lim_{\l\to\infty}Z(\l)\sim \l^4
\ee
The scale $Z_a$ determines the topological susceptibility,
while the strong coupling asymptotics are dictated from glueball universality.

At zero temperature the axion solution that is compatible with known properties of large-$N_c$ YM is
\be
a(r)=(\theta_{UV}+2\pi k_0)~{\int_r^{r_0} {dr \over e^{3A}Z(\l)}
\over \int_0^{r_0} {dr \over e^{3A}Z(\l)}} \sp k_0\in Z
\ee
where $\theta_{UV}$ is the UV value of the $\theta$ parameter defined as an angle in the range $[0,2\pi)$, $k_0$ is an integer that labels
oblique confining vacua and is determined by minimizing the $\theta$-dependent vacuum energy
\be
 E(\theta_{UV},k)=-{M_p^3\over 2}~{(\theta_{UV}+2\pi k)^2\over   \int_0^r {dr} {e^{-3A}\over Z(\l)}}
\ee
Finally $r_0$ is the position of the singularity of the zero-temperature geometry in the radial coordinates.

{}From this solution we can extract the topological susceptibility and topological density condensate as
\be
\chi={Z(0)M_p^3\over \int_0^{r_0}{dr Z(0)\over e^{3A}Z(\l)}}\sp
{\langle Tr[F\wedge F]\rangle }=-{32\pi^2\over \ell^3}{(\theta_{UV}+2\pi k_0)\over
\int_0^{r_0}dr{Z(0)\over e^{3A}Z(\l)}}
\ee

At finite temperature, below the deconfining phase transition the situation is similar to $T=0$,
 since it is the same vacuum solution describing the physics.
Above $T_c$ however, once should switch to the black-hole solution.
In the black-hole solution  the axion background satisfies,

\be
\ddot a+\left(3\dot A+{\dot f\over f}+(\partial_{\l}\log Z)\dot\l\right)\dot a=0
\label{31s}
\ee
an equation that can be integrated once to
\be
\dot a={C_a\over f~e^{3A}~Z(\l)}
\label{d12}\ee
where $C_a$ is a constant.
Integrating once more we obtain
\be
a(r)=\theta_{UV}+C_a\int_0^{r}{dr' \over f~e^{3A}~Z(\l)}
\ee
Unlike the zero temperature solution, the non-trivial solution here has a singularity at the horizon.
Indeed, $f(r)\sim f_0(r-r_h)$ while both $Z(\l)$ and $e^A$ are regular there. Therefore the function
$\int_0^{r}{dr' \over f~e^{3A}~Z(\l)}$ diverges logarithmically as $r\to r_h$.
Since the background solution must be regular everywhere, we must necessarily impose $C_a=0$.

Therefore in the deconfined phase, the axion background is constant
\be
a(r)=\theta_{UV}
\ee
and the topological density condensate (proportional to $C_a$) vanishes
\be
{\langle F\wedge F\rangle }_{\rm deconfined}=0
\ee
In fact this can be generalized to higher derivative terms containing the axion.
This shows that in the deconfined phase, at large $N_c$ all moments of the topological density vanish.
This is in agreement with general expectations at large $N_c$, since in the deconfined phase such moments
obtain contributions only from instantons, that give vanishing $e^{-N_c}$ contributions as $N_c\to \infty$.
This expectation is in accordance with lattice calculations , \cite{ltheta}.

\section{Reduction and solution of the system in scalar variables}\lab{covvar}

Einstein's equations are hard to solve for a generic dilaton potential.
This is the case even in numerical evaluation. Here we shall present
a method to reduce the degree of the system of equations from 5 to 2 by introducing
variables that are explicitly invariant under general coordinate transformations.

There are a number of nice features of this reduction. First,
the thermodynamics will only depend on the reduced system. Therefore all thermodynamic observables
are determined by solving a coupled system of first order equations. This also
clarifies the number of physical parameters in the theory, see Section \ref{parasol}. It also allows us to find analytic
solutions (Appendix \ref{AnalyticSolutions}).

Aa another bonus, the UV expansion of the finite temperature metric and dilaton used throughout
the paper, eqs. (\ref{b-bo}-\ref{l-lo}) , are most easily derived with this formalism. This is done
in Subsection \ref{soluv}.

Finally, this form of Einstein's equation is used in Appendix \ref{conUVIR} to show
the existence of black-hole solutions with arbitrary $\l_h$ and $AdS$ UV asymptotics.

This section can be read independently of the rest of the paper. All of the ingredients necessary to derive the thermodynamics
from the Einstein's equations in a diffeomorphism-invariant manner are presented in this section in a self-contained way.

\subsection{Scalar variables}

The basic idea was introduced in \cite{ihqcd} for the zero-$T$ case and reviewed in Section \ref{sec2}.
Here we present a generalization of \cite{ihqcd} to the black-hole ansatz.
We propose to solve the Einstein's
equations by introducing the following scalar variables:
\be\lab{psv} X(\f) = \frac{\f'}{3A'},\qquad Y(\f) = \frac{g'}{4A'}.
\ee
Note $X$ and $Y$ are invariant under radial coordinate transformations. These variables obey
the following first order equations:
\bea\label{Xeq}
\frac{dX}{d\f} &=& -\frac43(1-X^2+Y)\le(1+\frac{3}{8X}\frac{d\log V}{d\f}\ri),\\
\frac{dY}{d\f} &=& -\frac{4}{3}(1-X^2+Y)\frac{Y}{X }. \label{Yeq}
\eea
As shown in Section \ref{freeXY}, the thermodynamics of the dual field theory are
completely determined by knowledge of $X$ and $Y$ as a function of $\f$. Roughly speaking,
$Y$ is dual to the entropy and $X$ to the energy of the gluon fluid.

It is crucial that the system above always admits a special solution, $Y=0$. This corresponds
to the {\em thermal gas solution} (or the zero-$T$ solution for non-compact time),
that is present for any $V$. This is because, for the solution $Y=0$,
the equation (\ref{Xeq}) reduces to the corresponding zero-$T$ equation (\ref{X0eq})
whose only solution with fixed IR asymptotics is the thermal gas.

The metric functions are given in terms of $X$ and $Y$ by integrating eqs. (\ref{psv}). Let us introduce a  cut-off
$\f_0$, that plays the role of  the regularized UV boundary.
We call the value of the scale factor $A$ at this point as $A_0$.
On the other hand, the black-hole asymptotics requires that $g$ vanishes ($f\to 1$) on the boundary.
With these initial conditions, integration of (\ref{psv}) gives
\bea
A &=& A_0 + \int_{\f_0}^{\f} \frac{1}{3X}d\tilde{\f},\lab{Adet}\\
g &=&  \int_{\f_0}^{\f} \frac43\frac{Y}{X}d\tilde{\f}.\lab{gdet}
\eea
We note that one does not need the cut-off $\f_0$ in solving (\ref{Xeq}), (\ref{Yeq}). As we prove in the sequel,
the physical observables only depend on the functions $X$ and $Y$. Therefore they will be independent
of the cut-off $\f_0$.

In appendix \ref{XYeq} we prove that the reduced $X$-$Y$ system solves the full equations
of motion (\ref{eom1app}-\ref{eom4app}) in the original $u$-variable.
There, we also provide formulas for the derivatives
of the metric functions
$A'$, $g'$ and $\f'$ in eqs. (\ref{Ap}), (\ref{Fp}) and (\ref{gp}),
 hence completing the full five-degree system of equations.

Let us finally note that, one can
invert the equations (\ref{Xeq}) and  (\ref{Yeq}) for the dilaton potential:
\be\lab{ftpot} V(\Phi) =
V_0(1+Y-X^2)e^{-\frac83\int^{\f}_{-\infty}dt
\le(X-\frac{Y}{2X}\ri)}.
\ee
This equation will be useful later on. We note that for $Y=0$ it reduces to the corresponding
zero-$T$ equation in \cite{ihqcd}.

\subsection{UV Asymptotics}\lab{soluv}

In what follows we prefer working in the $\l=\exp(\f)$ variable
instead of $\f$\footnote{For notational simplicity,
we shall allow for an abuse of notation by referring to, in fact,
{\em different} functions when we write e.g. $V(\l)$ and $V(\f)$, related by $V(x) \to V(e^x)$.}.
The black-hole deformation is obtained by turning on $Y$ near the
boundary. In other words $Y$ should vanish as one approaches the boundary.
In addition, the condition that the BH solution approaches the thermal
gas solution requires that $X\to X_0$ on the boundary.
We recall that the UV asymptotics of the function $X_0$
is presented in eq. (\ref{x0}). Let us now write,
\begin{equation}\label{Xas}
X(\l) = X_0(\l) + \delta X(\l), \qquad \l\to 0,
\end{equation}
where $\delta X \ll X_0$ for small $\l$. Studying the small $\l$ asymptotics of
the explicit solution for $Y$, given in eqs. (\ref{defs},\ref{soly}), one learns that $Y$
vanishes non-perturbatively in $\l$; to be precise $Y\sim e^{-4/b_0\l}\l^{-4b}$. This can
also be seen by assuming that $Y$ is exponentially small and than solving (\ref{Yeq})
in the vicinity of $\l=0$. Then, it follows from (\ref{Xeq}) that
$\delta X$ also vanishes with the same exponential factor.

One derives the asymptotic behavior of the functions $Y$ and $\delta X$ by solving (\ref{Yeq})
and (\ref{Xeq}) near $\l=0$. We spare the details of this calculation to appendix \ref{UVasympApp}.
The result is
\bea\lab{Y0}
Y(\l) &=& \Y0\,\, e^{-\frac{4}{b_0\l}}(b_0\l)^{-4b}, \\
\lab{C0}
\delta X(\l) &=& \le[\frac{\Y0/2-\C0}{X_0}+\C0 X_0\ri]e^{-\frac{4}{b_0\l}}(b_0\l)^{-4b}.
\eea

Here $\Y0$ and $\C0$ are integration constants.
They retain finite values as the cut-off is removed by sending $\l_0\to 0$. These
values can be computed by matching the solutions above to the
full solution of (\ref{Xeq}) and (\ref{Yeq}). Generally these integration constants are non-trivial
functions of temperature and this dependence is determined by the regularity condition
at the horizon. This is explained in Section \ref{reghor} below.

The physical meaning of these integration constants will become clear
below: {\em $\C0$ determines the energy, $\Y0$ determines the
entropy and the combination, \be\lab{G0} \G0  = \C0
-\frac{\Y0}{2}, \ee determines the vev of the gluon condensate
in the gluon plasma.}

Finally, we would like to know the UV expansions of the metric functions $A,~f$ and $\f$ in the radial
variable $u$ or $r$. These can be determined using the asymptotic expressions for $X$ and $Y$ given by (\ref{Y0}) and (\ref{C0})
 in the formulae for the derivatives $A',~f'$ and $\f'$  given by (\ref{Ap},\ref{Fp},\ref{gp}).

To make use of these equations,  we first define the fluctuations in $A$ and $\f$ in the domain wall frame  as,
\be\lab{flucAf}
\delta A(u) = A(u) - A_o(u),\qquad \delta\f(u) = \f(u)-\f_o(u).
\ee
One then finds:
\bea
\delta A &=&  \frac12 \G0 e^{-\frac{4}{b_0\l}}(b_0\l)^{-4b} +\cdots =  \frac12 \G0 (\Lambda \ell)^4  e^{4u/\ell}(-u/\ell)^{-16/9} +\cdots\label{adet2} \\
f &=& 1 - \Y0 e^{-\frac{4}{b_0\l}}(b_0\l)^{-4b} +\cdots = 1- \Y0 (\Lambda \ell)^4 e^{4u/\ell}(-u/\ell)^{-16/9}+\cdots\label{fdet2} \\
\delta \f &=&  \frac94 \G0 e^{-\frac{4}{b_0\l}}(b_0\l)^{-4b-1}
+\cdots =  \frac94 \G0 (\Lambda \ell)^4
e^{4u/\ell}(-u/\ell)^{-16/9+1}+\cdots\label{phidet2}. \eea In the last
equations we used the results in Appendix \ref{XYeq} to convert
the expression in the $u$ variable, with the scale $\Lambda$
defined in eq. (\ref{LQCD-2}).

Finally, we want to write an expression for the fluctuation in the
conformal frame scale factor and dilaton. To this end,  we can use
the relation (valid close to the UV boundary $r\to 0$):
\be\label{rtou} {r\over \ell} = e^{u/\ell}
\left(-u /\ell\right)^{-4/9},
\ee
see Appendix \ref{XYeq}.  However, it is not enough to just re-express
eq. (\ref{adet2}) and (\ref{phidet2}) in terms of $r$ through
(\ref{rtou}): as shown in Appendix (\ref{fluctap}) one gets an
extra shift in $\delta A$ when changing the frame. The final
result is: \be \delta A(r) = {2\over 5} \G0 (\La r)^4, \quad
\delta \Phi = \frac94 \G0 (\La r)^4 \log r\Lambda, \quad f(r) = 1
- \Y0 (\La r)^4 \ee

Comparison of these expansions to (\ref{fUV}) and (\ref{b-bo})   relates the coefficients $\Y0$ and $\C0$ defined quantities
$\cG$ and $C$ that enter in the free energy, eq (\ref{free-2}) as
\be\lab{C0Y0rel}
\Y0 = \frac{C\ell }{4(\ell \Lambda)^4},\qquad \G0 =\frac52 \frac{\cG \ell}{ (\Lambda\ell)^4}.
\ee

\subsection{Asymptotics near the horizon}\lab{reghor}

The dependence on $T$ on the constants  $\C0$ and $\Y0$ is
determined by the geometry of the black-hole near the horizon.
Near the horizon, the black-hole solution should be regular. In particular one
requires that the various metric functions and the dilaton behave
as,
\begin{eqnarray}
  f(r) &=& f_1(r_h-r)+\cO(r_h-r)^2\lab{fash} \\
  A(r) &=& A_h + A_1(r_h-r)+\cO(r_h-r)^2 \lab{Aash} \\
  \f(r) &=& \f_h + \f_1(r_h-r)+\cO(r_h-r)^2 \lab{pash}
\end{eqnarray}

In terms of $Y$ and $X$, the requirement of regular horizon translates into the following conditions:
\begin{eqnarray}
  Y(\f) &=& \frac{Y_h}{\f_h-\f} + Y_1 + \cO(\f_h-f) \lab{Yhor},\\
  X(\f) &=& -\frac43 Y_h + X_1(\f_h-\f) +\cO(\f_h-f)^2\lab{Xhor},
\end{eqnarray}
where $Y_1$ and $X_1$ are yet undetermined constants and
\begin{equation}\label{yh}
    Y_h = -\frac{\f_1}{4A_1}.
\end{equation}

When one solves the coupled system of eqs. (\ref{Xeq}) and
(\ref{Yeq}) near any point $\f_i$, the solution will generally be
parameterized by two integration constants. At first sight, it seems
that these two parameters can be taken as $Y_h$ and
$\f_h$. However a more careful look at the system of equations
reveals that demanding  a regular horizon reduces the number of
parameters to a single one, that fixes  $Y_h$ in terms of $\f_h$.
This can easily be seen by substituting (\ref{Yhor}) and
(\ref{Xhor}) in (\ref{Xeq}). One obtains:

\begin{equation}\label{detyh}
Y_h = \frac{9}{32}\frac{V'(\f_h)}{V(\f_h)}.
\end{equation}
Similarly $X_1$, $Y_1$, etc. are determined from the sub-leading terms in the expansion near the horizon.
See Appendix \ref{X1Y1App}.
One finds the higher order coefficients in the expansion of (\ref{Yhor}) and (\ref{Xhor}) order by order.
The important point is that there is no room for an arbitrary integration constant, the requirement of a regular
horizon completely determines all of the coefficients in terms of $\f_h$.
This shows that we
have a single parameter family of solutions of the $X$-$Y$ system,
parameterized  by the location of the horizon $\f_h$.

The near horizon solution presented here is continuously connected to the near UV solution that is presented in Section \ref{soluv}. This fact can easily be derived by using the analytic structure of the
equations of motion (\ref{Yeq}) and (\ref{Xeq}). See appendix  \ref{conUVIR}.

\subsection{Thermodynamic functions and relations}\lab{freeXY}

The thermodynamics is completely determined in terms of the integration constants $\C0$ and $\Y0$. In the following, we
present the derivations one by one.

\vspace{0.5cm}
{\em Temperature}
\vspace{0.5cm}

The temperature of the dual gauge theory is given by the derivative of $f$ at the horizon,
\be\lab{T1}
T = \frac{|\dot{f}(r_h)|}{4\pi}.
\ee
Here, we shall express it in terms of the solution of the $X-Y$ system. The computation is straightforward and
the details are presented in Appendix \ref{detailT}. One finds,
\be\lab{T2u}
T = \frac{\Lambda}{\pi}~\le(\Lambda~\ell \ri)^3~\frac{\Y0(\l_h)}{b^3(\l_h)}.
\ee
where $b(\l)$ is determined from $X$ by (\ref{Adet}).  This equation gives $T$ in the physical units of $\Lambda$.

\vspace{0.5cm}
{\em Entropy}
\vspace{0.5cm}

The entropy of the field theory is determined by the area of the horizon as in (\ref{entropy}):
\be\lab{ent}
S= 4\pi\sigma b^3(\l_h),
\ee
where $b(\l)$ is again determined from $X$ by (\ref{Adet}).

We note that, the equations (\ref{T2u}) and (\ref{ent}) combine to yield the entropic  contribution to the free energy density as follows,
\be\lab{ent2}
ST =  4\sigma \Y0 \ell^3 \Lambda^4.
\ee
This equation clarifies the physical meaning of the integration constant $\Y0$.

\vspace{0.5cm}
{\em Free Energy and Energy}
\vspace{0.5cm}

One can obtain an exact expression for the free energy in terms of the scalar variables. This is done by converting
the eq. (\ref{actionbh}) in $\l$ using the equations (\ref{Ap},\ref{Fp},\ref{gp}). This calculation is explained in detail in the appendix
\ref{freelambda}. The result is expressed very simply in terms of the constants of motion defined in the previous section:

\begin{equation}\label{freeCY}
{\cal F} = - p V_3 =  \sigma \Lambda^4 \ell^3 \le(6\C0 - 4\Y0\ri),
\end{equation}
where the second equation relates pressure to the free energy of the system.
 Using eq. (\ref{G0}) and (\ref{C0Y0rel}), the above expression coincides
with (\ref{free-2})

The energy follows directly from (\ref{freeCY}) and (\ref{ent2}) as,
\be\lab{energy}
E=  \rho V_3 = 6\sigma \C0 \Lambda^4\ell^3,
\ee
where we defined the {\em energy density} as $\rho$. This clarifies the physical meaning of the integration constant $\C0$.

\vspace{0.5cm}
{\em Specific heat}
\vspace{0.5cm}

The specific heat is given by,
\be\lab{cv}
C_v = \frac{d E}{dT} = 6 \sigma \Lambda^4\ell^3  \frac{d\C0}{dT} =
4 \sigma \Lambda^4 \ell^3 \le(\frac{d\Y0}{dT}-\frac{\Y0}{T}\ri).
\ee
where we used (\ref{CinY}), see below.

\vspace{0.5cm}
{\em Speed of sound}
\vspace{0.5cm}

The speed of sound in the medium is defined by $c_s^2 = d p/d\rho$, where the pressure $p$ is
given in eq. (\ref{freeCY}). By using thermodynamic relations,  one can show that:
\be\lab{sound2}
c_s^2 = \frac{S}{C_v}.
\ee
Using (\ref{cv}) we can derive a relation for $c_s$ directly in terms of $\Y0$:
\be\lab{soundspeed}
\frac{1}{c_s^2} = \frac{d\log \Y0}{d\log T}-1.
\ee
Using the high-$T$ asymptotics of $\Y0$ given in eq. (\ref{Y0hT}), we see that $c_s^2\to 1/3$ for $T/T_c\gg 1$ as required
from the conformality in this limit.

We refer the reader to the results presented in Appendix \ref{highT} for the high temperature behavior of the thermodynamical
functions discussed in this section.

\vspace{0.5cm}
{\em A relation between $\C0$ and $\Y0$}
\vspace{0.5cm}

One can also relate $\C0$ and $\Y0$, by using the definition of the entropy,
\be\lab{sdfdt}
S =   - \frac{d {\cal F}}{dT}.
\ee
It follows from (\ref{sdfdt}), (\ref{freeCY}) and (\ref{entropy}) that,
\be\lab{CinY}
\C0(T) = \frac23 \Y0(T) - \frac23\int_{T_c}^T\frac{\Y0(t)}{t} dt.
\ee
Therefore, knowledge of $\Y0$ as a function of $T$ determines the coefficient
$\C0$ analytically.

Equation (\ref{CinY}) also helps us determine the
thermodynamics at high-$T$. As $T$ increases, $\l_h$ approaches zero,
hence the geometry of the black-hole becomes the geometry of an
$AdS$ black-hole. For the $AdS$ black-hole one has, $b=\ell/r$,
$f=1-(r/r_h)^4$ and $T=1/\pi r_h$. Therefore from the definition
of $Y$ in {\ref{psv}), of $\Y0$ in (\ref{Y0C0}) and the conversion
between $\l$ and $r$ coordinates near the boundary (\ref{rl}) we
obtain,
\be
\lab{Y0hT} \Y0(T) \to \le(\frac{\pi
T}{\Lambda}\ri)^4\qquad \frac{T}{T_c}\gg 1. \ee Using this in
(\ref{CinY}) one obtains, \be\lab{c1hT} \frac{\C0 - \Y0/2}{T^4}
\to 0 \qquad \frac{T}{T_c}\gg 1.
 \ee
We present the explicit high-T behavior of this function in Appendix \ref{highT}. We also show
that this implies, through eq. (\ref{G0}), that   the gluon condensate divided by $T^4$ vanishes at high-$T$.

\vspace{0.5cm} {\em A useful equation for $s/T^3$} \vspace{0.5cm}

One can also obtain a relation between the dimensionless
thermodynamic observable $s/T^3$ and the dilaton potential as
follows: \be\lab{sT3} \frac{s}{T^3} = \frac{4\pi^2
(12M_p\ell)^3}{\ell^6} \frac{e^{-4\int_0^{\l_h}\frac{d\l}{\l}
X(\l)}}{V(\l_h)^3} = \frac{4\pi^2(16M_p\ell)^3}{27\ell^3}
\le(\frac{W(\l_h)}{V(\l_h)}\ri)^3. \ee In the second line we used
the definition of the thermal superpotential in terms of $X$, eq.
(\ref{relXW}). We refer to App. \ref{dersT3} for the derivation.

This relation is valid for an arbitrary background. In our case,
where the small extreme high T limit is free YM theory, the Planck
mass is given by (\ref{Planck}). As a consistency check we find in
the case of a conformal theory, where $V=12/\ell^2$ and $W$ are
constants and $X=0$, that (\ref{sT3}) reduces to the conformal
thermodynamics: \be\lab{confsT3} \frac{s}{T^3} = 4\pi^4
(M_p\ell)^3, \ee which again agrees with the free YM gas upon use
of (\ref{Planck}).

Using (\ref{sT3}) together with the relation between $T$ and
$\f_h$ (\ref{Tder3}), one can directly relate the thermodynamic
observable $s/T^3$ as a function of $T/\Lambda$ to the dilaton
potential $V(\f)$. We note that this relation side-steps the
construction of the constants of motion $\C0$ and $\Y0$ described
above, a procedure that is numerically quite non-trivial. As
$s/T^3$ as a function of $T/\Lambda$ is one of the main
observables that can be used in comparison with the lattice data
in principle, (\ref{sT3}) is quite utile\footnote{However, one
usually uses $s/T^3$ as a function of $T/T_c$ to compare with the
lattice data. In this case the utility of (\ref{sT3}) is reduced
as one needs additional knowledge of $T_c/\Lambda$, which in turn
requires the free energy, i.e. knowledge of $\C0$ and $Y0$.}.

\subsection{Matching the zero-$T$ solution}\lab{matchcov}

As we discussed above, the zero-$T$ solutions of the $X$-$Y$ system correspond to the special case  $Y=0$.
We analyzed the entire set of solutions in this case, in Appendix \ref{fixedXY}. The conclusion
of this analysis is a rephrasing, in terms of the scalar variables,
of the general classification in terms of the superpotential that we have
discussed in Section \ref{superIR} and in Appendix \ref{superapp}. For any dilaton potential
asymptotic freedom in the UV \ie $V(\l)\to V_0 + V_1 \l+\cdots$, as $\l\to 0$
and which exhibits exponential asymptotics in the IR, $V(\l)\to \l^{2Q}\le(\log\l\ri)^P$ as $\l\to\infty$,
there are three different classes of solutions to $X$, with different IR behavior (as $\l\to\infty$):
\begin{enumerate}
\item[i.] Solutions with ``special'' type of IR asymptotics: $X\to -3Q/4$.
We denote it $X_*(\f)$,
\item[ii.] Solutions with ``generic'' type of IR asymptotics:  $X\to -1$,
\item[iii.] Solutions with ``bouncing'' type of IR asymptotics: $X\to 0$.
\end{enumerate}
The second case is not desired because
 the fluctuations of the bulk fields fall into the singularity, hence it is not of repulsive type,
whereas the last case corresponds to $\l$ being a non-monotonic function of the RG scale, hence it can not
yield a sensible RG flow. Therefore we based our holographic
construction on the special class, case i.

Now, we consider the black-hole solutions to the same potential, with $Y\ne 0$. {\it A priori}, there is
no guarantee that a regular black-hole solution does not correspond to cases ii or iii as the deformation
is taken away \ie $Y\to 0$ (by sending the BH horizon to the singularity).
However, as presented in Section \ref{superIR},  we have  the following:\\
\noindent
{\bf No-go theorem:} {\em The only vacuum solutions of Einstein-dilaton gravity, with exponential
asymptotics given above, that can be continuously lifted to a regular black-hole correspond
to the special class of solutions, case i.  }\\

In the language of scalar variables, the proof is simple. In the previous subsection, we gave the condition for regularity of the horizon, eq. (\ref{detyh}). We can write this in terms of $X(\Phi)$ using eq. (\ref{Xhor})
\be\lab{reg2}
X(\f_h) = -\frac38 \frac{V'(\f_h)}{V(\f_h)}.
\ee
For all regular BHs one should be able to push the horizon down to the singularity of the
zero-$T$ solution $r_0$,
by continuously sending $r_h\to r_0$.
First, we rule out case iii: Since $V(\f),\, V'(\f)>0$ for all $\f$, one finds from (\ref{reg2}) that $X(\f_h)<0$. On the other hand,
in the region close to the singularity in case iii, one finds $X>0$, see Appendix \ref{fixedXY}. Thus, it is not possible
that regular black-holes can be continuously connected to case iii, as the horizon is taken close to the singularity, $\Phi_h\to \infty$.
In this limit, (\ref{reg2}) clearly fixes $X\to -3Q/4$. Therefore, the function
$X$ exhibits the desired asymptotics of case i.

On the other hand, one can solve for $Y$ in this asymptotic region,
using the analytic solution of (\ref{soly}). One finds,
\be\lab{soly2} Y(\f)\to \frac{e^{c\f}}{C_1 - \frac{d}{c} e^{c\f}},
\qquad as\,\,\, \f_h\gg 1 \ee Here $c>0$ and $d>0$ are given in
terms of $Q$ and the location of the horizon is given by the
integration constant $C_1$ as $$C_1=\frac{d}{c}e^{c\f_h}.$$ We can
show that, in the limit $\f_h\to\infty$ the entire $Y$ function
becomes a spike centered at $\f=\infty$: From (\ref{Yeq}) we
observe that the RHS is positive definite, hence $Y$ is a
monotonically increasing function of $\f$.
From (\ref{soly2}), it is also clear that $Y(\f_h)\to 0$ as $\f_h\to\infty$, for any $\f\gg1$ but $\f<\infty$, in the asymptotic region.
Combining these two facts, we learn that $Y(\f)$ should vanish for {\em all} $\f\ne\infty$, including the UV region. On the other hand,
it diverges exactly at $\f_h=\infty$.\footnote{In fact, one can prove that $Y(\f)$ becomes proportional to $\delta(\f_h-\f)$ in the limit $\f_h\to\infty$ by using a
limit representation of the delta function. We will not need this here however. }

Thus we proved that, in the limit $\f_h\to\infty$ of {\em any regular BH}, $X$ limits to $X_*$ and $Y(\f)\to 0$ for any $\f<\infty$.
This corresponds to the zero-$T$ solution $Y=0$, with the integration constant of $X$ equation tuned to $X(\infty)= -3Q/4$. In other words, $X(\f)\to X_*(\f)$
in the entire range of $\f$ as $\f_h\to\infty$.

\subsection{Parameters of the solutions}\lab{parasol}

Here we examine the integration constants  in the Einstein equations for a generic black-hole solution.
We solve the system by requiring the asymptotic behavior of $X$ and
$Y$ near the horizon, as discussed in Section \ref{reghor}. This solution flows in the UV to
$X\to X_0$ and $Y\to 0$ as described in Subsection \ref{soluv}. Thus
one has a single integration constant $\f_h$ from eqs. (\ref{Xeq})
and (\ref{Yeq}). It determines the temperature by eq. (\ref{T2u}).

 After $X$ and $Y$ are determined, the  metric and dilaton are obtained using eqs.
(\ref{Ap})-(\ref{gp}). The condition $g\to0$ near the boundary
fixes the integration constant in  (\ref{gp}). The two remaining
constants are $\l_0$ and $A_0$. As described in \cite{ihqcd}, these
combine to determine the mass scale $\Lambda$ of the physical
system, (\ref{LQCD-2}). This combination can be viewed as one
integration constant of the two equations (\ref{Ap}) and
(\ref{Fp}). (The other one is irrelevant due to a shift symmetry
in $r$, see (\cite{ihqcd}).) As we require that the finite-$T$
solution approaches to the zero-$T$ on near the boundary, the value
of $\Lambda\ell$ is determined by the corresponding value at zero
$T$.

Thus, we conclude that  {\em one has only two parameters in the
solutions, $\Lambda$ and $\l_h$ which corresponds to the $\Lambda_{QCD}$
and the temperature}. Furthermore, one of them i.e. $\Lambda$ is completely fixed
by the zero-$T$ solution. It is practically set by the mass of the lowest glueball in
the spectrum, see \cite{ihqcd}.

\section{Outlook}

In this paper  we have analyzed in detail the equilibrium thermodynamics of the 5D Einstein-dilaton
system that was proposed in \cite{ihqcd} as a phenomenological holographic dual of
4D large-$N_c$ pure Yang Mills.
There is a variety of possible directions to extend our work.

An example of an explicit background with similar asymptotics, in critical (IIB compactified to 5D)
or a non-critical string theory would be desirable. In addition to justifying our phenomenological set-up based on
the principles of AdS/CFT correspondence, this would allow more detailed studies on the $\alpha'$ corrections
and how they can affect our results especially in the UV. For the thermodynamics of QCD, the higher
derivative corrections are desirable also for a phenomenological reason: It is well-known \cite{bl}
that $\eta/s$ is constant in any gravitational theory based on a two-derivative action. However,
this quantity is expected to be a non-trivial function of $T$ in QCD that becomes asymptotically large for large temperatures.
The higher derivative corrections may provide
the desired $T$ dependence.

Having set the general construction in this paper, a natural step forward is to compute dynamical observables
 (bulk viscosity, drag force, jet quenching parameter) that are important for the physics of
the RHIC collider and the upcoming LHC collider . We will address this problem in the near future, \cite{GKMicN}.
Another related issue is the computation of the various Debye screening masses, where a better comparison with lattice data can emerge.

Another important
direction involves the meson sector, that should be introduced through probe $D_4+\overline{D_4}$ branes in the background.
Introduction of baryon chemical potential is the next very interesting step to analyze. We expect that this
will involve the study of charged black-holes  under the overall  $U(1)$ gauge field of the flavor branes.

\vskip 1cm
\centerline{\bf Acknowledgements}
\addcontentsline{toc}{section}{Acknowledgements}

We would like to thank  Ofer Aharony, Luis Alvarez-Gaum\'e, Adi Armoni, Massimo Bianchi, Francesco Bigazzi, Richard Brower, Aldo Cotrone, Frank Ferrari,  Steve Gubser,
Gary Horowitz, Thomas Hertog, Edmond Iancu, Frithjof Karsch, David  Kutasov, Hong Liu, Biagio Lucini,
 David Mateos, Carlos Nunez,  Andrei Parnachev, Ioannis Papadimitriu,
Edward Shuryak,
Kostas Skenderis, Dam Son, Jacob Sonnenschein,  Shigeki Sugimoto, Marika Taylor,  Michael  Teper, Mithat
 Unsal, Urs Wiedemann,  and Lawrence Yaffe for discussions.
This work was partially supported by ANR grant NT05-1-41861,
RTN contracts MRTN-CT-2004-005104 and MRTN-CT-2004-503369, CNRS PICS  3059 and 3747,
Marie Curie Intra-European Fellowships
MEIF-CT-2006-039962 and MEIF-CT-2006--039369, INFN, and by the VIDI
grant 016.069.313 from the Dutch Organisation for Scientific Research
(NWO).

Elias Kiritsis is on leave of absence from CPHT, Ecole Polytechnique (UMR du CNRS 7644).

\newpage
\appendix
\renewcommand{\theequation}{\thesection.\arabic{equation}}
\addcontentsline{toc}{section}{Appendices}

\section*{APPENDIX}

\section{Various forms of Einstein's Equations }
We use a metric signature $(-,+,+,+,+)$. We start from the action:
\begin{equation}
   S_5=-M^3\int d^5x\sqrt{g}
\left[R-{4\over 3}(\partial\Phi)^2+V(\Phi) \right]+2M^3\int_{\partial M}d^4x \sqrt{h}~K
    \label{app1}\end{equation}
    with
    \be
K_{\m\n}\equiv  -\nabla_\mu n_\nu = {1\over 2}n^{\rho}\partial_{\rho}h_{\m\n}\sp K=h^{ab}K_{ab}
\label{app2}\ee
where $h_{ab}$ is the induced metric on the boundary and $n_{\m}$ is the (outward directed) unit
normal to the boundary. E.g. if $r$ denotes the $AdS$ conformal coordinate,
\be
n^{\mu}=-{1\over \sqrt{ g_{rr}}}\left({\partial\over
\partial r}\right)^{\mu}={{\delta^{\mu}}_{r}\over \sqrt{ g_{rr}}}.
\label{app3}\ee

The sign of the bulk term is chosen in such a way that 1) in the Euclidean regime,
the scalar field kinetic term is positive definite and 2) the curvature
of Euclidean $AdS$ is negative. With this choice, the sign  of the Gibbons-Hawking term
is fixed, as usual, by the requirement that the variation of the action does not contain metric derivatives.

Einstein's equations are:

\bea
&& E_{\m\n}-{4\over 3}\left[\pa_{\m}\Phi
\pa_{\nu}\Phi-{1\over 2}(\pa\Phi)^2 g_{\m\n}\right]-{1\over 2}g_{\m\n}V=0,
    \label{app4}\\
&& \Box_5 \Phi + {\de V\over \de\Phi} =0, \label{app41}
\eea
with the Einstein tensor defined as,
\begin{equation}
E_{\m\n}=R_{\m\n}-{1\over 2}Rg_{\m\n}.
    \label{app5}\end{equation}

\subsection{Conformal Frame}

Consider the following ansatz for the metric and dilaton:
\be
ds^2=b(r)^2\left[{dr^2\over f(r)}-f(r)dt^2+dx^idx^i\right], \qquad \Phi = \Phi(r).
 \label{app7}\ee
with   Einstein tensor,
\be
E_{rr}={3\dot{b}(4f\dot{b}+b\dot{f})\over 2b^2 f}\sp E_{tt}=-{3f(\dot{b}\dot{f}+2f\dot{b})
\over 2b},\sp E_{ij}={6\dot{b}\dot{f}+6f\ddot{b}+b\ddot{f}\over 2b}\delta_{ij},
 \label{app8}\ee
Laplacian,
\be
\square\Phi={f\over b^2}\ddot{\Phi}+{f\over b^2}\left({\dot{f}\over f}+3{\dot{b}\over b}\right)\dot{\Phi},
 \label{app9}\ee
and Dilaton stress tensor,
\be
T_{rr}={2\over 3}\dot{\Phi}^2+{b^2\over 2f}V\sp T_{tt}={2f^2\over 3}\dot{\Phi}^2-{b^2f\over 2}V\sp T_{ij}=-{2f\over 3}\dot{\Phi}^2+{b^2\over 2}V.
 \label{app10}\ee

The equations of motion are:
\begin{equation}
6{\dot{b}^2\over b^2}+3{\ddot{b}\over b}+3{\dot{b}\over b}{\dot{f}\over f}={b^2\over f} V,
\sp
6{\dot{b}^2\over b^2}-3{\ddot{b}\over b}={4\over 3}\dot{\Phi}^2,
\label{app11}\end{equation}

\begin{equation}
{\ddot{f}\over \dot{f}}+3{\dot{b}\over b}=0\sp \ddot{\Phi}+
\left({\dot{f}\over f}+3{\dot{b}\over b}\right)\dot{\Phi}+{3\over 8f}b^2 \frac{dV}{d\Phi}=0.
\label{app12}
\end{equation}

The second equation in (\ref{app12}) is not independent of the other three, and
it can be dropped.

 The Ricci scalar is:
 \be
 R=-{2\over 3}E=-{f\over b^2}\left[{\ddot f\over f}+8{\ddot b\over b}+8{\dot b\over b}{\dot f\over f}+4{\dot b^2\over b^2}\right]
 \ee

\subsection{Domain-wall frame}\lab{DWframe}

We define, \be\lab{bgapp} b=e^A,\qquad f = e^g, \ee and use the
domain-wall parametrization of the metric, \be\lab{covapp} dr =
e^{-A}du. \ee In this coordinate frame the metric has the following
form:
\begin{equation}\label{fTdwapp}
    ds^2 = e^{2A}\le(-f dt^2 + dx^2\ri) + \frac{du^2}{f}.
\end{equation}
The equations of motion (\ref{a11}) and (\ref{a12})
 in the variable $u$ take the following form:
\bea
12 {A'}^2+3A'g'-\frac43 {\f'}^2-e^{-g} V &=&0,\label{eom1app}\\
A'' +\frac49 {\f'}^2 &=& 0,\label{eom2app}\\
g'+\frac{g''}{g'}+4A' &=& 0,\label{eom3app}\\
\f''+4A'\f'+g'\f'+\frac38 e^{-g}\frac{dV}{d\f} &=& 0.\label{eom4app}
\eea

\subsection{Dilaton frame}\label{l-frame}

This frame uses $\Phi$ (or $\l\equiv \exp \Phi$) as the radial coordinate, and it
is in some sense ``maximally gauge fixed,'' since only the physical integration
constants of Einstein's equations appear in the metric. To change variables from $u$ to $\l$
it is useful to define a {\em  superpotential} $W(\Phi)$, such that eq. (\ref{eom2app}) is written as:
\be\label{eom5app}
A'= -{4\over 9} W(\Phi), \qquad \Phi' = {d W\over d \Phi}.
\ee
The coordinate change  $u=u(\Phi)$  is obtained by inverting the second equation  in (\ref{eom5app}).
The existence and properties of the superpotential in the zero-temperature and  black hole case will be
extensively discussed in Appendices \ref{superapp} and \ref{superfiniteT} respectively.
 The solution of Einstein's  equation in this frame is:
\be\label{eom6app}
ds^2 = {e^{-g(\Phi)}d\Phi^2\over (\de_{\Phi}W)^2 } + e^{2A(\Phi)} \left(-e^{g(\Phi)}dt^2 + dx^2\right),
\ee
where with a slight abuse of notation we have written $g(u(\Phi)) \equiv g(\Phi)$ and $A(u(\Phi)) \equiv A(\Phi)$.
One interesting property of the setup we are discussing is that two different dilaton potentials $V(\Phi)$ and
$\tilde{V}(\Phi)$, related by  $\tilde{V}(\Phi) = V(\Phi+ K)$ for some constant $K$ give essentially
the same set of solutions and the same physics. First, given a solution $(A(u), g(u), \Phi(u))$ of
eqs. (\ref{eom1app}-\ref{eom4app}) with potential $V(\Phi)$, it is straightforward to show that
the functions $(\tilde{A}(u), \tilde{g}(u), \tilde{\Phi}(u)) =  (A(u), g(u), \Phi(u) - K)$ solve
the system  with potential   $\tilde{V}(\Phi)$. The two solutions  are physically
equivalent (except for a change in initial conditions, which as we know \cite{ihqcd} only affects the
overall scale). This is easily seen writing the  second solution in the $\tilde{\Phi}$-frame:
\be\label{eom7app}
d\tilde{s}^2 = {e^{-\tilde{g}(\tilde{\Phi})}d\tilde{\Phi}^2\over (\de_{\tilde{\Phi}}\tilde{W})^2 }
+ e^{2\tilde{A}(\tilde{\Phi})} \left(-e^{\tilde{g}(\tilde{\Phi})}dt^2 + dx^2\right),
\ee
where $\tilde{W}$ is the appropriate superpotential. However, since $\tilde{A}(u) = A(u)$ and $\tilde{g}(u) = g(u)$, it follows that
\be
\tilde{A}(\tilde{\Phi}) = A(\Phi), \quad \tilde{W}(\tilde{\Phi}) = W(\Phi), \quad \tilde{g}(\tilde{\Phi}) = g(\Phi).
\ee
Thus,  after  a change of coordinates $\tilde{\Phi} \to \Phi = \tilde{\Phi} +K$
 the metric (\ref{eom7app}) becomes identical to the solution of the original system with potential $V(\Phi)$,  eq. (\ref{eom6app}). The initial
conditions for the two systems, such that the solutions coincide, are related by
$\tilde{A}(\Phi_0 -K) = A(\Phi_0)$.

Thus, there is an ``accidental degeneracy'' in  the classification of Einstein-dilaton gravity by the
dilaton potential, since the two potentials $V(\l)$ and $V(\kappa \l)$ lead to the same
physical results. For this reason, the value of the proportionality constant
between the dilaton $\l$ and  the physical Yang-Mills coupling $\l_t$ is irrelevant.

\subsection{Relating fluctuations in different frames}\label{fluctap}

In this appendix we work out the relation between the scale factor and dilaton fluctuations close to the boundary
in different frames.
For simplicity we set $\ell=1$.
Suppose we start with the zero-temperature and black-hole metrics,  both in the domain wall frame:
\be\label{metricsu}
ds^2 = {du^2\over f} + e^{2A^{u}}\left(f dt^2 + d x_3^2\right), \quad ds^2_o = du^2 + e^{2A^{u}_o}\left( dt^2 + d x_3^2\right)
\ee
For clarity,  we have added a label $(u)$ to the warp factor.
If the two solutions   obey the same boundary conditions at $u=-\infty$, then as shown in Section \ref{soluv} the
 two scale factors are related, to lowest order,  by:
\be \label{deltaAu}
\delta A^{u}\equiv A^{u} (u) - A_o^{u}(u) \simeq {\cal G}^u \,e^{4u}
\ee
for some constant ${\cal G}^u$. The difference between the dilatons, $\delta \Phi^u \equiv \Phi^u(u) -  \Phi^u_o(u)$,
  can be related to $\delta A$ by perturbing equation (\ref{eom2app}), which gives:
\be \label{pertu}
(\delta A^{u})'' + \frac89(\Phi_o^{u})'(\delta \Phi^{u})'
\ee
which can be integrated to give:
\be \label{deltaPhiu}
\delta\Phi^u \simeq \frac92\,  {\cal G}^u\,(-u)e^{4u},
\ee
in agreement with eq. (\ref{phidet2}).

Now we want to obtain the same quantities, namely $A-A_o$ and $\Phi-\Phi_o$, in conformal coordinates. Naively,
one may think that it should be enough to make the replacement $u\to \log r$ in eqs. (\ref{deltaAu}) and
(\ref{deltaPhiu}), since any correction to the relation between $u$ and $r$ would only affect higher orders
in $\delta A$ and $\delta \Phi$. This is however incorrect, as a careful analysis reveals.

What we want to obtain  is the difference $\delta A^r = A^r(r) - A^r_o(r)$, where the conformal warp factors
are such that the metrics have the form:
\be\label{metricsr}
ds^2 =  e^{2A^{r}}\left({dr^2\over f} +f dt^2 + d x_3^2\right), \quad ds^2_o =  e^{2A^{r}_o}\left(dr^2 + dt^2 + d x_3^2\right).
\ee
Now, it is clear that to bring the two metrics in this form one needs two  {\em different} coordinate transformations. We can first define $e^{-A^u(u)}du = dr$: if we perform this  coordinate  transformation  on both metrics we get:
\be  \label{metricsr2}
ds^2 =  e^{2A^{u}(u(r))}\left({dr^2\over f} +f dt^2 + d x_3^2\right), \quad ds^2_o =  e^{2A^{u}_o(u(r))+ 2 \delta A^u(u(r)) } dr^2 +  e^{2A^{u}_o(u(r))}\left(dt^2 + d x_3^2\right).
\ee
So the black-hole metric is in the conformal form, but the zero-temperature one is not. From the above
expression we read off that the function $A^r(r)$ in (\ref{metricsr})  is given by $A^u(u(r))$, but a similar relation  does not hold for $A_o^r$.

Let us define $\tilde{A}^r_o (r)\equiv A^{u}_o(u(r))$.
 To get the correct scale factor we have to perform a further coordinate
transformation to bring $ds_o^2$ in conformal form. We thus define:
\be \label{rtortilde}
\left(1 + \delta A^u(u(r))\right) dr = d\tilde{r}\quad \Rightarrow \quad \tilde{r} =  r + {1\over 5}\, {\cal G}^u\, r^5.
\ee
This transformation brings the metric $ds^2_o$ in conformal frame, parametrized by the coordinate $\tilde{r}$,
 with scale factor given $A^r_o(\tilde{r}) = \tilde{A}^r_o(r(\tilde{r}))$. Using the explicit form of $r(\tilde{r})$, we can write this as:
\be \label{shift0}
A^r_o (\tilde{r}) = \tilde{A}^r_o( \tilde{r} - {\cal G}^u\tilde{r}^5/5 ) \simeq \tilde{A}^r_o( \tilde{r}) - {{\cal G}^u\over 5}\, \tilde{r}^5 \de_{\tilde{r}}\tilde{A}^r_o( \tilde{r}) = A^{u}_o(u(\tilde{r})) +  {{\cal G}^u\over 5 }\tilde{r}^4
\ee
In the last step we have used the fact that, to lowest order, $\tilde{A}^r_o( \tilde{r}) \sim -\log r$.  The
first and last steps of the above equality mean that there is an extra shift in $A_o$,
compared to the naive change of variables. Renaming  $\tilde{r} \to r$ in (\ref{shift0}), we finally have:
\be\label{shift}
\delta A^r \equiv A^r(r) - A^r_o (r) = A^r(r) -  A^{u}_o(u(r)) -  {{\cal G}^u\over 5 }r^4 = {4\over 5} {\cal G}^u r^4
\ee
This result could have been guessed from the fact that $\delta A$ is not invariant under
linearized diffeomorphisms $r\to r + \xi$, but rather it transforms as $\delta A \to \delta A +  \dot{A}_o\xi$.
For $\xi = {\cal G}^u r^5/5$, this gives exactly eq. (\ref{shift})

Following the same procedure, one can write $\delta \Phi$ in conformal coordinates, and one would
find that this time the leading order is simply given by the change of variables $u\to \log r$. The
reason is that, under  $r\to r + \xi$, the dilaton fluctuation transforms as $\delta \Phi \to \delta \Phi +
\dot{\Phi}_o \xi = \delta\Phi + G (r^4/5) (\log r)^{-1}$. Thus the shift induced by the extra diffeomorphism
is subleading with respect to the first term, which behaves as $r^4 \log r$. Thus, we have to leading order:
\be\label{deltaPhir}
\delta \Phi^r (r) \simeq \delta\Phi^u(u(r)) =  \frac92 {\cal G}^u r^4 \log r\Lambda.
\ee

Finally, to connect this discussion with the definition of ${\cal G}$ given in Section 3,  let
us define ${\cal G} \equiv 4{\cal G}^u/5$. Then we arrive at:
\be
\delta A^r(r) = {\cal G} r^4, \qquad \delta \Phi^r = \frac{45}{8}\,{\cal G}  r^4 \log r\Lambda.
\ee

For consistency, one can check that the above fluctuations solve the linearized
 Einstein's equations in conformal coordinates, obtained by perturbing  eq. (\ref{app11}).

 \section{The $AdS_5$ case revisited\label{ads}}
In this appendix we will reconsider the Einstein system plus a scalar in the conformal case, with a view of exploring all
potential boundary conditions at infinity and their effect in the bulk physics both at zero and finite temperature. This situation is
simpler than the one we are studying but some of the lessons are similar. Although what we
present here is mostly understood in the literature (see for example \cite{Kehagias,DF}),
 they are not widely known and we would like to put them in the right perspective.

\subsection{Zero temperature}

We reconsider the zero temperature field equations (in the conformal coordinate system) of the Einstein-dilaton system with a potential
\be
6{\dot{b}^2\over b^2}+3{\ddot{b}\over b}={b^2} V(\Phi)
\sp
6{\dot{b}^2\over b^2}-3{\ddot{b}\over b}={4\over 3}\dot{\Phi}^2
\label{D1}
\ee
\be
\ddot{\Phi}+3{\dot{b}\over b}\dot{\Phi}+{3\over 8}b^2 \frac{dV}{d\Phi}=0
\label{D6}\ee
We will set the potential to be a constant $V={12\over \ell^2}$
and we will find the UV asymptotics of solutions for arbitrary initial conditions.

The first equation can be integrated once to yield
\be
\dot b=-\sqrt{{C^2\over b^4}+{b^4\over \ell^2}}
\label{D2}\ee
where we have chosen the minus sign branch so that b decreases with increasing $r$, and $C^2$ is an integration constant that we set
to be positive.
When $C=0$ the solution is $AdS_5$
\be
b={\ell\over r-r_0}
\label{D222}\ee
The constant $C$, and two extra boundary conditions for the two first order equations (\ref{D2}) and the second one in (\ref{D1}) viewed as a
first order equation for the dilaton  are the full set of boundary conditions necessary near the boundary.

For general $C$, the first order equation (\ref{D2}) can be integrated as
\be
\int_{1\over b_*}^{1\over b}{du\over \sqrt{1+C^2\ell^2 u^8}}={r-r_0\over \ell}
\label{D3}\ee
giving
\be
 {1\over b}F\left[{1\over 8},{1\over 2},{9\over 8}, -{\ell^2 C^2\over  b^8}\right]=
 {1\over b_*}F\left[{1\over 8},{1\over 2},{9\over 8}, -{\ell^2 C^2 \over b_*^8}\right]+{r-r_0\over \ell}
 \label{D4}\ee
 where $F$ is the standard hypergeometric function and $b_*=b(r_0)$.
 The equation for the dilaton becomes
 \be
 \dot \Phi={3C\over b^3}~~~\to~~~\Phi(r)=\Phi_*+3C\int_{r_0}^r{dr'\over b^3}=\Phi_*-
 3C\ell\int_{b_*}^{b}{db/b\over\sqrt{{\ell^2C^2}+{b^{8}}}}
\label{D5} \ee
$$
~~~\to~~~\Phi=\Phi_*+{3\over 4}{\rm ArcTanh}\sqrt{1+{b^8\over (\ell C)^2}}~\Bigg |_{b_*}^{b}
$$
 where the sign ambiguity is hidden in the sign of $C$.
 Note that for the AdS solution the dilaton is constant.
 The three integration constants $C,b_*,\Phi_0$ are in one-to-one correspondence with the three boundary conditions at the boundary.
 As explained in \cite{ihqcd}, one of them, that we can take to be $b_*$ is a gauge artifact, related to the position of the boundary in the radial coordinate.
 We will therefore choose $r_0=0$ to be the position of the boundary, $b_*=\infty$.
 Then the solution becomes
 \be
 {1\over b}F\left[{1\over 8},{1\over 2},{9\over 8}, -{\ell^2 C^2\over  b^8}\right]={r\over \ell}\sp
 \Phi(r)=\Phi_*+{3\over 4}{\rm ArcSinh}{C\ell\over b^4}
 \label{DD4}\ee

Near the boundary ${b^8\over (C\ell)^2 }\to \infty$ and we can expand F around zero to find
 \be
 b\simeq {\ell\over r}\left[1-{C^2r^8\over 18\ell^6}+{\cal O}\left(r^{16}\right)\right]
 \label{D7}\ee
 valid when $r\to 0$.
 In the same region
 \be
 \Phi=\Phi_*+{3\over 4}{C\over \ell^{3}}r^4+{\cal O}\left(r^{12}\right)
\label{D9}\ee
Therefore for non-zero $C$ there is a non-zero vev of the operator dual to the scalar $\Phi$.

 Consider now the region $b\to 0$. We use the transformation rule
 \be
 F\left[{1\over 8},{1\over 2},{9\over 8}, -{\ell^2 C^2\over  b^8}\right]={\Gamma\left[{1\over 8}\right]\Gamma\left[{3\over 8}\right]
 \over 8\Gamma\left[{1\over 2}\right]}{b\over (C\ell)^{1\over 4} }-{1\over 3}{b^4\over C\ell }
 F\left[{1\over 2},{3\over 8},{11\over 8}, -{b^8\over \ell^2 C^2 }\right]
 \label{D10}\ee
 to obtain
 \be
 b\simeq \left(3C{(\hat r_0-r)}+\cdots\right)^{1\over 3}\sp \hat r_0=
 {\Gamma\left[{1\over 8}\right]\Gamma\left[{3\over 8}\right]
 \over 8\Gamma\left[{1\over 2}\right]}\ell(C\ell)^{-{1\over 4}}
\label{D111} \ee
 The scalar there diverges as
 \be
 \Phi\sim \log(\hat r_0-r)
 \label{D12}\ee

 We may use the relations
 \be
 {\ddot b\over b}={2b^2\over \ell^2}-{2C^2\over b^6}\sp {\dot b^2\over b^2}=
 {b^2\over \ell^2}+{C^2\over b^6}
\label{D13} \ee
 to calculate the curvature invariant for the metric. We obtain
 \be
 R=-{1\over b^2}\left[4{\dot b^2\over b^2}+8 {\ddot b\over b}\right]={12C^2\over b^8}-{12\over \ell^2}
\label{D14} \ee
Near the boundary, $b\to \infty$ and we obtain constant negative curvature as expected.
In the interior, as $b\to 0$, we observe that the space has a curvature singularity, if $C\not=0$ at
a distance $\delta r\sim (C\ell)^{-{1\over 4}}$ from the boundary.
Imposing regularity in the bulk imposes $C=0$. Therefore the dynamics of the theory does not
allow for a non-trivial vev associate to the operator dual to $\Phi$.

\subsection{The black-hole solution}
We will now solve again the equations with constant potential seeking a black-hole type solution.
\begin{equation}
6{\dot{b}^2\over b^2}+3{\ddot{b}\over b}+3{\dot{b}\over b}{\dot{f}\over f}={12b^2\over \ell^2f}
\sp
6{\dot{b}^2\over b^2}-3{\ddot{b}\over b}={4\over 3}\dot{\Phi}^2
\label{D56}\end{equation}

\begin{equation}
{\ddot{f}\over \dot{f}}+3{\dot{b}\over b}=0\sp
\ddot{\Phi}+
\left({\dot{f}\over f}+3{\dot{b}\over b}\right)\dot{\Phi}=0
\label{D57}
\end{equation}
We may integrate once the two last equations as
\be
\dot{f}=-{8C\over b^3}\sp
\dot \Phi={3D\over b^3f}
\ee
and we will take the two constants of integration to be positive
$D>0,C>0$.
Using this  may derive the following equations for $b$
\be
\left(8{\dot b^3\over b^3}-{\dot b\ddot b\over b^2}-{\dddot b\over b}\right)=
{2C\ell^2\over b^5}\left({\dot b^2\ddot b\over b^3}-2{\ddot b^2\over b^2}+{\dot b\dddot b\over b^2}\right)
\label{D58}\ee
\begin{equation}
f\left(2{\dot{b}^2\over b^2}+{\ddot{b}\over b}\right)=8C{\dot{b}\over b^4}+{4b^2\over \ell^2}
\sp
f^2\left(2{\dot{b}^2\over b^2}-{\ddot{b}\over b}\right)={4D^2\over b^6}
\label{D59}\end{equation}

We now introduce an auxiliary variable
\be
\z=\sqrt{1+{4D^2\dot b^2\over \left({b^6\over \ell^2}+2C\dot b\right)^2}}\sp \sqrt{\z^2-1}=
-{2D\dot b\over \left|{b^6\over \ell^2}+2C\dot b\right|}={2\e D\dot b\over \left({b^6\over \ell^2}+2C\dot b\right)}
\label{D60}\ee
where $\e=1$ iff ${b^6\over \ell^2}+2C\dot b<0$ and $\e=-1$ if ${b^6\over \ell^2}+2C\dot b>0$, ($\dot b$ is always negative).
In terms of this new variable we obtain
\be
{\ddot b\over b}=2{\dot b^2\over b^2}\left[{4\over \z+1}-1\right]\sp \dot b={b^6\sqrt{\z^2-1}\over 2\ell^2(\e D-C\sqrt{\z^2-1})}
\label{D61}\ee
which may translated as a first order equation for $\z$
\be
b\z'={8\over D}(1-\z)(D-\e C\sqrt{\z^2-1})
\label{D62}\ee
where the prime stands for derivative with respect to $b$.
This can be integrated
 as
\be
{(D-\e C\sqrt{\z^2-1})\over \z-1}\left({\sqrt{C^2+D^2}-\e C-D\sqrt{{\z-1\over \z+1}}\over
\sqrt{C^2+D^2}+\e C+D\sqrt{{\z-1\over \z+1}}}\right)^{\e C\over \sqrt{C^2+D^2}}=\tilde C b^8
\label{D63}\ee
Using the relation
\be
\left(\sqrt{C^2+D^2}-\e C-D\sqrt{{\z-1\over \z+1}}\right)\left(\sqrt{C^2+D^2}+\e C+D\sqrt{{\z-1\over \z+1}}
\right)={2D\over \z+1}(D-\e C\sqrt{\z^2-1})
\label{D64}\ee
the solution can be written in the following alternative form
\be
{1\over 2D}{\z+1\over  \z-1}{\left(\sqrt{C^2+D^2}-\e C-D\sqrt{{\z-1\over \z+1}}\right)^{1+{\e C\over \sqrt{C^2+D^2}}}\over
\left(\sqrt{C^2+D^2}+\e C+D\sqrt{{\z-1\over \z+1}}\right)^{-1+{\e C\over \sqrt{C^2+D^2}}}}=\tilde C b^8
\label{D65}\ee
In particular, for $\e=1$
\be
{1\over 2D}{\z+1\over  \z-1}\left(\sqrt{C^2+D^2}- C-D\sqrt{{\z-1\over \z+1}}\right)^{a_+}
\left(\sqrt{C^2+D^2}+C+D\sqrt{{\z-1\over \z+1}}\right)^{a_-}=\tilde C b^8
\label{D66}\ee
with
\be
a_+=1+{C\over \sqrt{C^2+D^2}}\geq 0\sp a_-=1-{C\over \sqrt{C^2+D^2}}\geq 0
\label{D67}\ee
while for $\e=-1$
\be
{1\over 2D}{\z+1\over  \z-1}\left(\sqrt{C^2+D^2}+C-D\sqrt{{\z-1\over \z+1}}\right)^{a_-}
\left(\sqrt{C^2+D^2}-C+D\sqrt{{\z-1\over \z+1}}\right)^{a_+}=\tilde C b^8
\label{D68}\ee
We will now investigate several special cases of this general solution.

\subsubsection{$C=0$}
The solution becomes
\be
{D\over \z-1}=\tilde C b^8
\label{D69}\ee
which using (\ref{D60}) becomes
\be
\dot b^2={1\over 4\tilde C^2\ell^4}{1\over b^4}+{1\over 2\tilde CD\ell^4}b^4
\ee
Compatibility with the other equations determines $\tilde C={1\over 2D\ell^2}$.
This solution has no horizon (f=1). It is the same solution found in the previous subsection, where $D$ plays the role of the $\Phi$ condensate.

\subsubsection{$D=0$}

The equation (\ref{D65}) becomes the trivial one $\z=1$.
{}From (\ref{D59}) we obtain ${\ddot b\over b}=2{\dot b^2\over b^2}$ which is solved by the AdS scale factor $b\sim {1\over r}$.
Finally the rest of the equations give
\be
b={\ell\over r}\sp f=1-{2C r^4\over \ell^3}
\ee

This is the standard AdS black-hole solution with a flat horizon.

The two previous cases indicate that the constant $C$ controls
 the temperature of the solution while the constant $D$ controls the $\Phi$ condensate.

\subsection{Analysis of the general solution}

The function $f$ is given by
\be
f={2D\ell^2(D-\e C\sqrt{\z^2-1})\over b^8(\z-1)}
\label{D70}\ee
The boundary $b\to \infty$, is always at  $\z=1$.
To test whether we have a regular horizon we need the
the trace of the Einstein tensor which is given by
\be
E=-{f\over b^2}\left[{\ddot f\over f}+8{\ddot b\over b}+8{\dot b\over b}{\dot f\over f}+4{\dot b^2\over b^2}\right]=
{2\over \ell^2}{D(3\z -13)+10\e C\sqrt{\z^2-1}\over D-\e C\sqrt{\z^2-1}}
\label{D72}\ee
To continue we distinguish two cases:

\subsubsection{$\e=-1$}

There is no horizon in this case as $f$ never vanishes.
There is also no singularity as $E$ never blows up.
The scale factor becomes a constant asymptotically

\subsubsection{$\e=1$}

The horizon ($f=0$) is at
\be
\z_h=\sqrt{1+{D^2\over C^2}}
\label{D71}\ee
It is singular unless $D=0$. At the horizon $b\to 0$.
Therefore, even at finite temperature the $\Phi$ condensate must vanish in order to have a regular solution.

\section{The Black hole action and ADM mass}

\subsection{The on-shell action}\label{onshell}

We want to compute the regularized action evaluated on a solution
of Einstein's equations, (\ref{app4}).
 We start from eq. (\ref{a1}) :
\be
   S_5=S_E+S_{GH},  \quad S_E=-M^3\int d^5x\sqrt{g}
\left[R-{4\over 3}(\partial\Phi)^2+V(\Phi) \right], \quad  S_{GH}=2M^3\int_{\partial M}d^4x \sqrt{h}~K.
    \label{app1a}\ee
We work in conformal frame, with the metric given by eq. (\ref{app7}).

Taking the trace of equation (\ref{app4}) we obtain for the Ricci scalar:
\be\label{ricci}
R =  {4\over 3} (\de \Phi)^2 - {5\over 3}V.
\ee
This leads to the on-shell Einstein action:
\bea
   S_E&& ={2\over 3}M^3\int d^5x\sqrt{g}~~
V(\Phi)={2\over 3}M^3V_3\int_0^{\beta} dt\int_{0}^{r_h} dr~b^5~~
V(\Phi)\nn \\
&&=2M^3V_3\int_0^{\beta} dt\int^{r_h}_0 dr~\frac{d}{dr}(fb^2\dot{b})
    \label{app13a}\eea

From (\ref{app3}) the components of the  unit normal to the boundary are $n_r = - {b/\sqrt{f}}$, $n_i=0$.
The trace of the extrinsic curvature is: 
\be
K={\sqrt{f}\over 2b}\left[8{\dot{b}\over b}+{\dot{f}\over f}\right]
\ee
We find:
\be
{\cal S}^{\e}_E=2M^3V_3\int_0^{\beta} dt\int_{\e}^{r_h} dr~\frac{d}{dr}(fb^2\dot{b})
=2\beta M^3V_3 (f(r_h)b^2(r_h)\dot{b}(r_h)-f(\e)b^2(\e)\dot{b}(\e))\label{sein}
\ee

\bea
{\cal S}^{\e}_{GH}&&=2M^3V_3\int_0^{\beta} dt{b^3(\e)f(\e)\over 2}\left[8{\dot{b}\over b}+{\dot{f}\over f}\right]_{\e}
=2M^3V_3
\beta~{b^3(\e)f(\e)\over 2}\left[8{\dot{b}\over b}+{\dot{f}\over f}\right]_{\e}\nn\\
 && =2M^3V_3
\beta~{b^3(\e)f(\e)\over 2}\left[8{\dot{b}\over b}+{\dot{f}\over f}\right]_{\e} \label{sgh}
\eea
Putting together eqs. (\ref{sein}) and (\ref{sgh}) we obtain for the full result:
\be\label{action-os}
{\cal S}^{\e}={\cal S}^{\e}_{E}+{\cal S}^{\e}_{GH}=2\beta M^3V_3 \left[3b^2(\e)f(\e)\dot{b}(\e)+{1\over 2}\dot{f}(\e)b^3(\e)\right]
\ee
where we used that $f(r_h)=0$.

The calculation for the zero-temperature background is exactly the same, except that the integral of the
Einstein-Hilbert action extends on the region $(0,r_0)$, where $r_0$ is the singularity. Thus
in evaluating (\ref{sein}) one gets:
\be
{\cal S}^{\e}_E=
2\beta M^3V_3 (b^2_o(r_0)\dot{b}_o(r_0)-b^2_o(\e)\dot{b}_o(\e))\label{sein0}
\ee
The IR contribution vanishes whenever $b_o^2 \dot{b}_0 \to 0$ as $r\to r_0$. This is always true
for good singularities.

\subsection{Evaluation of the free energy}\label{free-en}

We start from eq.  (\ref{free}), which we rewrite below:
\be\label{free-app}
{\cal F}=  \sigma \,\left\{ 6 b^2(\e) \sqrt{f(\e)} \left[\dot{b}(\e) \sqrt{f(\e)}
 - {b^2(\e)\over b^2_o(\tilde{\e})}\dot{b}_o(\tilde{\e})\right] + \dot{f}(\e) b^3(\e)\right\}
\ee
where  the limit  $\e\to 0$ is understood. In terms of  $\db=b-b_o$, and $\delta\eps=\tilde{\e}-\e$, the
previous equation  reads:
\be
{{\cal F}\over \sigma} =
6b^2(\e)\sqrt{f(\e)}\left[(\dot b_o+\dot\db)(\e)\sqrt{f(\e)}-{(b_o+\db)^2(\e)\over
(b_o + \delta\e\, \dot{b}_o)^2(\e)} (\dot b_o + \delta \e \,\ddot{b}_o)(\e)\right]+\dot{f}(\e)(b_o+\db)^3(\e).
\label{free1}\ee
For small $\e$, $b_o(\e)\sim \e/r$, and  by eqs. (\ref{b-bo}), (\ref{fUV}) and (\ref{deltaeps}) we have:
\be
\delta b(\e) \simeq {\cal G}{\e^3\over \ell^2},\; \;\delta \dot{b}(\e) \simeq {\cal G}{3\e^2\over \ell^2},\;\;
f(\e)\simeq 1 - {C\over 4}{\e^4\over\ell^3}, \;\;\delta \e = -{45\over 8}{{\cal G}\over \ell^3}\e^5 (\log \e \Lambda)^2.
\label{free3}\ee
We see that the only divergent contribution inside the square brackets, i.e. $\dot{b}_o$, cancels. What is left is
of order $\e^2$ times eventually some logarithmic corrections. Therefore, to this order we can
replace the overall prefactor $ b^2(\e)\sqrt{f(\e)}$ by $\ell^2/\e^2$.
 Thus, to lowest non-vanishing order:
\be
{{\cal F}\over \sigma} = 6 {\ell^2\over \e^2}\left[ \left(5 {\cal G} +{C\over 8}\right){\e^2\over \ell^2} + b_o(\e)\delta\e\left(2{\dot{b}_o^2\over b_o^2}(\e) - {\ddot{b}_o \over b_o}(\e)\right)\right] - C
\ee
The last term in the parenthesis requires more care: due to the extra logarithm in $\delta\e$ we cannot just
replace $b_o(r)$ by $\ell/r$. On the other hand we can use the zeroth order Einstein's equation (\ref{app11})
to write it as:
\be
b_o(\e)\delta\e\left(2{\dot{b}_o^2\over b_o^2}(\e) - {\ddot{b}_o \over b_o}(\e)\right) = b_o(\e)\,\delta\e \,{4\over 9}\dot{\Phi}_o^2(\e) = - {5\over 2}\,{\cal G}\,{\e^2\over \ell^2}
\ee
where in the last line we used $\dot{\Phi}_o(\e)= -(\e \log \e\Lambda)^{-1}$ (cfr. eq. (\ref{lUV})). Notice
that the logarithm in $\delta \e$ has canceled. Finally, we get:
\be\label{free4}
{{\cal F}\over \sigma}=15\, {\cal G}-{C\over 4}
\ee

\subsection{The black-hole mass}\label{ADM}

The mass of a  solution, with respect to a reference background, can be defined following the
procedure: first consider a time slicing of the 5-dimensional metric,
of the ADM form
\be
ds^2 = -N^2 dt^2 + \gamma_{ij}( dx^i - N^i dt) ( dx^j - N^j dt)  \qquad i,j = r,1,2,3
\ee
where  $\gamma_{ij}$ is the induced metric on each 4D slice $\Sigma_t$. Then the mass of the
solution, with respect to a reference background with the same asymptotic behavior at spatial
infinity,  is given by \cite{HH}
\be  \label{mass}
E = -{1\over 8\pi G_5} \int_{\Sigma_{\infty}} N \left(\sqrt{\gamma^{ind}}\, ^{(3)} K  -  \sqrt{\gamma^{ind}_o}\, ^{(3)} K_o\right)
\ee
where $\Sigma_{\infty}$ is a 3-dimensional surface at   spatial infinity embedded in the  4D constant-time
slice  $\Sigma_t$,  $\gamma_{ind}$ is the three-dimensional induced metric, $^{(3)} K$ is its extrinsic
curvature, and $\gamma^{ind}_o$ and  $^{(3)} K_o$ the analogous quantities for the reference background. The latter should be
chosen so that the geometry of the 3-dimensional surface at infinity and the value of the scalar field on
that surface match.

In our case,  the 5D solutions are static and  in conformal coordinates they   are of the form:
\be
ds^2 = b(r)^2 \left( -f(r)dt^2 + {dr^2\over f(r)} + dx^mdx_m\right), \quad ds^2_o =b_o(r)^2 \left( -dt^2 + dr^2 + dx^mdx_m\right).
\ee
The boundary at infinity is  at $r=\e$, with $\e\to0$.   Thus, we have:
\be
N  = b(r)\sqrt{f}, \qquad \gamma_{ij}dx^idx^j = b(r)^2\left( {dr^2\over f(r)} +  dx^mdx_m\right),
\ee
and on the surface $r=\e$ we have
\be
\gamma_{mn}^{ind} = b(\e)^2 \delta_{mn},  \qquad n^i  =- {\sqrt{f}(\e)\over b(\e)}\left({\de\over \de r}\right)^i
\ee
The 3D  extrinsic curvature is given by:
\be
 ^{(3)} K  = \nabla_i n^i = {1\over \sqrt{\gamma}}\de_j \left(\sqrt{\gamma} \gamma^{ij} n_j\right)  =
-3\sqrt{f(\e)}{\dot{b}(\e)\over b^2(\e)}.
\ee
The reference background has  $f=1$,
boundary at $\tilde\e = \e + \delta \e$ (as in \ref{deltaeps}) so that $\l(\e) = \l_o(\tilde\e)$,  and rescaled volume
$\tilde{V}_3 = V_3 b^3(\e)/b_o(\tilde{\e})$. Also, the time slicing has to be the same, i.e. $N(\e)=N_o(\tilde\e)$ in the  ADM decomposition.
Thus from eq. (\ref{mass})
we obtain:
\be\label{mass2}
 E  = {3V_3\over 8\pi G_5} b^2(\e)\sqrt{f(\e)} \left(\sqrt{f}(\e) \dot{b}(\e) -  {b^2(\e)\over b_o^2(\tilde{\e})}\dot{b}_o(\tilde{\e})\right)
\ee
Using eqs. (\ref{free3}) and performing similar steps  to those that led to eq. (\ref{free4}) we obtain:
\be\label{mass3}
E = M_p^3 N_c^2 V_3 \left(15 {\cal G} +{3\over4}C\right)
\ee
where we have used $16 \pi G_5 = M_p^{-3}N_c^{-2}$.

\section{The gluon condensate asymptotics} \label{free-en-2}

We want to show that the gluon condensate $\cal G$ obeys to the asymptotics \refeq{CT} at high temperatures. To do this we need to compute
 explicitly the relation between $T$ and
$r_h$ in the UV limit $r_h\to0$.

\paragraph{Temperature} Knowing that the expansion for the metric reads as in \refeq{bUV} (this is the zero temperature solution, however the $r^4$ correction of the finite temperature solution is
subleading w.r.t. logarithms) we find  that the expansion for the thermal factor $f(r)$ in the UV is given by solving the second equation in \refeq{a11}

\bea
f(r)=1-\frac{r^4}{r_h^4}\left[1+\frac{4}{3}\frac{\log\frac{r}{r_h}-\frac{4}{3}}{\log\Lambda r\,\log\Lambda r_h}+{\cal
O}\left(\frac{\log(-\log \Lambda r)}{\log^{2}\Lambda r}\right)\right]
\eea
The derivative of $f$ w.r.t. $r$ then evaluates to
\bea
\dot f(r) = -4 \frac{r^3}{r_h^4}\left[1+\frac{4}{3}\frac{\log\frac{r}{r_h}-\frac{4}{3}}{\log\Lambda r\,\log\Lambda r_h}
+\oneover{3}\frac{1+{\cal O}\left(\oneover{\log \Lambda r}\right)}{\log\Lambda r\,\log\Lambda r_h}
+{\cal O}\left(\frac{\log(-\log \Lambda r)}{\log^{2}\Lambda r}\right)\right] \nn\\
\eea

The temperature is obtained by evaluating the above expression at the horizon

\bea
T = \frac{|\dot f(r_h)|}{4 \pi} = \oneover{\pi r_h}\left[1 - \frac{4}{9} \oneover{\log^2 \Lambda r_h} + {\cal O}\left(\frac{\log(-\log \Lambda r)}{\log^{3}\Lambda r}\right) \right]
\eea

\paragraph{Entropy} We now want to calculate the entropy density $s=4\pi M^3 b^3(r_h)$.
Inverting this relation we get $r_h$ as a function of the temperature

\bea
r_h = \oneover{\pi T}\left[1 - \frac{4}{9} \oneover{\log^2 \frac{\Lambda}{\pi T}} + {\cal O}\left(\frac{\log(-\log \frac{\Lambda}{\pi T})}{\log^{3} \frac{\Lambda}{\pi T}}\right) \right]
\eea
Plugging this expression into the expansion for the scale factor \refeq{CT} we obtain $b$ as a function of the temperature

\bea
b(T) = \pi \ell T \left[ 1 + \frac{4}{9} \oneover{\log \frac{\Lambda}{\pi T}} + {\cal O}\left(\frac{\log(-\log \frac{\Lambda}{\pi T})}{\log^{2} \frac{\Lambda}{\pi T}}\right) \right]
\eea
The subleading term ${\cal O}\left(\log^{-2}(\Lambda/\pi T)\right)$ will not  enter into the leading calculation of the gluon condensate asymptotics.
The entropy density evaluates to

\bea
s(T) = 4\pi^4 \ell^3 T^3 \left[ 1 + \frac{4}{3} \oneover{\log \frac{\Lambda}{\pi T}} + {\cal O}\left(\frac{\log(-\log \frac{\Lambda}{\pi T})}{\log^{2} \frac{\Lambda}{\pi T}}\right)
\right] \equiv 4 \pi^4 \ell^3 T^3 \xi(T)
\eea
where the last equality is the definition of the function $\xi(T)$.

\paragraph{Gluon condensate asymptotics}
Putting together the information relating the gluon condensate to the free energy on the one hand through eq. \refeq{free-2} and, on the other hand, the free energy to the entropy
through ${\cal F}=-\p S/\p T$ we arrive to an equation for the gluon condensate

\bea\begin{split}
12 {\cal G}(T) =& \frac{Ts(T)}{4} - \int^T_{T_c} \intd t \; s(t) \\
=& \pi^4 \ell^3 T^4 \xi(T) - 4 \pi^4 \ell^3 \int^T_{T_c} \intd t \; t^3 \xi(t) \\
=& \pi^4 \ell^3 \int^T_{T_c} \intd t \; t^4 \xi'(t) \;.
\end{split}\eea
The last line uses integration by parts. The expansion of the derivative of the function $\xi(T)$ reads

\bea
\xi'(T) = \frac{4}{3} \oneover{T} \oneover{\log^2 \frac{\Lambda}{\pi T}} \left(1+\dots\right)
\eea
The ellipsis indicates subleading terms in the log expansion.

So that finally the gluon condensate expansion at high temperatures $T\gg\Lambda$ at leading order can be written as

\bea
{\cal G}(T) \approx \frac{\pi^4}{36} \ell^3 \frac{T^4}{\log^2 \frac{\Lambda}{\pi T}}
\eea

\section{The Superpotential at zero-$T$} \label{superapp}

Here we analyze the general solution of the zero-temperature superpotential equation,
eq. (\ref{super1}), (below, $\l=e^\Phi$).
\be\label{spot1}
-{4\over 3}\l^2 (W'(\l))^2 + {64\over 27}W^2(\l) = V(\l).
\ee
we assume $V(\l)>0$
First let us observe some general properties:
\begin{enumerate}
\item The  solution can only exist as long as $|W(\l)|> \sqrt{(27/64) V(\l)}$;
\item The equation has a symmetry $W \to -W$, so we can
limit the analysis to $W>0$.
\item For any  $\l_0\neq 0$, there are two solutions of
(\ref{spot1}), $W_+(\l),W_-(\l) $ passing through the point $\l_0$,
such that $W_+(\l_0) =  W_-(\l_0)$, and $W_+'\l_0) = - W_-'(\l_0)$.
In other
words there are two branches of solutions: one where   $W$ and $W'$
have the same sign (i.e. $W_+'(\l)>0$) , another where they have opposite
sign ($W_+(\l)<0$)
\item At any $\l_*\neq 0$ where  $|W(\l_*)|= \sqrt{27/64 V(\l_*)}$, $W'=0$.
\item A solution can go past such a point  $\l_*$ {\em only} if $V'(\l_*)=0$.
Indeed, suppose that $V'(\l_*)> 0$.
  if the solution exists for $\l<\l_*$, at the point $\l_*$ we have:
\be
W(\l_*) = \sqrt{(27/64)V(\l_*)}; \;\; W'(\l_*)=0 \;\;; V'(\l_*) >0 \quad \Rightarrow
W(\l_* + \epsilon ) < \sqrt{(27/64)V(\l_*+\epsilon)}
\ee
therefore the solution does not exists  for $\l>\l_*$.
\item By the same
argument, if $V(\l_*)<0$, the solution does not exist for $\l<\l_*$.
\item It follows from points 3,4 and 5 that,  if $V(\l)$ is positive and monotonic, the two branches $W_+(\l)$ and $W_-(\l)$
(see point 4)  are completely
disconnected, since neither $W'$ nor $W$ can change sign. However two solutions belonging to different branches
can be glued together at a point $W'=0$.
\item All solutions that reach $\l=0$ have either $W(0) = \sqrt{(27/64)V(0)}$,
or $W'(0)=\infty$.
\end{enumerate}

In what follows we assume $V(\l) >0$ and without loss of generality we take $W(\l)>0$.

\subsection{Solution close to a critical point}\label{critical}
Let us see how the solution approaches the critical points $W'(\l_*)=0$.
For definiteness, consider $V(\l)$ monotonically increasing,
and $W(\l)>0$ (as
in our model). The solution exists only for $\l<\l_*$.
As we said there are
two disconnected branches with opposite sings
of $W'$.  Let us  are analyze them separately.

\subsubsection*{$W_-$ Branch}
In this case, $W'(\l)<0$ for all $\l<\l_*$.
Then eq. (\ref{spot1}) can be written as:
\be\label{spot2}
W'(\l) = -{4\over 3 \l}\sqrt{W^2 -{27\over 64}V}
\ee
Let us look at this equation
 close to a point $\l_*\neq 0$ where $W^2(\l_*) = 27/64 V(\l_*)\equiv W^2_*$.
Write $W =  W_* + w(\l)$, with $w(\l_*)=0$,
and expand (\ref{spot2}) to linear order in $w(\l)$:
\be\label{spot3}
w'(\l) = -{4\over 3\l_*} \sqrt{2W_*w(\l) - V'(\l_*) (\l-\l_*)}
\ee
this is still hard to solve explicitly, but we can carry out the analysis  by making
further assumptions
about the possible behavior of $w(\l)$ close to $\l_*$. There are only three
possibilities:
\begin{enumerate}
\item $|w(\l)| >  O(\l-\l_*)$ as $\l\to \l_*$. \\
then we can neglect the second term under the square root,
and the equation becomes
\be
w'(\l) = -{4\over 3\l_*} \sqrt{2W_*w(\l)}
\ee
which is solved by $w(\l) \sim (\l-l_*)^2$. This is inconsistent with
the assumption  $|w(\l)| >  O(\l-\l_*)$, so this case is ruled out.
\item  $|w(\l)|\simeq -w_1(\l-\l_*)$ as $\l\to \l_*$, with $w_1>0$ .\\
Eq. (\ref{spot3}) becomes:
\be
 w'(\l)= -{4\over 3\l_*}\sqrt{(2W_*w_1+V'(\l_*))(\l_*-\l)},
\ee
which is solved, for $\l<\l_*$, by $w \simeq const (\l_*-\l)^{3/2}$ ,
in contradiction with our assumption. So this case is ruled out too.
\item  $|w(\l)| <  O(\l-\l_*)$ as $\l\to \l_*$. \\
In this case we can neglect the first term in the square root, and
obtain:
\be\label{spot4}
w'(\l) = -{4\over 3\l_*}\sqrt{V'(\l_*) (\l_*-\l)}
\ee
which integrates to:
\be\label{spot5}
w(\l) \sim  {8\over9 \l_*}\sqrt{V'(\l_*)} (\l_*-\l)^{3/2} \qquad \l<\l_*
\ee
This time the solution is consistent with the hypothesis.
\end{enumerate}

The three possibilities listed above exhaust all possible behaviors of $W(\l)$ close
to a critical point $\l_*$, and the only one which does not lead to  a contradiction
is the last one. Thus, we can conclude that
the  behavior close to a
critical point is given by eq. (\ref{spot5}). in other words,
$W_-(\l)$ is positive and decreasing for $\l<\l_*$, and it  reaches
a finite value $W_*$ at the critical point, where it behaves  like
\be\label{spot66}
W_-(\l) \sim W_* + W_1 (\l_*-\l)^{3/2}, \qquad W_1 >0
\ee

The behavior of $W_-(\l)$ close to a critical point is exemplified  in
figure \ref{super5} (a)
\begin{figure}[h]
 \begin{center}
\includegraphics[scale=0.7]{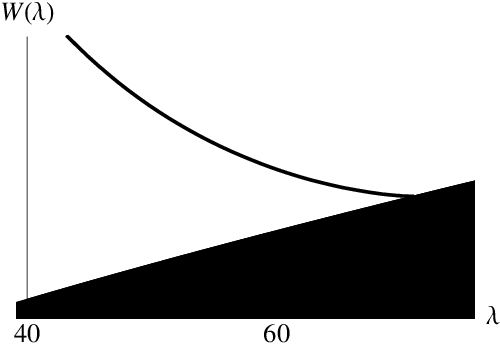} \hspace{1cm}
\includegraphics[scale=0.7]{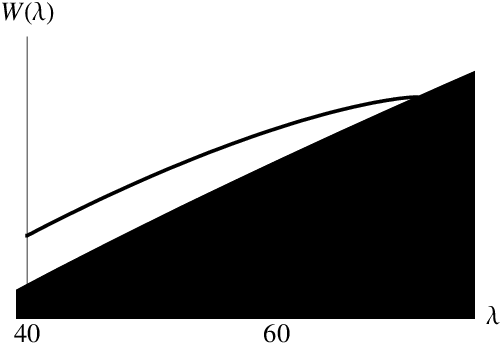}\\
(a) \hspace{5cm} (b)
 \end{center}
 \caption[]{Superpotential on (a) the $W_-$ branch  and (b) the
$W_+$ branch,  close to
a critical point. The black area is the ``forbidden'' region below
the critical curve $\sqrt{27 V/64}$, where $W'$ would become imaginary. The solution stops where
it meets the critical curve.}
\label{super5}\end{figure}

\subsubsection*{$W_+$ Branch}

The analysis is the same as for the $W_-$ branch: the solution
is defined only for $\l\leq\l_*$, except that $W(\l)$ is increasing and
 close to $\l_*$ we have:
\be\label{spot77}
W_+(\l) \sim W_* -  W_1 (\l_*-\l)^{3/2}, \qquad W_1 >0
\ee
this type of solution is shown in figure \ref{super5} (b)
\subsubsection*{Metric and dilaton  close to a critical point}

Although $W'(\l_*)=0$, the metric is {\em not} $AdS$ close to $\l_*$. In fact,
the equation for $A(\l)$ reads:
\be\label{spot6}
\l {d A \over d\l}  = -{3\over X(\l)}, \qquad X = -{3\over 4} \l {W'\over W}.
\ee
Close to the critical point $\l_*$, $X \sim \mp (3W_1/2W_*)(\l_*-\l)^{1/2}$,
so the scale factor close to $\l_*$ is finite, and  behaves as
\be\label{spot7}
A_{\pm}(\l) \sim A_* \pm A_1 (\l_* -\l)^{1/2}, \qquad A_1>0
\ee
Notice that the upper sign corresponds to the $W_+$ branch ($X<0$), in which
the scale factor {\em decreases} towards the endpoint $\l_*$,
which is therefore in the IR.
The other branch $W_-(\l)$ ($X>0$) has $\l=\l_*$  as the UV.

Finally, we can integrate the dilaton equation as a function of the coordinate $r$,
\be
{\dot{\l} \over \l} = \l {dW\over d\l} e^A.
\ee
Using the form  of  $A(\l)$ and $W(\l)$ close to $\l_*$, eqs. (\ref{spot7}), we arrive at:
\be\label{spot777}
\l(r) \sim \l_*  -\l_1 (r_* - r)^2, \quad
 A_{\pm}(r)   \sim A_* \pm A_1 |r_* - r| , \qquad \l_1 > 0, A_1>0
\ee
Here the upper sign holds for $r<r_*$, the lower for $r>r_*$.
From the last equation, we see that in the $r$ coordinate the point $r_*$ where $\l(r_*) = \l_*$ is perfectly
regular, and we can obtain a full solution describing both branches by simply removing the absolute value,
\be\label{spot778}
\l(r) \sim \l_*  -\l_1 (r_* - r)^2, \quad
 A(r)   \sim A_* + A_1 (r_* - r) , \qquad \l_1 > 0, A_1>0
\ee

At the critical point the dilaton reaches its maximum value $\l_*$, then
reverts the direction of running, and is not a good coordinate globally.
Instead $A(r)$ is monotonic along the full solution. The UV corresponds to $r<r_*$, the IR to $r>r_*$.

\subsection{Solutions close to $\l=0$} \label{SuperappUV}

Here we still have two branches, but we have two completely different behaviors in each branch. Assume
$V>0$, $V'>0$ close to $\l=0$, and a power expansion of the form:
\be
V = V_0 + V_1 \l + V_2 \l^2 +\ldots
\ee

\subsubsection*{$W_-$ Branch}

In this case the general solution of eq. (\ref{spot1}) close to $\l=0$
is:
\be\label{spot8}
W_-(\l) \sim W_0\left({C\over \l^{4/3}} + {\l^{4/3}\over C} + W_1
{\l^{4/3+1}\over C} + \ldots\right), \quad C>0
\ee
where $W_0$ and $W_1$ are completely fixed by the expansion coefficients
of $V(\l)$ around zero. Since $W_-$ is a decreasing function for small $\l$, from
our general considerations we know that if  $V(\l)$ is monotonic then
$W_-(\l)$ is monotonically decreasing globally, therefore the solution will hit
a critical point at some $\l_*>0$ and terminate.

Solving the metric and dilaton equations in the region $\l\sim 0$ with the superpotential (\ref{spot8}) gives
 a singularity at a finite value $r=r_0$ of the conformal coordinate, where both the scale factor $a(r)\to 0$ and the
dilaton $\l\to 0$.
\be\label{spot888}
 e^{A(r)} \sim (r_0 - r)^{1\over 3}, \qquad \l(r) \sim(r_0 - r)^{{1\over 2}}.
\ee
We see that $ e^{A(r)}$ decreases to zero as $\l(r) \to 0 $, and $\dot{l}/\dot{A} > 0$. With our holographic dictionary,
($\log A \leftrightarrow  E$ and $\beta(\l) =  \dot{l}/\dot{A}$, we conclude that
Therefore in this case the small $\l$ region is in the
IR, and the theory is IR-free.

\subsubsection*{$W_+$ Branch}

In this branch, any
solution
of (\ref{spot1}) necessarily satisfies
\be
W_+(0) = \sqrt{{27\over 64} V_0}
\ee
(one can show that any ansatz  with $W'(0)=+\infty$ cannot solve the
equation).

Moreover, if $W$ is written as  a power series
expansion around  $\l=0$,
\be\label{spot88}
W(\l) = W_0 + W_1\l + W_2 \l^2 + \ldots
\ee
 then {\em all the coefficients $W_i$  are uniquely determined by
the expansion coefficients $V_i$ of $V(\l)$}.   However,  it is incorrect to conclude that
that the solution in this branch is unique.  To see this, take any function $\hat{W}(\l)$
that solves (\ref{spot1}) to all orders in powers of $\l$.
Then, consider  a function $W(\l)$  that, close to $\l=0$,
behaves as
\be\label{spot9}
W(\l) =  \hat{W}(\l) + w(\l), \qquad {w(\l)\over \hat{W}(\l)} \to 0 \,\,as\,\,
\l\to 0.
\ee
 Inserting  this in the eq. (\ref{spot1}),
and expanding  to linear order in $w(\l)$ gives, close to $\l=0$:
\be\label{spot11}
-{4\over 3} 2 \l^2 \hat{W}'(0) w'(\l) + {64\over27}2 \hat{W}(0)w(\l) =0
\ee
i.e.  a homogeneous, linear equation, whose  general solution is
\be\label{spot12}
w_C(\l) = C \l^{16/9-4b} \exp\left[-{16\over 9} {\hat{W}(0)\over\hat{W}'(0)}{1\over
    \l}\right]
\ee
Therefore, also in
the $W'>0$ branch we  have  a one-parameter family of  solutions,
that   close to $\l=0$  all have the same power expansion, and  look like:
\be
W(\l) = \hat{W} + w_c(\l) + \ldots
\ee
where $\hat{W}$ is a fixed power series
(say, with no exponential part), $w_c(\l)$ is given in eq. (\ref{spot12}),
and the dots represent even more subleading terms ($\sim w^2$).

Due to the expasion (\ref{spot88}), the solution close to $\l=0$ is, to leading order,  an $AdS_5$ spacetime with
logarithmic running,
\be\label{spot1313}
b(r) = {\ell\over r}\left[1 + {4\over 9} {1\over \log r \Lambda}  + \ldots\right], \quad \l(r) = -{1\over b_0  \log r \Lambda} + \ldots
\ee
Notice that the exponent in (\ref{spot12}) is fixed by the first two
expansion coefficients of $V(\l)$, and one can easily show that, in terms
of the $\beta$-function coefficient $b_0$ :
\be\label{spot13}
 {16\over 9} {\hat{W}(0)\over\hat{W}'(0)} = {4\over b_0} \quad \Rightarrow
 \quad w_c \sim
 \l^{16/9-4b} e^{-{4\over b_0\l}},
\ee
 Using the perturbative asymptotics $b_0 \l\sim (\log r)^{-1}$ , this corresponds to a power-law  correction to the
 logarithm expansion in  (\ref{spot1313}),
that scales  like  $r^4$ close to the $AdS$ boundary $r=0$. Since the power series expansion of $W(\l)$ around
$\l=0$ is independent of the integration constant $C$ in (\ref{spot12}), we conclude that metric that correspond
to different solutions on the $W_+$ branch differ only by {\em non-perturbative} $O(r^4)$  terms,
which correspond to different values for the gluon condensate.

\subsection{Solutions close to $\l=\infty$}

Finally we analyze the solution of (\ref{spot1}) in the asymptotic region of large $\l$. We assume
for the potential a power-law behavior
\be\label{spot14}
V(\l) \sim V_{\infty} \l^{2Q} (\log \l)^{P}\qquad \l \to \infty
\ee
for some constant $V_{\infty}$ and $Q>0$. We are interested in $V_\infty>0$, since this case
corresponds to a potential which is bounded from below.
There are two kinds of solutions:
\begin{enumerate}
\item a continuous one-parameter family of  the form:
\be  \label{spot15}
W_C(\l) = W_\infty \left[C \l^{4/3}  + {C^{-1}\over (4-3Q)}\l^{2Q - 4/3}(\log \l)^P + \ldots\right], \qquad W_{\infty} = \sqrt{27 V_\infty \over 64}
\ee
where $C$ is an arbitrary constant of integration;
\item a {\em single} solution that asymptotes as
\be\label{spot16}
W_s(\l)  = \tilde{W}_\infty \l^Q (\log \l)^{P/2}, \qquad \tilde{W}_\infty = \sqrt{27 V_\infty \over 4(16 - 9Q^2)}
\ee
\end{enumerate}
Notice if $V_\infty>0$,   both types of solutions exist only if $Q<4/3$: for $Q>4/3$ the l.h.s. of the
differential equation is asymptotically negative. In this case there is no solution that reaches arbitrarily
large values of $\l$, but rather all solutions to (\ref{spot1}) are of the type described in section
(\ref{critical}): they reach a maximum value $\l_*$ where a $W_+$ and a $W_-$ solutions join.

With the
restriction $Q<4/3$  the first term in eq. (\ref{spot15}) is the dominant one, and the singular solution
grows slower than any of the solutions in the continuous family.

For all superpotentials in the continuous family the metric and dilaton exhibit the same kind of IR
singularity at finite $r$ (where $a(r) \to 0$, $\l(r) \to \infty$) , regardless of the value of $Q$ and $P$:
\be
a(r) \sim (r_0 - r )^{1/3} , \qquad \l(r) \sim {1\over (r_0 - r )^{1/2}}.
\ee
This is similar to the singularity in eq. (\ref{spot888}), up to $\l \to 1/\l$. These singularity always fall
in the pathological class, as discussed in (\cite{ihqcd}): the singularity is not screened from physical
fluctuations, and one can have an infalling flux of particles. Moreover, as shown in Appendix \ref{superfiniteT},
these singularity do not appear as continuous limits of black-hole solutions with regular horizons.

On the ohter hand the singular solution (\ref{spot16}) is the most interesting from a physical point of view: the
singularity is repulsive for $Q<2\sqrt{2}/3$ and it can be cloaked by a horizon. Solutions of this
kinds are the ones that give rise to the most interesting holographic constructions from the QCD perspective.

\subsection{General Classification of the solutions} \label{general}

The results of this Appendix can be summarized as follows:
 For any positive and  monotonic potential $V(\l)$ with  the asymptotics :
\bea
&&V(\l) = V_0 + V_1 \l + V_2\l^2 +\ldots \quad V_0>0, \qquad \l\to 0 \nn\\
&&V(\l) = V_\infty  \l^{2Q} (\log \l)^{P}, \quad  V_\infty >0, \qquad \l \to \infty\nn
\eea
the zero-temperature superpotential equation has three types of solutions, that we name the {\em Generic},
the {\em Special}, and the {\em Bouncing} types:
:
\begin{enumerate}

\item  A continuous one-parameter family  that has a fixed power-law expansion near $\l=0$, and
 reaches the asymptotic large-$\l$ region where it  grows as
\be
W \simeq C_b \l^{4/3} \qquad \l \to \infty
\ee
where $C_b$ is an arbitrary  positive real number
These solutions lead to backgrounds with ``bad'' (i.e. non-screened) singularities at finite $r_0$,
where $b(r) \to 0$ and $\l \to \infty$ as
\be
a(r) \sim (r_0 -r)^{1/3}, \qquad \l(r) \sim (r_0-r)^{-1/2}
\ee
We call  this  solution {\em generic}. An example is shown in figure \ref{superbad}
\begin{figure}[h]
 \begin{center}
\includegraphics[scale=0.9]{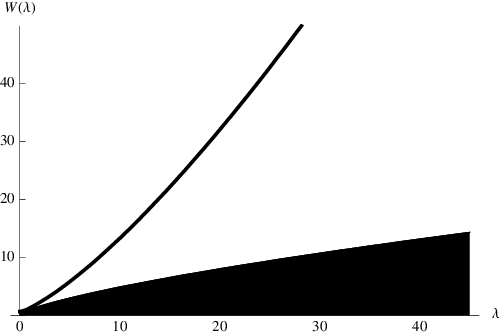}
\caption{Superpotential of the ``generic'' kind. The black area is the forbidden region below
the curve $\sqrt{27 V/64}$ } \label{superbad}
\end{center}
\end{figure}

\item A  unique   solution,    which also reaches the large-$\l$ region, but slower:
\be\label{singular}
 W(\l) \sim W_\infty \l^Q (\log \l)^{P/2}, \qquad W_\infty = \sqrt{{27 V_\infty  \over 4(16- 9Q^2)}}
\ee
This leads to  a repulsive  singularity, provided $Q<2\sqrt{2}/3$ \cite{ihqcd}.
We call  this  the {\em special} solution. An example is shown in figure \ref{supergood}

\begin{figure}[h]
 \begin{center}
\includegraphics[scale=0.9]{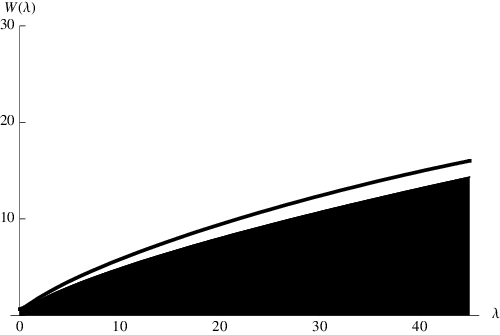}
\caption{Superpotential of the ``special'' solution. The black area is the forbidden region below
the curve $\sqrt{27 V/64}$ } \label{supergood}
\end{center}
\end{figure}

\item A second continuous one-parameter family where $W(\l)$  does not reach the asymptotic region.
 These solutions have two branches that both reach $\l=0$ (one in the UV, the
other in the IR)  and merge at a point $\l_*$
where $W(\l_*) = \sqrt{27 V(\l_*)/64}$. The IR branch is again a  ``bad'' singularity
at a finite value $r_0$, where $W\sim \l^{-4/3}$,  and
\be
b(r) \sim (r_0 -r)^{1/3}, \qquad \l(r) \sim (r_0-r)^{1/2}.
\ee
We call  this  solution {\em bouncing}. An example is shown in figure \ref{superugly}
\begin{figure}[h]
 \begin{center}
\includegraphics[scale=0.9]{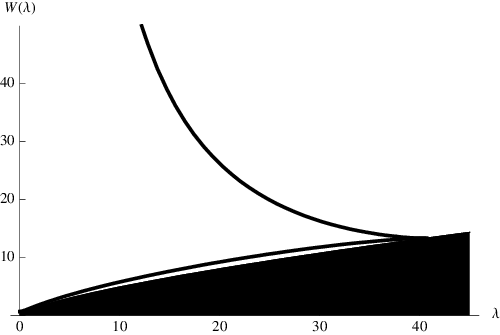}
\caption{Superpotential of the ``bouncing''  kind. The black area is the forbidden region below
the curve $\sqrt{27 V/64}$ } \label{superugly}
\end{center}
\end{figure}
\end{enumerate}

Notice that,  as two solutions with positive derivative cannot cross, the
special  solution (figure \ref{supergood}) marks the
boundary between the generic solutions, that reach the asymptotic large-$\l$   region
as $\l^{4/3}$ (figure \ref{superbad}) and the bouncing  ones,  that don't reach it,
figure \ref{superugly}.
Notice that, if $Q>4/3$, only  bouncing  solutions exist.

In all types of solutions the UV
corresponds to the region $\l\to 0$ on the $W_+$ branch.
There  the behavior of $W_+$ is universal: a power series in $\l$ with
{\em fixed} coefficients, plus a subleading non-analytic piece
which depends on an arbitrary integration constant $C_w$:
\be
W  = \sum_{i=1}^{\infty} W_i\l^i + C_w \l^{16/9} e^{-{16 W_0\over 9 W_1}} \left[1 + O(\l)\right]
\ee
All the power series coefficients $W_i$ are completely determined by the
coefficients in the small $\l$ expansion of $V(\l)$, the first few being:
\be
W_0 =  {\sqrt{27 V_0}\over 8}, \quad W_1 = {V_1\over 16}\sqrt{{27\over V0}}, \quad W_2 = {\sqrt{27}(64V_0V_2 -7V1^2)\over 1024 V_0^{3/2}}
\ee

\section{The Superpotential at finite  T}\label{superfiniteT}

\subsection{The thermal  Superpotential} \label{thermalsuper}

A useful way of  counting  the integration constants and to parametrize the
different black-hole solution is to extend the notion of superpotential to
the black-hole backgrounds.
We will call the resulting {\em thermal superpotential} $W(\Phi)$,
and its zero-temperature precursor $W_o(\Phi)$.

First,
we define \be\lab{bg} b=e^A,\qquad f = e^g, \ee and use the
domain-wall parametrization of the metric, \be\lab{cov} dr =
e^{-A}du. \ee In this coordinate frame the metric has the following
form:
\begin{equation}\label{fTdw}
    ds^2 = e^{2A}\le(-e^g dt^2 + dx_mdx^m\ri) + e^{-g}du^2.
\end{equation}
The equations of motion (\ref{a11}) and (\ref{a12})
 in the variable $u$ take the following form (a prime denotes derivative w.r.t. $u$):
\bea
A'' +\frac49 {\f'}^2 &=& 0,\label{eom2}\\
g'+\frac{g''}{g'}+4A' &=& 0,\label{eom3}\\
12 {A'}^2+3A'g'-\frac43 {\f'}^2-e^{-g} V &=&0,\label{eom1}
\eea

Equation (\ref{eom2}) is the same as in the zero-temperature
system, cfr. eqs. (\ref{eqs2}). This has an interesting
consequence: we recall that (\ref{eom2}) is the equation
that guarantees a well-defined arrow of the RG flow in the gauge theory. Here
we see that one can define the RG flow in the same way in the gauge theory at
finite-$T$. In other words, the conclusion that $A(u)$ is monotonically decreasing
still holds at finite temperature.

Since eq. (\ref{eom2}) takes the same form as when $g(r)=0$, we can use this equation
to define a superpotential just as for  $T=0$: the second order equation (\ref{eom2}) can
be replaced by the two first order equations, and the system becomes:
\bea
&& A'=-\frac{4}{9}W\left(\Phi\right), \qquad  \Phi'=\frac{\intd W\left(\Phi\right)}{\intd\Phi}\label{supPhi-A}, \\
&& \frac{g''+g'^2}{g'}=\frac{16}{9}W\left(\Phi\right) \label{supg}\\
&& -\frac{4}{3}\left(\frac{\intd W\left(\Phi\right)}{\intd\Phi}\right)^2+\frac{64}{27}W\left(\Phi\right)^2 -{4\over 3}W\left(\Phi\right)g'=
\ex^{-g} V(\Phi). \label{supW}
\eea
Equations (\ref{supPhi-A}) provide the definition of the thermal superpotential $W(\Phi)$.

\subsection{Counting integration constants: uniqueness properties of  BH solutions}\label{counting}
The system (\ref{supPhi-A}-\ref{supW}) has to be solved for the functions $A, \Phi, g, W$ once $V(\Phi)$ is given
as input. As in the $T=0$ case, once $W(\Phi)$ is given the scale factor and dilaton
are uniquely fixed up to a single physical integration constant (a choice of scale)
However, now $W(\Phi)$ is not simply a solution of a first order equation,
as in (\ref{super1}). Since we cannot decouple the equation for  $W$, we cannot
choose it as an input as we did in the $T=0$ case.

As it turns out, the most natural place where to fix the integration constants  is the horizon,
rather than the boundary. In other words, the general solution of the fifth order system
(\ref{supPhi-A}-\ref{supW}) is most easily parametrized by the horizon value of the
functions involved.

Consider a black-hole with a regular horizon located at $u=u_h$.
As we approach the horizon, $u\to u_h$,  the dilaton, the scale factor and the superpotential
have regular expansions,
\bea
&& A\to A_h - (u_h-u) A'(u_h) +\ldots, \quad \l\to \l_h -(u_h-u) \l'(u_h)+\ldots, \nn \\
&& \quad W \to W_h -(u_h-u)\l'(u_h)\de_\l W(\l_h),
\eea
where $W_h\equiv W(\l_h)$.
On the other hand close to the horizon we must have have:
\bea
g = \log(u_h-u) + g_h + O\left((u_h-u)^2\right) \,,\qquad u\to u_h
\eea
where $g_h = \log[-f'(u_h)]$.
Substituting these values in equation \refeq{supW} we see that regularity
at the horizon requires the condition:
\be \label{constraintW}
W_h={1\over 3} e^{-g_h} V(\l_h).
\ee
On the other hand, the quantities  $A_h$, $\l_h$ and $g_h$ are free.
The differential equation determines the following terms in the expansion
around $u_h$ in terms of these quantities, and of $V(\l)$ and its derivatives
at $\l_h$.

We can use the horizon quantities to fix all integration
integration constants of the system (\ref{supPhi-A}-\ref{supW}).
The system is first order in $A(u)$,$\Phi(u)$
and $W(u)$,  and second order in $g(u)$, therefore it contains five integration constants. One of them is un-physical, and it is due to reparametrization
invariance. It can be eliminated by using $\l$ in place of $u$ as a coordinate,
or it can be fixed by setting  $u_h$ to an arbitrary value.
The four quantities ${\l_h, A_h, g_h, W_h}$ provide the remaining four
integration constants. For an arbitrary choice the solution will be
singular at the horizon, whereas regularity is assured by imposing
the constraint (\ref{constraintW}).

Notice that  the value of the potential
at the horizon has to be positive in order to get a well-behaved black-hole solution.
This is similar  to  Gubser's  criterion identifying good singularities \cite{gub2}.\\

What we have  shown above means that the theory has  a three-parameter
family of regular black-hole solutions, characterized by the three
real numbers ${\l_h, A_h, g_h}$.
However, since the initial conditions
were set at the horizon, these solutions will not all have the same
UV asymptotics. To understand what happens in the UV,  $u\to -\infty$
in these coordinates), notice that close to $u_h$ we
have  $W(u)>0$, $g'(u)<0$. Moreover by eq. (\ref{supg}),   $g''(u)<0$
as long as $g'(u)<0 $ and $W>0$.
 Thus, $|g'(u)|$ decreases as we go further away
from the horizon. As a consequence, the extra term $-Wg'$ in (\ref{supW}),
that did not appear in the zero-temperature equation (\ref{super1}),
become less and less important as we move away from the horizon towards
the asymptotic region.  At the same time
due to the relative signs of $g'$ and $g''$, $g(u)$ approaches
 a constant value $g_0$, and the r.h.s. of eq. (\ref{supW})
approaches the r.h.s. of (\ref{super1}) up to a multiplicative
constant. Therefore  the solution will get closer and
closer to one of the zero-temperature solutions, up to a rescaling of $W(\Phi)$: $W(\l)\to e^{g_0/2} W_o(\l)$ as $\l \to 0$.
The existence of a solution that connects the horizon to the UV boundary will be proved more rigorously in Appendix \ref{conUVIR}.

This is not the end of the story, since we want the the black-hole
solution to have the same UV asymptotics as the zero-temperature solutions.
Notice however that eqs. (\ref{supPhi-A}-\ref{supW}) are invariant under
the two  following independent transformations:
\bea
&& {\l(u), A(u), g(u), W(u)} \to   {\l(u e^{-\delta_1}), A(u e^{-\delta_1}), g(u e^{-\delta_1})- 2\delta_1, e^\delta_1 W(u e^{-\delta_1})}  \\
&&  {\l(u), A(u), g(u), W(u)} \to   {\l(u), A(u) + \delta_2, g(u),  W(u)}
\eea
where $\delta_{1,2}$ are arbitrary real numbers.
These transformations map a solution into another solution, and also
preserve the regularity condition (\ref{constraintW}). Therefore one can
use these transformations to move in the space of solutions, and reach the
one with the desired asymptotics. Specifically, one can
construct a regular UV black-hole solution with  UV asymptotics matching a
given $T=0$ background in the following way:
\begin{enumerate}
\item choose an arbitrary horizon position $u_h$ and fix arbitrary
values for the initial data at the horizon, namely $A_h,g_h,\l_h$
\item fix the fourth initial data, $W_h$, according to the
regularity condition.
\item Evolve the solution from the horizon to the UV. In general, as $u\to -\infty$, $g(u)$ will go to a constant $g_{UV}\neq 0$.
\item Use a symmetry transformation with parameter $\delta_1 = g_{UV}/2$. In
the new solution $g$ goes to the correct UV limit (namely, zero).
\item Use a $\delta_2$ transformation to reset the overall scale of the
solution to the desired value (e.g. to match a given $T=0$ solution). (This
does not affect $g(u)$)
\end{enumerate}
At the end of this procedure, the only free parameter remains the initial choice
of $\l_h$. Thus, for each $\l_h$, the solution with given UV boundary conditions
is unique.  For a given choice of UV asymptotics,  the black-hole metric and
temperature depends only on $\l_h$, which can then be used as an unambiguous quantity to parametrize the different solutions.

\subsection{Asymptotic form of the solution}

Here we determine the behavior of the solution
of the finite-T generalization of the superpotential equations,
(\ref{supW}), in the small $\l$ and large-$\l$ region, respectively.

The nontrivial part of eqs.(\ref{supPhi-A}-\ref{supW}) is the one that determines $W$ and $g$; then, $A$ and
$\Phi$ follow as in the zero-temperature case. We can decouple the $W-g$  system from the rest, as follows:
First
using  (\ref{supPhi-A}), we rewrite eqs. (\ref{supg}) and \refeq{supW}
keeping $\Phi$ as the independent variable. In this way, the four equations (\ref{supPhi-A}--\ref{supW}) split
in two  independent sets of equations. For the scale factor we have (as for zero-temperature)
\be
\de_\Phi A = -{4\over 9}{W\over \de_\Phi W};
\ee
 for
$W(\Phi)$ and $g(\Phi)$ we find:
\bea
&&\left(\de_\Phi g + {\de^2_\Phi g\over \de_\Phi g}\right)\de_\Phi W + \de^2_\Phi W = {16\over 9} W \label{supg2app}\\
&&-{4\over 3}W(\de_\Phi W) (\de_\Phi g )- {4\over 3}\left(\de_\Phi W\right)^2 + {64\over 27}W^2 =
e^{-g} V(\l) \label{supW2app}
\eea

Eq. (\ref{supg2app}) can be integrated to a
closed expression for $g(\Phi)$ in terms of $W(\Phi)$. Write

\be\label{c1}
{\de_\Phi} \left[g + \log (-\de_\Phi g) \right] = {{16\over 9} W - \de^2_\Phi W  \over \de_\Phi W} \equiv F_{W}(\Phi)
\ee
Thus,
\be\label{c2}
 g + \log (-\de_\Phi g) = \int d\Phi\, F_{W}(\Phi),
 \ee
and exponentiating:
\be\label{c3}
(-\de_\Phi g) e^g = e^{\int d\Phi\, F_{W}(\Phi)}.
\ee
 Integrating one more time we obtain an explicit expression for $g(\Phi)$ in terms
 of the (still unknown) function $F_{W}(\Phi)$:
 \be \label{c4}
 g(\Phi) = \log \int d\Phi e^{\int d\Phi' F_{W}(\Phi')}.
 \ee

\subsubsection{Solution  of the $W(\Phi)$-$g(\Phi)$ system  in the $\l\to 0$ limit} \label{superfinTUV}

As usual, we assume a power series expansion of $V(\l)$ as $\l\to 0$.
We take a power series ansatz for $W(\l)$ as well, as for its
zero temperature counterpart,
\be   \label{c5}
W(\l) = \tilde{W_0} +  \tilde{W_1}\l  + \tilde{W_2}\l^2 + \ldots
\ee
where the expansion coefficients $\tilde{W_i}$ are {\it  a priori}
temperature dependent.

Using this ansatz in eq. (\ref{c4}), written in terms of $\l = e^\Phi$, we
can obtain  the form of $g(\l)$ for small $\l$. We have from eq (\ref{c1})
\be\label{c6}
F_W (\l) =  {16 \tilde{W_0} \over 9  \tilde{W_1}}{1\over \l} + {1\over 9}
\left(7-32 {\tilde{W_0}\tilde{W_2}\over \tilde{W_1}^2}\right) + O(\l), \qquad \l\to 0.
\ee
Using this expression in  eq. (\ref{c4}) we obtain:
\be \label{c7}
g(\l) = \log \left\{ g_1 +  g_2 \l^\gamma e^{-{16\tilde{W_0} \over 9  \tilde{W_1}}{1\over \l}} \left[1 + O(\l)\right]\right\}
\qquad \gamma \equiv {16\over 9}-32{\tilde{W_0}\tilde{W_2}\over \tilde{W_1}^2}
\ee
where $g_1$ and $g_2$ are integration constants.
To recover the UV boundary condition $g(\l)\to 0$ as $\l\to 0$, we must
fix $g_1=1$. $g_2$ ultimately determines the temperature of the solution,
and is only required to be negative.

Next, we insert the asyptotics  (\ref{c7}) in  the equation
(\ref{supW2app}), in order to determine the coefficients.
For small $\l$ it reads:
\be\label{c8}
 -{4\over 3}g_2 W(\de_\l W) \l^{\gamma+2}e^{{-16\tilde{W_0} \over 9  \tilde{W_1}}
 {1\over \l}}\left[1+ O(\l)\right] - {4\over 3}\l^2 \left(\de_\l W\right)^2 + {64\over 27}W^2 =
 V(\l)\left[1 - g_2 \l^{\gamma}e^{{-16\tilde{W_0} \over 9  \tilde{W_1}}{1\over \l}} + \ldots \right]
\ee
To any finite order in powers of $\l$, {\em  this equation is the same
as the zero-temperature superpotential equation}, (\ref{spot1}). It follows
that the power series expansion of $W(\l)$ is not affected by the temperature,
and all the coefficients are the same as in the zero temperature solution:
$\tilde {W_i} = W_i$. The difference between the finite-$T$ and zero-$T$ solutions
are of order $\l^\gamma e^{-4/(b_0\l)} \sim r^4$, and imply a temperature dependent
value for the gluon condensate, cfr. Appendix E.

Since the series coefficients of $W(\l)$ completely determine the UV series
in inverse logarithms of $r$
of the metric and dilaton, it follows
that such series has the same form for any temperature.

\subsubsection{Solution of the $W(\Phi)$-$g(\Phi)$ system in the asymptotic
large $\Phi$ region} \label{appIRas}

Next we want to solve the system of eqs (\ref{supg2}--\ref{supW2}),
for in the asymptotic region of large  $\Phi$. This is defined as the
region beyond some $\Phi_0$
where the potential can be well approximated by its leading asymptotic\footnote{the actual value of $\Phi_0$ is immaterial, and will depend on
the specific choice for the potential.},
\be
V(\Phi) \simeq  e^{2Q\Phi} \Phi^P \qquad \Phi \gg \Phi_0.
\ee
 We assume the horizon is situated in this region, i.e. we work in the limit
$\Phi_h\geq \Phi > \Phi_0$.
The IR asymptotics of the zero temperature
superpotential   $W_o(\Phi)$  giving rise to the
well behaved solutions with the repulsive singularity are of the form:
\be\label{W0asapp}
W_o \sim \Phi^{P/2}\,e^{Q\,\Phi}, \Phi \gg \Phi_0.
\ee

Let us  try the ansatz for the large $\Phi$ behavior of $W$:
\be\label{c5b}
W(\Phi)\sim e^{\tilde{Q}\Phi}\Phi^{\tilde{P}/2}, \qquad \Phi\gg \Phi_0.
\ee
With this ansatz we can directly calculate the r.h.s. of eq. (\ref{c1}).
To leading and first subleading order we have:
\be\label{c6b}
 F_{W}(\Phi) \simeq K + {R\over \Phi} + O\left({1\over \Phi^2}\right),
 \quad K\equiv \left({16\over 9 \tilde{Q}} - \tilde{Q}\right),
 \quad R \equiv -{\tilde{P}\over 2}\left(1+ {16 \over 9\tilde{Q}^2}\right),
\ee
and from eq. (\ref{c3}) we obtain:
\be\label{c7b}
g(\Phi) = \log \left\{C_2 - C_1 \Phi^R e^{K\Phi}\left[1+O\left({1\over \Phi}\right)\right]\right\},
\ee
where $C_1$ and $C_2$ are two integration constants. This solution is supposed to be
valid in the whole region $\Phi_0\ll\Phi\leq\Phi_h$, and we can relate $\Phi_h$ to the
integration constants by going to the horizon,  $g \to -\infty$:
\be\label{c8b}
C_2 - C_1 (\Phi_h)^R e^{K\Phi_h} = 0
\ee
so we can write eq. (\ref{c7b}) as:
\be\label{c9b}
g(\Phi) = \log C_2 \left\{1  - \left({\Phi\over \Phi_h}\right)^R e^{K(\Phi-\Phi_h)}
\left[1+O\left({1\over \Phi}\right)\right]\right\}.
\ee
Finally, requiring $g\simeq 0$ for $1\ll\Phi\ll\Phi_h$, fixes $C_2=1$.

The appropriate solution for $f\equiv\ex^g$ is therefore:
\be\lab{solf}
f\left(\Phi\right)=1- \left({\Phi\over \Phi_h}\right)^R e^{-K\left(\Phi_h-\Phi\right)}.
\ee
which correctly interpolates between the desired
behavior at $\Phi\ll\Phi_h$ and $\Phi=\Phi_h$. This
solution is valid in the whole region $\Phi\gg \Phi_0$ up
to the horizon $\Phi=\Phi_h$\footnote{For $\Phi>\Phi_h$
the solution is the same, but the definition of $g$
changes: $f=-e^{-g}$.}.

Now let us look at equation (\ref{supW2}). Using (\ref{solf}), and
neglecting $O(1/\Phi^2)$ terms,  it becomes:
\bea\lab{supW3}
 && \left\{{4\over 3}\left(\tilde{Q}+ \tilde{P}/2\Phi\right)\left(K + {R\over \Phi}\right)
 {\left(
 {\Phi\over \Phi_h}\right)^R e^{K(\Phi-\Phi_h)}  \over 1 - \left(
 {\Phi\over \Phi_h}\right)^R e^{K(\Phi-\Phi_h)}}
- {4\over 3}\left(\tilde{Q}^2+ {\tilde{P}\over 2\Phi}\right)^2 + {64\over 27}\right\}
e^{2\tilde{Q}\Phi} \Phi^{\tilde{P}} \non\\
= && {1\over 1 -\left({\Phi\over \Phi_h}\right)^R e^{K(\Phi-\Phi_h)}}
\left\{-{4\over 3}\left(Q^2 +{P\over 2\Phi}\right)^2 +
{64\over 27}\right\}e^{2Q\Phi} \Phi^{P}.
\eea
For $\Phi\ll\Phi_h$ this requires:
\be
\tilde{Q} = Q, \quad \tilde{P} = P,
\ee
i.e. the superpotential must have  the same large $\Phi$ asymptotics as the zero-temperature special
solution. {\it  Then,
eq. (\ref{supW3}) is equivalent to  the  $\Phi$-independent  algebraic equations}:
\be
-{4\over 3}Q K = -{4\over 3}Q^2 + {64\over 27}, \qquad \left({PK\over 2} + RQ\right) = -QP
\ee
which are identically satisfied, due to the definitions of $K$ and $R$ in (\ref{c6b})!

Therefore the asymptotic solution with horizon at $\Phi=\Phi_h$ is , to this order:
\bea
&& W(\Phi) \simeq W_o (\Phi) \simeq e^{Q\Phi} \Phi^{P/2} \\
&& f(\Phi) \simeq 1 - \left(\Phi\over \Phi_h\right)^R\exp[- K(\Phi_h - \Phi)],
\eea
valid in the whole asymptotic region $\Phi_h > \Phi \gg \Phi_0$.

In the particular case of power-law behavior, $A_o(r)\sim -Cr^\a $,
corresponding to $Q=2/3$, $P=(\a-1)/\a$ (see \cite{ihqcd}),
 the solution in $r$-coordinates is, for large
$r$ :
\bea
&&A (r) \sim -C r^\a, \label{solAr2app}\\
&& \Phi(r) \simeq {3\over 2}C r^\a + {3\over 4}(\a-1)\log r, \label{solPhir2app}\\
&& f \sim  \Phi^{-{5\over 2}P } e^{2\Phi} \sim r^{-{5\over 2}(\a-1)} e^{3 C r^\a + {3\over 2} (\a-1) \log r} \sim r^{1-\a}  e^{3 C r^\a}
\eea
The horizon position is obtained by inverting $\Phi(r_h) = \Phi_h$. 

We can verify directly that these asymptotics solve eqs. (\ref{a11}--\ref{a12}), for
$V\sim \ex^{\frac{4}{3}\Phi}$. Using the asymptotic form
$b(r)=b_0\ex^{-C r^\a}$, we get, integrating eq (\ref{a17b})
\bea
f(r)&\sim&C_2+C_1\int^r\intd t\,b^{-3}(t)=C_2+\frac{C_1}{\a C}\int^{3C r^\a}\intd s\,s^{\oneover{\a}(1-\a)}\ex^s \non\\
&\simeq& C_2+\frac{C_1}{3\a C}r^{1-\a}\ex^{3Cr^\a} \label{bcube}
\eea
Substituting into the equation for the potential yields
\bea\label{Vcorr2}
V(r)&=&\frac{3}{b^2}\left\{ f\left[2\frac{\dot b^2}{b^2}+\frac{\ddot b}{b}\right]+\dot f\frac{\dot b}{b}\right\} \non\\
&\simeq& \frac{3}{b^2}\left\{\left(C_2+\frac{C_1}{3\a C } r^{1-\a} \ex^{3C r^\a}\right)
\left(3 \a^2 C^4 r^{2(\a-1)}\right)+C_1 \ex^{3C r^\a}\left(-\a C r^{\a-1}\right)\right\} \non\\
&\simeq& \frac{9\a^2 C^2 }{b_0n^2} C_2 r^{2(\a-1)} \ex^{2r^\a} \sim \Phi^{\a-1\over \a} e^{{4\over 3}\Phi}.
\eea
This is the expected leading behavior for the potential. The contributions proportional to $C_1 \ex^{5C r^\a}\simeq C_1\ex^{{10\over 3}\Phi}$
cancel between the two terms containing $f$ and $\dot f$ respectively. V is independent of $C_1$, which encodes the horizon position.
Furthermore, comparing eq.  \refeq{Vcorr2}
with the expression for $V(r)$ one obtains with $f \equiv 1$, we see that
 $C_2$ must be set to $C_2=1$ if the potential has to be the same as in the zero-temperature
 solution, in agreement with eq. (\ref{solfr6}).

\section{Multiple Big Black-holes}\label{multiX}

\begin{figure}
 \begin{center}
 \leavevmode \epsfxsize=12cm \epsffile{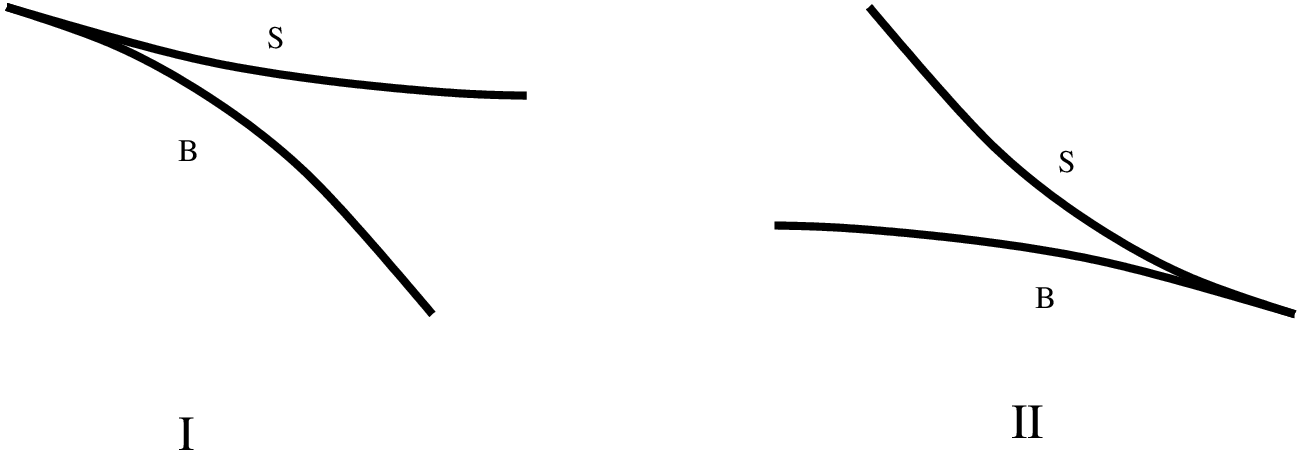}
 \end{center}
 \caption[]{Two possible types of cusps at which the small and big BH in pair can merge. These are denoted by "S" and "B" , respectively.
 The full curve $F(T)$ should be constructed by various combinations of these vertices. }
 \label{cusps}\end{figure}

In this section we generalize the proposition of sec. \ref{conftrans} to a much larger class of geometries,
for which the functions $F(\l_h)$ and $T(\l_h)$ may acquire multiple extrema
\footnote{We assume that these functions are $C^{\infty}$ in $\l_h$.},
see figs. \ref{Tlhmulti} and \ref{Flhmulti} for an example.
In these cases there are more than just two BHs. However for the generic confining theories\footnote{except the
borderline case of $\a=1$.},
they still come in pairs of one small ($T'(\l_h)>0$) and one big ($T'(\l_h)<0$) BH, connected at an
extremum of $T(\l_h)$. Therefore there is an even number of BHs in total. This follows from the fact that there always exists
an asymptotically AdS big BH for $\l_h\to 0$ and a small BH for $\l_h\to\infty$ and one extremum of $T(\l_h)$ creates
two branches (a small and a big), by definition.

In particular we want to prove the following:
\begin{enumerate}
\item Existence of a deconfinement phase transition,
\item Finite latent heat (first order transition),
\item Continuity of $F$ as a function of $T$,
\item Uniqueness of the deconfinement transition.
\end{enumerate}

We shall first present a graphical proof of these points.
For illustration purposes we make  the following
additional assumption (that seems natural and satisfied by a large class of potentials):
$C_v = T dS/dT$ is negative (positive) for small (big) BHs\footnote{as in the case of the $AdS_5$
BH with spherical horizon.}.
After the graphical demonstration below, we provide an analytic proof
which applies to cases when this additional assumption  is weakened.
The  assumption about the specific heat is actually not needed for point 1 in the above list,
so that the main result (the existence of a phase transition) is valid for an arbitrary behavior
of the specific heat.
The function $F(T)$ is determined from $F(\l_h)$ and $\l_h(T)$. Since the latter is
generally multi-valued, so is $F(T)$. As a multivalued function it can have a very complicated
form with cusps and crossings, see fig. \ref{FTmulti} for an example. Although complicated,
the form of $F(T)$ is restricted by certain rules:

\begin{figure}
 \begin{center}
 \leavevmode \epsfxsize=12cm \epsffile{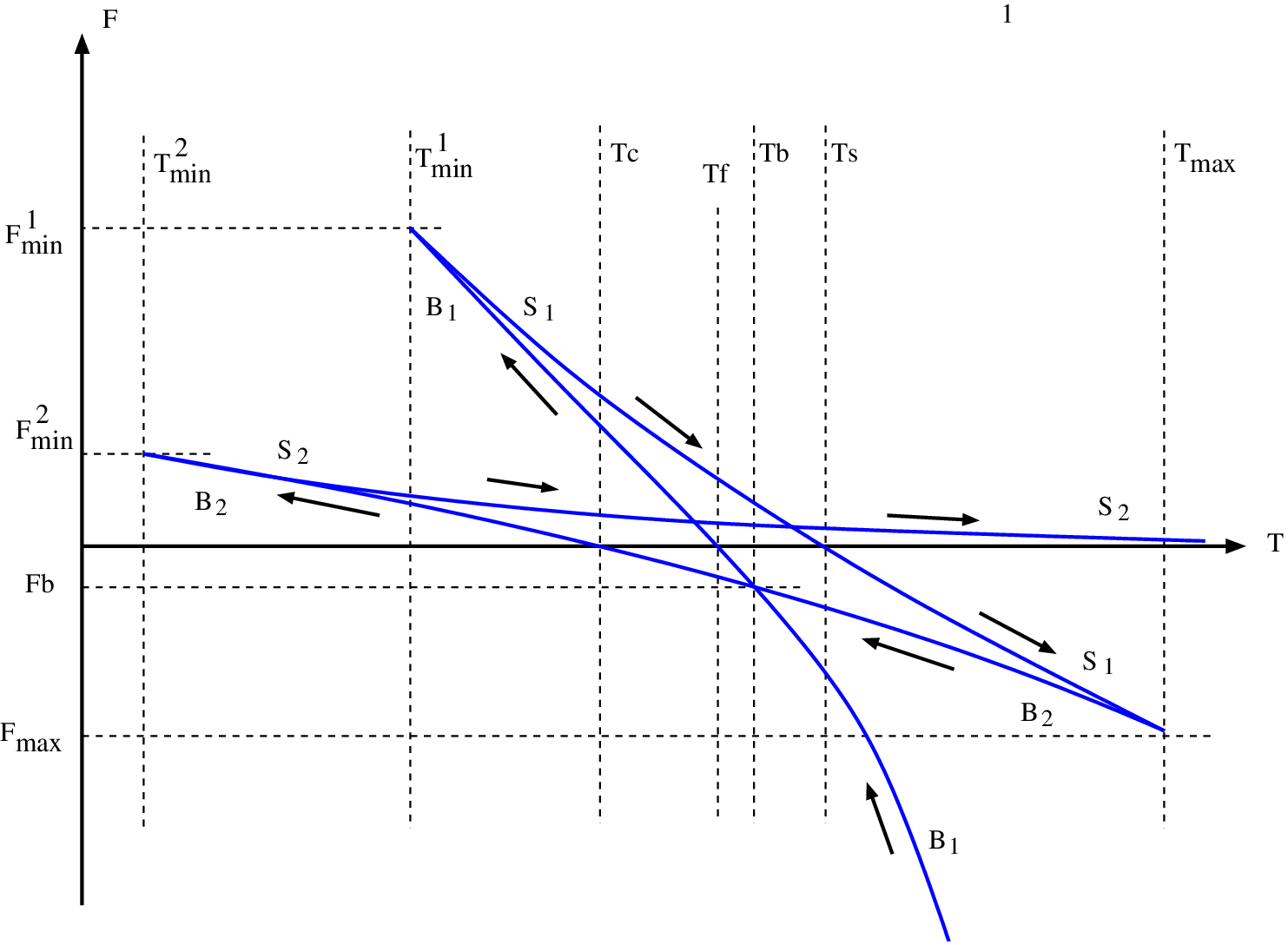}
 \end{center}
 \caption[]{Example of a curve $F(T)$ that exhibits multiple extrema. $S_1$ and $S_2$ denote small BHs whereas $B_1$ and $B_2$ denote big BHs.
 The arrows represent direction of increasing $\l_h$. }
 \label{FTmulti}\end{figure}

\begin{enumerate}

\item On every piece of the $F(T)$ curve, $F'(T)<0$. This follows from the positivity of
entropy.

\item On the small black-hole branches $F''(T)>0$, and on the big black-hole branches $F''(T)<0$.
This follows from our assumption above and from  $F'' \propto - C_v$.

\item  There should always be
a big black-hole branch
(which asymptotically becomes the AdS black-hole at high-$T$) in the high-$T$ (small $\l_h$) region, on which
$F(T)\to -\infty$ as $T\to\infty$.

\item There should always be a small black-hole branch in the high-$T$ (large $\l_h$ ) region, on which
$F(T)\to 0$. This follows from the discussion in section \ref{IRasymps}.

\item The small and the big BHs always come in pairs, hence there are equal numbers of branches on the $F(T)$ curve,
with negative and positive $F''(T)$. This is clear from the fact that one small and one big black-hole branches off
 at an extremum of $T(\l_h)$.

\item These merging points of a pair of big and small BHs are represented by a cusp. There are two possible types
of cusps as shown in fig. \ref{cusps}.
These particular shapes follow from the first two properties above. Since at the merging points the entropy
is the same, ${\cal F}'$ is the same on the two branches of the cusp.

\end{enumerate}

\begin{figure}
 \begin{center}
 \leavevmode \epsfxsize=12cm \epsffile{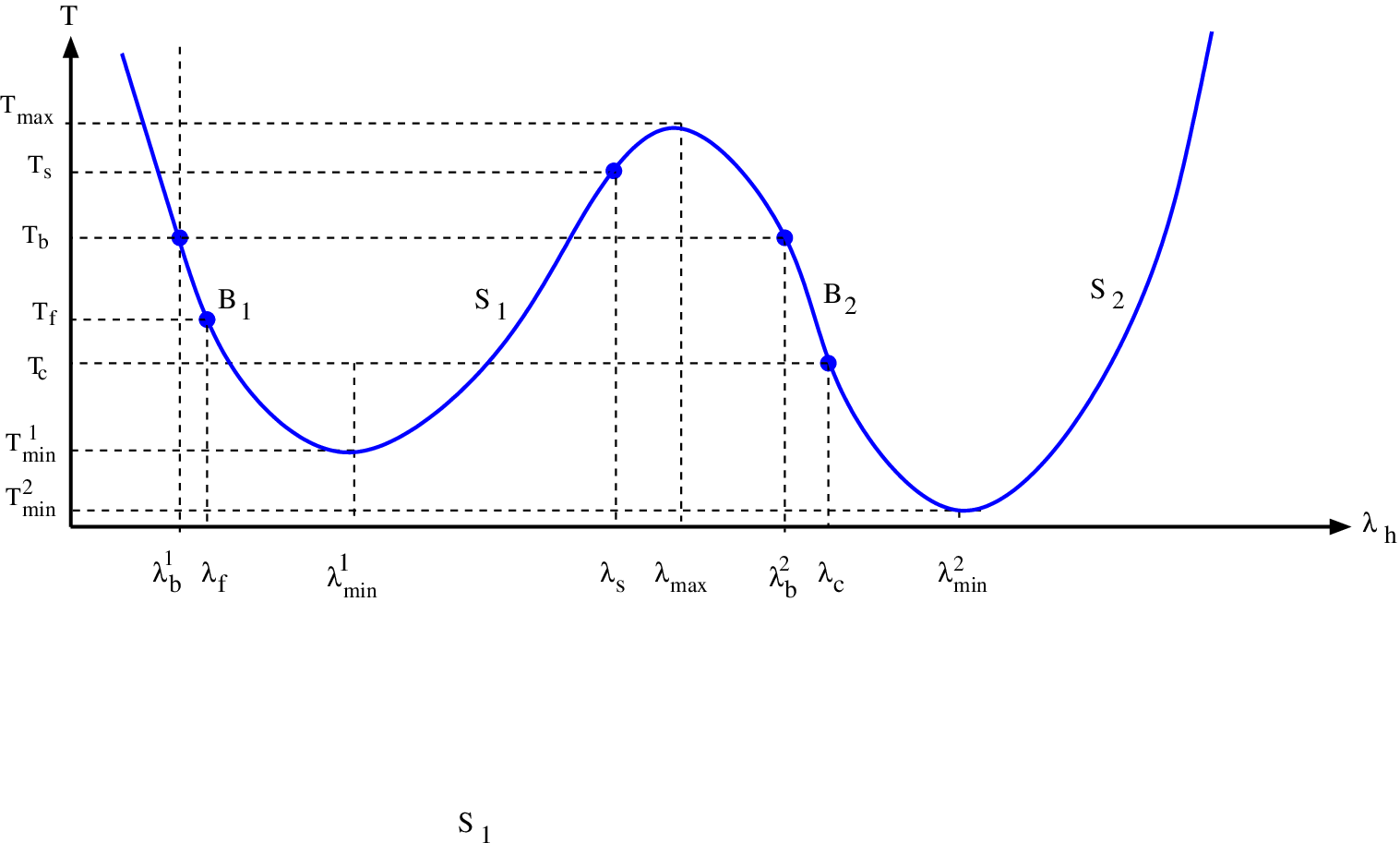}
 \end{center}
 \caption[]{$T$ as a function of $\l_h$ in the example of fig. \ref{FTmulti}. }
 \label{Tlhmulti}\end{figure}

For an example of a curve $F(T)$ with these properties, see fig. \ref{FTmulti}.
Given these properties, it is not hard to show that the {\em minimum energy configuration}
for $T>T_c$ \footnote{There may be more than one $T_c$ on which $F$ vanishes and our statement applies
to all of these points.} is always a big BH. This is because, the entire curve $F(T)$ should be formed out of
the vertices given in fig. \ref{cusps} connected with small and big BH legs. Clearly for any small BH, there exist
a big BH that stems from the same vertex which has lower energy. Therefore, in the entire curve for $F<0$,
the lowest energy configuration should be a big black-hole.

\begin{figure}
 \begin{center}
 \leavevmode \epsfxsize=12cm \epsffile{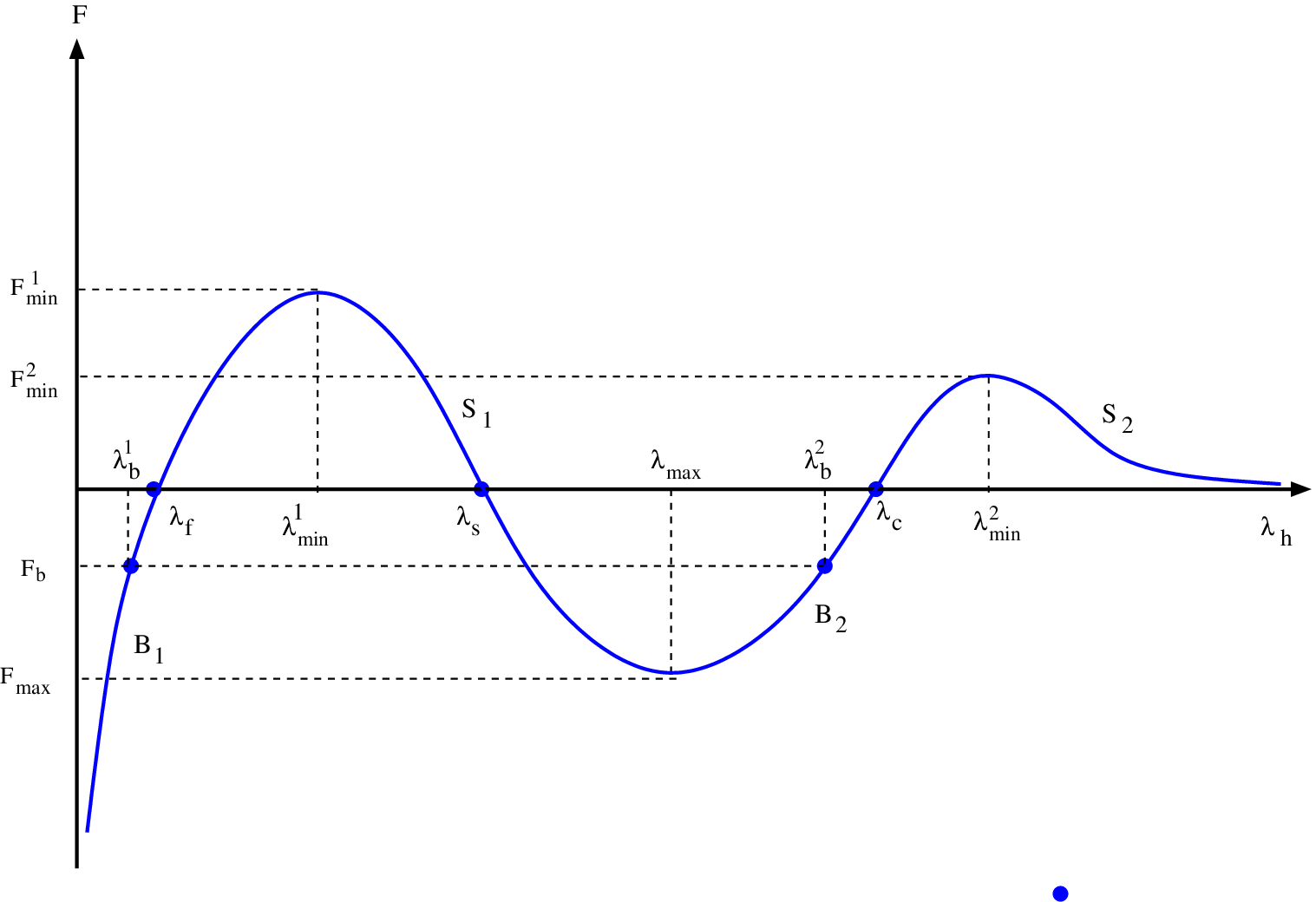}
 \end{center}
 \caption[]{$F$ as a function of $\l_h$ in the example of fig. \ref{FTmulti}.}
 \label{Flhmulti}\end{figure}

Of course, below $T=T_c$ the lowest energy configuration becomes the thermal gas and the curve $F_{min}(T)$ for the
minimum energy configuration always looks like in fig. \ref{FTmin}. For the example given in fig \ref{FTmulti}, the corresponding free energy diagram is constructed in \ref{FTmultimin}.
The reason that $F(T)$ should cross zero (point 1 in the list above),
hence exhibit a phase transition, follows from  the 2nd and the 3rd properties above. Similarly, the points 2, 3 and 4 simply follow
from the fact that one cannot draw a function $F(T)$ that violates these points with the properties listed above.
Thus, we demonstrated what we wanted: {\em the proposition of section \ref{conftrans} apply to the multiple extrema cases as well.}

Before going into the analytic proof of these statements,
let us analyze the example of fig. \ref{FTmulti} in more detail. In this particular case, there are two small and big BH pairs that are denoted by $(S_1, B_1)$
 and $(S_2,B_2)$.  The black-holes in each pair merge  at two local minima of $T(\l_h)$ (that are denoted by $T_{min}^1$ and $T_{min}^2$)) and the two pairs
 are connected at a local maxima of $T(\l_h)$ (denoted by $T_{max}$). See fig. \ref{Tlhmulti} which plots $T(\l_h)$, that corresponds to fig. \ref{FTmulti} for a clear
 demonstration of these facts. We also present the function $F(\l_h)$ that corresponds to this example in fig. \ref{Flhmulti}. Given fig. \ref{Tlhmulti} and fig. \ref{Flhmulti},
 fig. \ref{FTmulti} follows by solving for $F$ as a function of $T$ parametrically. The arrows in fig. \ref{FTmulti} point to the direction of increasing $\l_h$.

Fig. \ref{FTmulti} exhibits various {\em first order} transitions between different branches. First of all there are
transitions between thermal gas (with $F(T)=0$) and $B_1$, $B_2$ and $S_1$.
We denote these points as $T_f$, $T_c$ and $T_s$ respectively in fig. \ref{FTmulti}. However, not all of these points
correspond to actual phase transitions. In order to obtain the
true  free energy of the system, one should draw $F(T)$ {\em on the minimum energy configuration}.
This looks is much simpler and the one corresponding  to fig. \ref{FTmulti} is given by
fig. \ref{FTmultimin}. In particular, we see that out of $T_f$, $T_c$ and $T_s$, only $T_c$ is a real phase transition.
It is in fact the confinement-deconfinement phase transition in this example.
$T_f$ corresponds to a ``fake'' deconfinement transition, because $B_1$ has higher energy than $B_2$ at the point it crosses $F=0$. Similarly, at $T_s$,  there is a fake transition between a {\em small}
BH and the thermal gas geometry. Although these are fake transitions, hence uninteresting for
the dual field theory point of view, they may bear some interest on the bulk as they describe possible
transitions between various different geometries in asymptotically AdS spaces. The fake transitions
 described here parallels the transitions found in $R^4$ and $F^4$ corrected AdS geometry in \cite{harvard}.
In addition, there are various transitions among small and big BHs ($S_2$ and $B_1$ in the figure)
and different small black-holes ($S_1$ and $S_2$ in the figure).

However there is another interesting possibility that is also present in this example: {\em a first order transition between
two different big black-holes at $T_b$}. This transition is not ``fake'' as others, because it corresponds to first order transition
between two minimum energy configurations, $B_1$ and $B_2$ in fig \ref{FTmulti}. See also fig. \ref{FTmultimin}.
It will be interesting to investigate whether or not a dilaton potential $V(\l)$ with this property exists; if so, whether or not
this corresponds to a meaningful phase transition in the dual gauge theory.

\subsection{Analytic demonstration}

Suppose the function  $T(\l_h)$ has an arbitrary  number $n$ of minima, with
$n$  (even) odd  for (non-)confining IR asymptotics\footnote{except in the borderline confining case where there is also an even number of minima}, corresponding
to certain values $\l_i$, $i=1\ldots n$.    Then, there will be $n+1$ black-hole branches, each corresponding to the ranges $(\l_i,\l_{i+1})$, with $\l_0 \equiv 0$ and
$\l_{n+1} \equiv +\infty$.   In each branch the free energy as a function of $T$ is single valued, and
is given
by the first law, ${\cal F}_i(T) = -\int S_i(T')  dT' + C_i$ where $S_i$ is the entropy function in
the $i$th branch (i.e. in the interval  $(\l_{i-1},\l_i)$), and $C_i$ are integration constants.
 We can still write the
free energy in compact form as
\be\label{g1}
{\cal F}(\l_h) = \int_{\l_h}^{+\infty}d\l'\, S(\l') {d T\over d\l'},
\ee
as this is the unique continuous function that satisfies $d{\cal F}/d T = -S $ on every
branch, and vanishes as $\l_h \to \infty$. Since $S$ is nowhere vanishing for finite $\l$,
a minimum for $T(\l)$ corresponds to a maximum for ${\cal F}(\l)$.
Using  the integral expression of the free energy, we can
give an analytic proof of the proposition in Section \ref{conftrans} relating confinement
with the existence of phase transitions.

 Let us consider
confining asymptotics. We want to show that these always exhibit a first order phase
transition. In particular we want to prove  the following three statements:
\begin{enumerate}
\item ${\cal F}$ changes sign at some finite  $\l$. Specifically,
${\cal F} \to -\infty$ as $\l\to0$, whereas ${\cal F}(\l_n) >0$;
\item For every $\l_i$ corresponding to a {\em minimum} of $T(\l)$ , either ${\cal F}(\l_i)$ is positive or
it is larger than the free energy of some big black-hole with the same temperature;
\item For every  $\l_i$  corresponding to a {\em maximum} of $T(\l)$, ${\cal F}(\l_i)$ there exists
a big black-hole with the same temperature and lower free energy.
\end{enumerate}
Step 1 by itself shows that there is indeed a phase transition.
It goes in  the right direction,  since at small temperature ($T<T_{min}$) there are no black-hole
solutions, and moreover
in any black-hole branch the free energy is  a decreasing function
of temperature. Therefore the black-hole free energy crosses from positive to negative
(with respect to the vacuum)  as the temperature increases.

Steps 2 and 3  imply
that the free energy never jumps in either directions: the branching points when a  large/small black-hole
pair appears or disappears have always higher free energy than some other state in the ensemble (which therefore
must have been dominant starting from some lower temperature). If this weren't true, it would be possible
to have discontinuous jumps in the free energy, i.e. phase transitions with infinite latent heat. In this step,
one needs an  extra assumption about the ordering of the black-hole branches.
Step 3 also implies that  ones the system is in the black-hole phase for some temperature, it
stays in a black-hole phase for all higher temperatures, so there cannot be any inverse ``reconfining'' transition.

The proof of Step 1 is straightforward, and identical to the argument we used in section 5.3
in the case of $n=1$, i.e.  when   $T(\l_h)$ has a single minimum. Thus, the existence
of a deconfining phase transition {\em per se} is easy to prove.
To prove  Steps 2 and 3 however, in the general case $n> 1$
we need to make an extra assumption about the nature  of the entropy $S(\l)$: \\

\noindent
{\bf Assumption:}\\
{\em
the black-hole branches  are ``ordered'' in the sense that $S_i(\l) > S_{i+1}(\l')$
for any $\l \in (\l_{i-1}, \l_{i})$ and any $\l' \in (\l_{i}, \l_{i+1})$}. \\

\noindent
This still allows some local violation of the monotonicity of $S(\l)$ within each branch.
Although we cannot exclude violation of this assumption from first principles, it is
satisfied in all  examples we have studied numerically, where we also found cases
with strict monotonicity of $S(\l)$ violated (these cases correspond to regions of
stable small black-holes, as discussed in section 5.2.2).  The above assumption might
be relaxed, since it is probably not  a necessary condition for the statements 2 and 3 to hold,
 but it allows the construction of a simple enough proof.

With the above assumption, the proof of the proposition  goes as follows.

\subsubsection*{Step 1}

The fact that ${\cal F}<0$ for small enough $\l$ follows from eq. (\ref{fe-highT}), whose validity does not
depend on the number of extrema of $T(\l_h)$; on the other hand, evaluating the free energy
on the extremum (a minimum) of   $T(\l_h)$ with larger $\l_h$, we obtain:
\be\label{g2}
{\cal F}(\l_n) = \int_{\l_n}^{+\infty} S {d T\over d\l'} >0,
\ee
since in the last branch $d T/ d\l >0$.
This is true  without additional assumptions about  the function $S(\l)$.

\subsubsection*{Step 2}

First, consider the extremum $\l_i$ corresponding to the absolute minimum of $T(\l_h)$, $T(\l_i) = T_{min}$.
Since the  last extremum $\l_n$ is also a minimum, there  will be $2k=n-i$ minima and $2k-1$ maxima in
the region $\l > \l_i$, corresponding to $2k$ maxima and $2k-1$ minima for ${\cal F}(\l)$. We are going to show
that ${\cal F}(\l_i) - {\cal F}(\l_n)>0$.
Let us evaluate the black-hole free energy at the point $\l_i$:
\bea\label{g3}
&&{\cal F}(\l_i) = \int_{\l_i}^{+\infty} S(\l'){d T\over d\l'} \nn\\=
&&\int_{\l_i}^{\l_{i+1}}S(\l'){d T\over d\l'} + \int_{\l_{i+1}}^{\l_{i+2}}S(\l'){d T\over d\l'}+ \ldots \int_{\l_{n-1}}^{\l_{n}}S(\l'){d T\over d\l'}+ {\cal F}(\l_n).
\eea
Each integral is extended over a different black-hole branch. In every branch, we can use the mean value theorem:
\be \label{g4}
 \int_{\l_{j}}^{\l_{j+1}}S(\l'){d T\over d\l'} = S_{j+1}(T_{j+1} - T_j),
\ee
where  $S_{j+1}\equiv S(\bar{\l}_{j+1})$, for some appropriate value $ \bar{\l}_{j+1}$ with
$\l_j<\bar{\l}_{j+1}<\l_{j+1}$.

Then, (\ref{g3}) becomes:
\be
{\cal F}(\l_i) -{\cal F}(\l_n) = S_{i+1}(T_{i+1} - T_i) +  S_{i+2}(T_{i+2} - T_{i+1}) + \ldots  S_{n}(T_{n} - T_{n-1}).
\ee
The sum has alternating sign since, for any $l>0$, $T_{i+2l}$ are local minima and $T_{i+2l+1}$ are local maxima.
By assumption, $ S_{i+1}> S_{i+2}> \ldots S_n$, therefore:
\bea\label{g5}
&&{\cal F}(\l_i) - {\cal F}(\l_n)   >  S_{i+1}(T_{i+1} - T_i) +  S_{i+1}(T_{i+2} - T_{i+1}) + \ldots
 S_{n-1}(T_{n} - T_{n-1}) \nn\\ = && S_{i+1}(T_{i+2} - T_i) +  S_{i+3}(T_{i+4} - T_{i+2}) + \ldots  S_{n-1}(T_{n} - T_{n-1})
\eea
Notice that now only temperature corresponding to local {\em minima} appear. Next, subtract
the combination $S_{n-1}(T_n-T_i)$ from both sides of the above inequality:
\bea\label{g6}
&& {\cal F}(\l_i) - {\cal F}(\l_n) -S_{n-1}(T_n-T_i) >  S_{i+1}(T_{i+2} - T_i) +  S_{i+3}(T_{i+4} - T_{i+2}) +
\ldots \nn\\ && \ldots + S_{n-3}(T_{n-2} - T_{n-4}) +  S_{n-1}(T_{n} - T_{n-2}) - S_{n-1}(T_n-T_i)
\nn\\ &&= S_{i+1}(T_{i+2} - T_i) +  S_{i+3}(T_{i+4} - T_{i+2}) +
 \ldots + S_{n-3}(T_{n-2} - T_{n-4}) +
 S_{n-1} (T_i -T_{n-2}) \nn\\ &&>
 S_{i+1}(T_{i+2} - T_i) +  S_{i+3}(T_{i+4} - T_{i+2}) + \ldots + S_{n-3}(T_{n-2} - T_{n-4}) + S_{n-3} (T_i-T_{n-2}) \nn\\ &&=
 S_{i+1}(T_{i+2} - T_i) +  S_{i+3}(T_{i+4} - T_{i+2}) + \ldots + S_{n-3} (T_i-T_{n-4}) \nn\\ && >  \ldots >
S_{i+1}(T_{i+2} - T_i) + S_{i+3}(T_i - T_{i+2}) = (S_{i+1} - S_{i+3}) (T_{i+2} - T_i) > 0
\eea
In each subsequent step we have used $S_{l+1} > S_{l+3}$ and $(T_{l+2} - T_i)<0$ since we took $T_i$ to be
the absolute minimum. Therefore, from the first and last side of (\ref{g6}) we get:
\be\label{g7}
{\cal F}(\l_i) > {\cal F}(\l_n) +S_{n-1}(T_n-T_i)  >0
\ee
This shows that at the minimum temperature when a black-hole pair appears, the free energy is positive, and
the thermal gas background still dominates the ensemble, so the global free energy of the system does not
jump abruptly.

To show that the free energy does not exhibits jump at the creation of subsequent black-hole pairs occurring
at $T > T_{min}$, we proceed as follows. Consider the case when $\l_i$ is  a generic
local minimum. If all the subsequent minima $\l_{i+2}\ldots \l_n$ have higher
temperature, then we can proceed exactly as above and show that $F(\l_i)>0$; Otherwise, if there is a local  minimum
 for some $\l_{i+2l} > \l_i$, with $T_{i+2l} < T_i$, then  there is also a big black-hole with
temperature  $T_B=T_i$, corresponding to a point $\l_B \in (\l_{i+2l-1}, \l_{i+2l})$. It is easy to show
that ${\cal F}(\l_i) > {\cal F}(\l_B)$, using the same procedure that led to (\ref{g6}):
\bea\label{g8}
&& {\cal F}(\l_i) -{\cal F}(\l_B) = \int_{\l_i}^{\l_B} d\l'\, S(\l') T'(\l') \nn\\ && =
 S_{i+1}(T_{i+1} - T_i) +  S_{i+2}(T_{i+2} - T_{i+1}) + \ldots  S_{B}(T_{B} - T_{i+2l-1})\nn\\ && >  S_{i+1}(T_{i+2} - T_i) +  S_{i+3}(T_{i+4} - T_{i+2}) + \ldots  S_{B}(T_{B} - T_{i+2l-1}) \nn\\
&& > S_{B}(T_B - T_i) = 0
\eea
Therefore,  the black-hole pair that appears at $T_i$ cannot dominate the ensemble right at $T_i$, and
the global free energy of the system does not jump. \\

\subsubsection*{Step 3}

Finally, we show that there cannot be an increase of the free energy back above zero, for temperatures higher
than the critical temperature. Since in each branch the free energy of single black-holes is monotonically
decreasing, the only way the global free energy can increase is by jumping up  at a point where a small/big black-hole
pair disappears, i.e. at a local {\em maximum } of $T(\l_h)$ (i.e. a local {\em minimum} of ${\cal F}(\l_h)$ ).
Therefore, it is sufficient to show that for any maximum $\l_i$ of  $T(\l_h)$, with temperature $T_i$,
 there exist a black  hole with the same  temperature $T_B = T_i$  but lower free energy, so that the system
does is not forced to jump  at $T=T_i$. Since $T(\l\to 0) = +\infty$,
there certainly exists at least one big black-hole with $T_B=T_i$ and $\l_B<\l_i$. Let us consider  the closest
one to $\l_i$. $\l_i$ will be separated from $\l_B$ by an odd number of extrema ($l$ minima and $l-1$ maxima).
Proceeding along similar lines as in Step 2, we
now  compute the difference ${\cal F}(\l_B) - {\cal F}(\l_i)$ and show it is negative:
\bea\label{g9}
&& {\cal F}(\l_B) - {\cal F}(\l_i) =  \int_{\l_B}^{\l_i} d\l'\, S(\l') T'(\l') \nn\\
&& = S_{i-2k-1}(T_{i-2k-1} -T_B) + S_{i-2k}(T_{i-2k} -T_{i-2k-1}) + \ldots +
 S_i(T_i-T_{i-1}) \nn\\
&& < S_{i-2k-1} (T_{i-2k} - T_B) +  S_{i-2k+1} (T_{i-2k+2} - T_{i-2k}) \ldots + S_{i-1}(T_i - T_{i-2})= \nn\\
&& S_{i-2k-1} (T_{i-2k} - T_i) +  S_{i-2k+1} (T_{i-2k+2} - T_{i-2k}) \ldots + S_{i-1}(T_i - T_{i-2})
\eea
In the last line, we replaced $T_B$ with $T_i$  in the first term. Now
only  temperatures of
local maxima appear, of which $T_i$ is the highest one. Using repeatedly the
inequalities $S_{i-2k-1} > S_{i-2k+1} > \ldots S_{i-1}$ we have:
\bea \label{g10}
&& {\cal F}(\l_B) - {\cal F}(\l_i)
 < S_{i-2k-1} (T_{i-2k} - T_i) +  S_{i-2k+1} (T_{i-2k+2} - T_{i-2k}) +
\ldots + S_{i-1}(T_i - T_{i-2})\nn\\ &&
< S_{i-2k-+1} (T_{i-2k} - T_i) +
S_{i-2k+1} (T_{i-2k+2} - T_{i-2k}) +\ldots + S_{i-1}(T_i - T_{i-2}) \nn\\
&& = S_{i-2k+1} (T_{i-2k+2} - T_{i}) + \ldots + S_{i-1}(T_i - T_{i-2}) \nn\\
&& < \ldots < S_{i-1}(T_{i-2} - T_{i}) + S_{i-1}(T_i - T_{i-2}) =0
\eea
Thus, when a black-hole pair disappears at temperature $T_i$, the system
is already on another big black-hole branch with lower free energy.\\

So far we have treated the case of confining asymptotics. The converse can also be proven
under the same assumptions:
 non-confining asymptotics do not exhibit a thermal gas/black-hole phase transition\footnote{One
  cannot talk about a deconfining transition here, since the zero-temperature solution
itself is not confining}, rather the system is in a (big) black-hole phase for any $T>0$.
The proof proceeds along the same lines as in the confining case.

\section{Details of the computations with scalar variables}\label{XYdetails}

\subsection{Equivalence to Einstein's equations}\label{XYeq}

Here we prove that the reduced system of equations presented in the section \ref{covvar} are
equivalent to the equations of motion in the u-variable in (\ref{eom1app}-\ref{eom4app}). For this purpose
we should supplement (\ref{Xeq}) and (\ref{Yeq}) by the equations that determine the metric functions.
These are given by the following first order equations:
   \begin{eqnarray}
  A' &=& -\frac{1}{\ell}e^{-\frac43\int^{\f}_{-\infty} X}, \lab{Ap}\\
  \f' &=& -\frac{3X}{\ell}e^{-\frac43\int^{\f}_{-\infty} X},\lab{Fp}\\
  g' &=&-\frac{4Y}{\ell}e^{-\frac43\int^{\f}_{-\infty} X}.\lab{gp}
\end{eqnarray}
Using $d/du = \f' (d/d\f)$, (\ref{Xeq}), (\ref{Yeq}) and the three equations above in (\ref{eom1app}-\ref{eom4app}),
it is straightforward to show that they are all solved.

To convert the system (\ref{Xeq}), (\ref{Yeq}), (\ref{Ap}-\ref{gp}) to the conformal
coordinate system, one uses,
\be\lab{dudr}
\frac{du}{dr} = e^A.
\ee
Now, we use this equation to show how $r$ and $\l$ related near the boundary. Converting
(\ref{Fp}) into the r-variable by (\ref{dudr}), changing variable to $\l=\exp(\f)$ and
using the equation (\ref{Adet}), one has:
\be\lab{dldr}
\frac{d\l}{dr} = -\frac{3X}{\ell} \l e^{A_0+\int^{\l}_{\l_0} (3X\l)^{-1}-\frac43\int_0^{\l} \frac{X}{\l}}.
\ee
We use the expansion of $X$ near the boundary in (\ref{x0}) to get,
 \be\lab{dldr2}
\frac{d(b_0\l)}{dr} = \frac{1}{\ell} (b_0\l)^2 e^{A_0-\frac{1}{b_0\l_0}}(b_0\l_0)^{-b}
e^{\frac{1}{b_0\l}}(b_0\l)^{b}\le(1+ \cO(\l)\ri) .
\ee
Now, we use the definition of the QCD scale $\Lambda$ in (\ref{LQCD-2}) and integrate (\ref{dldr2}) to obtain:
\be\lab{rl}
r~\Lambda = e^{-\frac{1}{\l b_0}}(b_0\l)^{-b}.
\ee
The corresponding relation involving the domain wall coordinates is obtained by integrating eq. (\ref{Fp}).
The result in terms of $\lambda(u)$  reads:
\be\label{ul}
{1\over b_0 \l} +  \left (\frac49 + b\right) \log (b_0 \l) = -{u\over \ell} - \log \Lambda \ell
\ee

\subsection{Solution of eq. (7.3) }

Here, we note that (\ref{Yeq}) can be solved in terms of $X$ explicitly.
Define, \be\lab{defs} c(\f)= \frac{4(X^2-1)}{3X},\qquad
d(\f) = -\frac{4}{3X}. \ee Then, the solution is, \be\lab{soly}
Y(\f) = e^{\int^{\f}c(\f')
d\f'}\le(C_1-\int^{\f}d\f'd(\f')e^{\int^{\f'}c(\tilde{\f})d\tilde{\f}}\ri)^{-1}.
\ee
This is the general solution of (\ref{Yeq}) for non-zero $Y$. As already mentioned, $Y=0$ is
also a consistent solution, which corresponds to the thermal gas.

\subsection{A fixed point analysis of the X-Y system}\label{fixedXY}

In order to understand the number of integration constants in the system,
one can  perform a fixed point analysis of the $XY$ system given by
eqs. (\ref{Yeq}) and (\ref{Xeq}).

It is obvious that $Y=0$ is a fixed line in the phase space. This line corresponds
to the zero temperature solutions discussed in \cite{ihqcd}. We focus
on this case first. The solution is determined by (\ref{Xeq}) at $Y=0$. Clearly, there
are four fixed points of the system for an arbitrary potential. These are given by
$X= \pm 1$ and $\pm\infty$. For the special class of exponential
potentials $V\sim \exp(\a \f)$, there is the additional fixed point $X = -3\a/8$.
Furthermore, $X=0$ is also a fixed point for the types of potentials we study in this paper.

We solve the system by specifying the boundary conditions (or the asymptotic behavior)
in the IR, for large $\l$. Suppose $X=X_f$ at $\l=\l_f$. Let us assume for the moment
that $X_f<0$ (as is the case for the ``special'' and  ``generic'' solutions discussed in this paper).
Then, one obtains the following:
\begin{enumerate}
\item If $-1<X_f<0$, then $X\to 0$ as $\l$ decreases, at $\l=0$.
\item If $-\infty<X<-1$ then $X\to -\infty$ as $\l$ decreases at $\l=\l_i \ne 0$.
\end{enumerate}

To prove the above statements, let us first
show that $X=-1$ is a repulsive fixed point in the direction of decreasing $\l$.
To show this, we substitute $X= -1 +\eps$, $1\gg\eps>0$ in  (\ref{Xeq}). One has,
\be\lab{X1fpe}
\frac{d\eps}{d\f} = \eps(-\frac83+\frac{d\log(V)}{d\f}),
\ee
with the solution,
\be
\eps = c\,\, V(\Phi) e^{-8\Phi/3}.
\ee
As $\Phi\to\infty$ $V\to e^{4\f/3}$ in our case. Hence one falls into $X=-1$
as $\f\to\infty$. Note that the fixed point would instead be attractive for decreasing $\l$,
were $V\to e^{a\f/3}$ with $a>8/3$.
The analysis is the same for $X= -1 -\eps$, $1\gg\eps>0$, as one obtains the same equation,
(\ref{X1fpe}).

Now let us focus on the vicinity of $X=0$, by writing $X=-\eps$, $1\gg\eps>0$. One obtains
\be\lab{X0fpe}
\frac{d\eps}{d\f} = \frac{4}{3}-\frac{1}{2\eps}\frac{d\log(V)}{d\f}.
\ee
Let us assume that one can reach $X=0$ at finite $\Phi = \f_f$. Then, one can ignore the
first term in the RHS of (\ref{X0fpe}) above and obtain the follwing solution:
\be\lab{indiba}
\eps^2 = -\log(V(\Phi)/V(\f_f))
\ee
as $\Phi\to \f_f$. This shows that $X=0$ {\em can never be reached} in the decreasing
$\l$ direction, in finite $\l$-time.
{\em Instead $X$ always runs into $X=0$, in the direction of decreasing $\l$}.

\begin{figure}
 \begin{center}
 \leavevmode \epsfxsize=12cm \epsffile{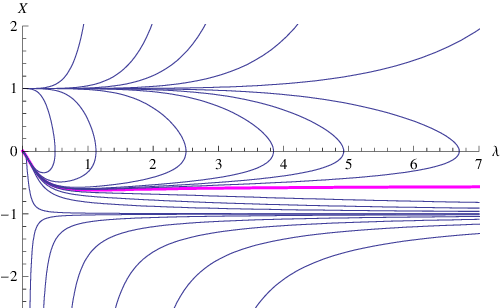}
 \end{center}
 \caption[]{All solutions to the $X$ equation of motion are shown. The thick (red) curve corresponds to
our solution $X_0$, that flows to the fixed point $X=-1/2$.
One clearly observes the fixed points $X=\pm 1,\pm\infty$ and $1/2$ in the figure. The direction of flow (as a function of r)
is towards the left for $X>0$ and towards the right for $X<0$.}
 \label{fxpts}\end{figure}

On the other,
hand (\ref{indiba}) allows to pass the $X=0$ point in the direction of {\em increasing} $\l$,
in finite $\l$-time. These solutions (as functions of $r$) continue to the positive $X$ region
and hit back to the $X=1$ fixed point as $r$ increases, see figure \ref{fxpts}.
However, in these solutions the derivative of $\l(r)$
changes sign  at the locus $X=0$. As our purpose is to find solutions dual
 to field theories with negative definite $\beta$-functions, these do not correspond to any
reasonable theories.

One can easily carry out a similar analysis in the vicinity of $X=-\infty$. For this purpose,
we define $Z= 1/X$, and focus on the vicinity of $Z\to 0^-$ by defining $Z=-\eps>0$.
To leading order,
\be\lab{Xinfpe}
\frac{d\eps}{d\f} = \frac{4}{3} - \frac{\eps}{2} \frac{d\log(V)}{d\f},
\ee
with the solution,
\be
\eps = \frac43 V[\f]^{-\half} \int_{\f_i}^{\f_f} V^{\half}[\f'] d\f'
\ee
where $\f_i$ finite, is an integration constant. It is clear that,
$\eps$ goes to 0 only at a finite point $\f_i$.

This completes the proof of the assertions above. A similar analysis can
be carried out in the region of $X>0$. We summarize the behavior of
flows in the direction of decreasing $\l$ in fig. \ref{flowchartX}.

\begin{figure}
 \begin{center}
 \leavevmode \epsfxsize=12cm \epsffile{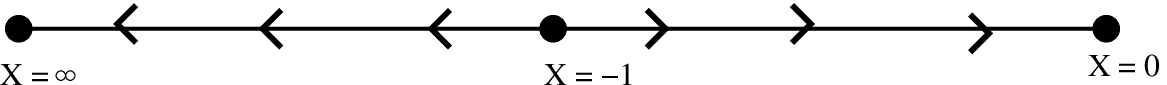}
 \end{center}
 \caption[]{Flow chart of the solutions for $Y=0$ case. The arrows show the direction of decreasing $\l$.}
 \label{flowchartX}\end{figure}

The class of solutions to the $X$ equation of motion for $Y=0$, are summarized in fig. \ref{fxpts}.
There are five classes:
\begin{itemize}

\item First of all there is our main solution, $X_0$ that flows to $-1/2$ as $l$ increases.
We recall that the fixed point $X=-1/2$ exists only for the potentials that has an asymptotic form $V\to \exp(4\f/3)$.
This is shown as the thick (red) curve in \ref{fxpts}.
\item Secondly, there are the solutions $-1<X<X_0$ that flow to $-1$. These are also asymptotically AdS
in the UV but have differ in the IR from $X_0$. The solutions described in \cite{DF} is
an example of this class.
\item The solutions $X_0<X<+1$. They also have an AdS fixed point in the UV.
As a function of $r$ the solutions continue in the region of positive $X$, hence positive
$\b$-function. This is not acceptable for an RG-flow in a field theory. The reason that such a behavior
can happen in a gravity dual is because $\l(r)$ is determined by solving a second order differential equation
whereas in the field theory it solves the first-order Gell-Mann-Low equation.
Therefore we discard these solutions as un-physical for holographic purposes.
\item Finally there are the solutions $-\infty<X<-1$ and $\infty>X>1$ that posses negative and positive $\b$-functions
respectively. They do not exhibit an asymptotically AdS fixed point, hence are not useful for our purposes here.
\end{itemize}

Now, it turns out that the above behavior of solutions easily generalize the case of $Y\ne 0$. For simplicity,
we focus only the region of the phase space that we are interested in, i.e. $X<0$, $Y>0$ and $Y>X^2-1$.
In this
case, the fixed point $X=-1$ is replaced by the fixed line $X = -\sqrt{1+Y}$. The flow chart in the
phase space is shown in fig. \ref{fxptsXY}. The thick (red) curve represents the fixed line of the system
$1-X^2+Y=0$. As described in section \ref{reghor}, $Y$ diverges at the horizon that corresponds to a
$X=X_h= -3/8V'(\l_h)/V(\l_h)$. Therefore each different curve corresponds to a different temperature.

\begin{figure}
 \begin{center}
 \leavevmode \epsfxsize=12cm \epsffile{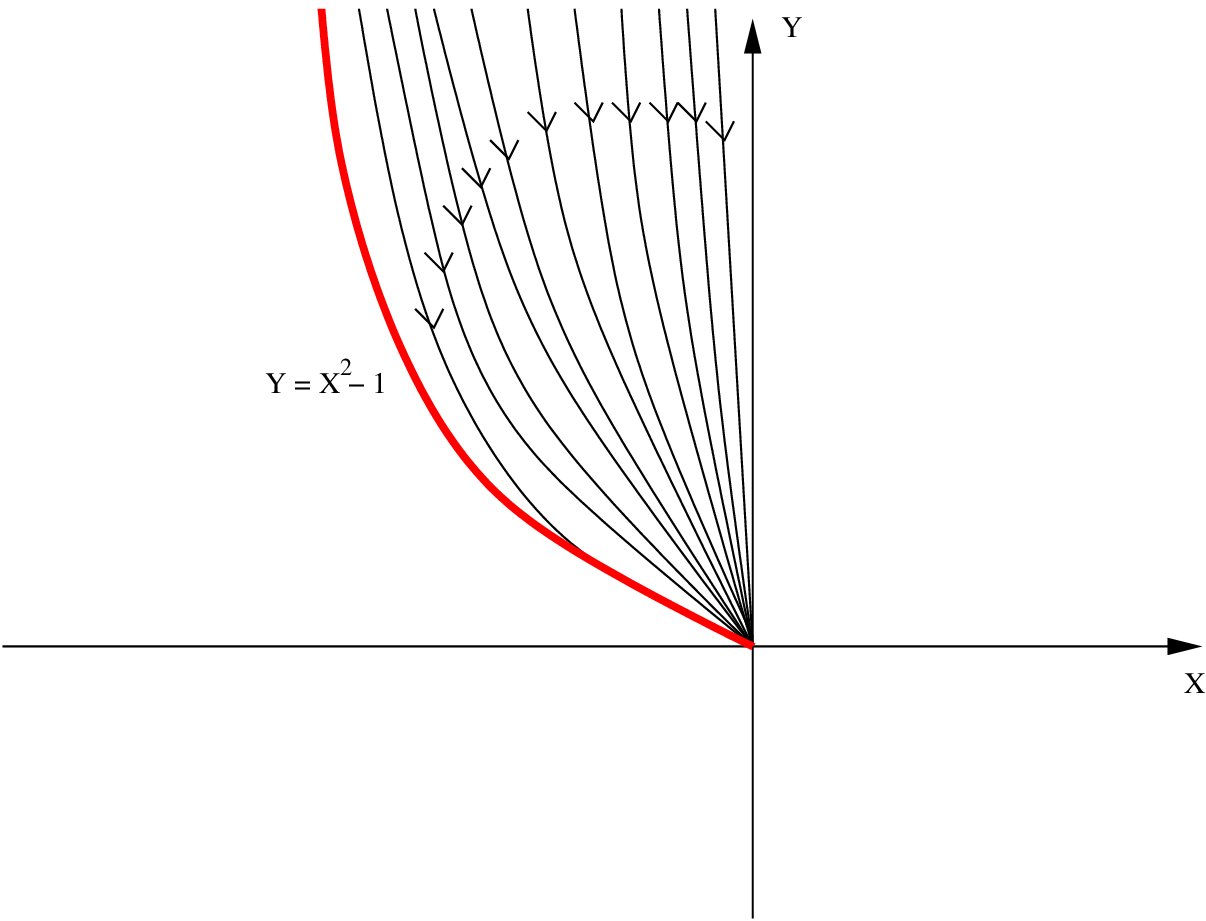}
 \end{center}
 \caption[]{Flow chart of the solutions for the general case for non-zero $Y$.
 The arrows show the direction of decreasing $\l$.}
 \label{fxptsXY}\end{figure}

It may be interesting to carry out an analysis as a function of $r$ and obtain a flow chart that generalizes fig.
\ref{fxpts} to the case of the full X-Y system. This would determine the entire family of solutions to
(\ref{Xeq},\ref{Yeq}), most of which would be un-physical. We leave this question to future work.

\subsection{Near-horizon continuously connects near-boundary}\lab{conUVIR}

We want to show that the BH solution that is given by the initial conditions for $X$ and $Y$ at the horizon
i.e. (\ref{Yhor}), (\ref{Xhor}) continuously extends over the solution near the boundary, i.e. $Y\approx 0$, $X\approx X_0$.
As shown by eqs. (\ref{Y0}) and (\ref{C0}), the solution of the system becomes $Y\approx 0$, $X\approx X_0$ as $\f\to-\infty$,
independently of the initial conditions at the horizon. We shall prove that this asymptotic UV solution is continuously connected
to the solution near the horizon. This is achieved by examining the equations is motion (\ref{Xeq}) and (\ref{Yeq}).

First, we see from (\ref{Yeq}) that
$X=X_0$, $Y=0$ cannot be reached at a finite point  $\f=\f_f$ with $-\infty<\f_f<\f_h$:
Suppose $Y=\eps$ near $\f_f$. Then, one obtains
$$\frac{d\eps}{d\f}= -\frac{4(1-X^2)}{3X}\eps$$ with the solution $$\eps = C~e^{-\frac{4(1-X_f^2)}{3X_f}\f},$$
where $X_f = X(\f_f)$. Thus, $\eps =0$ can only be reached as $\f\to -\infty$ (recall that $X<0$) in contradiction with our assumption.
Therefore we learn that $Y$ and $X$ should be finite at an arbitrary mid-point $\f_f$.

The only way
the near UV ($\f<\f_f$) and the near horizon ($\f>\f_f$) solutions are detached would be a divergence in the RHS of (\ref{Xeq}) and/or (\ref{Yeq}) at
$\f=\f_f$ as $\f$ decreases down from $\f_h$. In this region, $V$ is bounded, $X$ is also bounded as $-1<X<1$. From the exact solution for $Y$ in  (\ref{soly})
one finds that $Y$ is also bounded. Hence the only way one can get a divergence is as $X\to 0$ at $\f_*$. Now, we write $X=-\eps$ in (\ref{Xeq}) and
find that $\eps$ satisfies,
$$-\frac{d\eps^2}{d\f}=\eps_0 \frac{d}{d\f}\log(V(\f)),$$
  where $\eps_0 = 1+ Y(\f_f)>0$. The solution is,
  $$\eps^2 = -\eps_0\log\le(\frac{V(\f)}{V(\f_f)}\ri)$$
  which cannot be satisfied for any $\f>\f_f$, because we assumed $V(\Phi)$ monotonically increasing.
  Thus we proved that, given the initial conditions for $X$ and $Y$ at the horizon, the solution flows down to $\f=-\infty$ and connects continuously to
  the UV region where $Y\to 0$ and $X\to X_0$.

  Note however that $X=0$ {\em can} be reached at a finite $\f_f$ for $\f$ approaching $\f_f$ {\em from below}.
  This is analogous to the ``bouncing'' solution studied in the previous appendix. At finite temperature, this possibility is automatically ruled out by imposing the
  regularity condition at the horizon.

\subsection{Near boundary behavior of $\delta X$ and $Y$:}\lab{UVasympApp}

In order to derive the asymptotic behavior of $\delta X$ and $Y$, it proves useful to obtain  differential equations satisfied by
these perturbations. This is done by expanding (\ref{Xeq}) and
 (\ref{Yeq}) to first order in the perturbations and making use  of (\ref{ftpot}). One obtains:
 \begin{eqnarray}
  \l \frac{dY}{d\l} & = &  -\frac43 \frac{1-X_0^2}{X_0} Y,\label{npeqY}\\
  \l \frac{d \delta X}{d\l} &=& \delta X \le[
\frac{4}{3X_0}(X_0^2-1)+\frac{X_0^2+1}{X_0^2-1}\frac{\l}{X_0}\frac{dX_0}{d\l}
\ri] +\frac{\l}{1-X_0^2}\frac{dX_0}{d\l} Y.\label{npeqX}
\end{eqnarray}
The solution is straightforward:
\begin{eqnarray}
\label{Y02}
    Y(\l)  &=& Y(\l_0)~e^{-\frac43\int_{\l_0}^{\l}\frac{d\bar{\l}}{\bar{\l}}
\frac{1-X_0^2}{X_0}}, \\
\delta X(\l) &=& e^{-\frac43\int_{\l_0}^{\l}\frac{d\bar{\l}}{\bar{\l}}\frac{1-X_0^2}{X_0}}
\le[\le(\frac{Y(\l_0)}{2}-C_0(\l_0)\ri)\frac{1}{X_0}+C_0(\l_0) X_0\ri], \label{C02}\\
Y(\l_0) &=& \Y0\,\, e^{-\frac{4}{b_0\l_0}}(b_0\l_0)^{-4b}, \qquad C_0(\l_0) = \C0\,\,
 e^{-\frac{4}{b_0\l_0}}(b_0\l_0)^{-4b}.\label{Y0C0}
\end{eqnarray}
Here $\Y0$ and $\C0$ are integration constants. By expanding the above equations one obtains (\ref{C0Y0rel}).

\subsection{Free energy in $\l$: details}\label{freelambda}

We will compute the free energy in the $\lambda$-coordinate frame.
In this frame, the metric of the thermal gas and the black-hole
are as follows:
\bea\lab{lambdaTG}
ds^2_{TG} &=&  B_0^2(\l)\le( dt^2 + d{\vec{x}}^2 + \frac{d\l^2}{{D_0(\l)}^2}\ri),\\
ds^2_{BH} &=&  B^2(\l)\le( dt^2 F(\l) + d{\vec{x}}^2 + \frac{d\l^2}{F(\l)D(\l)^2}\ri).\lab{lambdaBH}
\eea
Here the various metric functions are defined as follows:
\bea\lab{B0B}
B_0(\l) &=& B_0(\l_0) e^{\frac13 \int_{\l_0}^{\l} \frac{d\tilde{\l}}{\tilde{\l}} \frac{1}{X_0}},\qquad
D_0(\l) = -\frac{3}{\ell}\frac{X_0(\l)}{\l}B_0(\l)e^{-\frac43 \int_0^{\l} \frac{d\tilde{\l}}{\tilde{\l}} X_0},\\
B(\l) &=& B(\l_0) e^{\frac13 \int_{\l_0}^{\l} \frac{d\tilde{\l}}{\tilde{\l}} \frac{1}{X}},\qquad
D(\l) = -\frac{3}{\ell}\frac{X(\l)}{\l}B(\l)e^{-\frac43 \int_0^{\l} \frac{d\tilde{\l}}{\tilde{\l}} X},
\lab{D0D}\\
F(\l) &=& e^{\frac43 \int_{0}^{\l} \frac{d\tilde{\l}}{\tilde{\l}} \frac{Y}{X}}.\lab{FL}
\eea
They are obtained directly from the the expressions for the metric functions defined in the text
 in terms of the radial variable $r$, viz. (\ref{Adet}),(\ref{gdet}) and (\ref{Fp}).
We call the metric functions in $\l$ with the capital letters to distinguish them from the
analogous functions of $r$. The relations are explicitly given by the following formulae:
\be\lab{rlamdarels}
B_0(\l) = b_0\le(r_0(\l)\ri), \qquad B(\l) = b\le(r(\l)\ri), \qquad F(\l) = f\le(r(\l)\ri),
\ee
where $r$ and $r_0$ are determined by,
\be\lab{r0r}
r_0(\l) = \int_0^{\l} \frac{d\tilde{\l}}{\tilde{\l}D_0(\tilde{\l})}, \qquad
r(\l) = \int_0^{\l} \frac{d\tilde{\l}}{\tilde{\l}D(\tilde{\l})}.
\ee
The expressions above completely determine the map between the r-frame and the $\lambda$-frame.

\subsubsection{Einstein contribution}

We first compute the Einstein contribution to the free energy. This is generally given by
the frame-independent expression,
\be\lab{E1}
S_E = \frac23 M^3 \int_{\cal M} \sqrt{g} V.
\ee
${\cal M}$ is the manifold with a boundary. We regulate the integral in the
$\l$-frame by placing a cut-off at $\l_0$. Thus, using the metric functions
defined above, one obtains the following expression in the lambda variable,
for the thermal gas solution,
\be\lab{E2}
S_E^{TG} = \frac23 M^3 \b'V_3' \int_{\l_0}^{\infty} B_0(\l)^5 V(\l) D_0(\l)^{-1}.
\ee
Here $\b'$ and $V_3'$ are the circumference of the Euclidean time and the volume of the
space-part. They are related to the analogous quantities in the black-hole geometry,
by matching the two solutions on the cut-off:
\be\lab{match1}
\b' B_0(\l_0) = \b B(\l_0)\sqrt{F(\l_0)},\qquad V_3' B_0(\l_0)^3 = V_3 B(\l_0)^3.
\ee

Now, we use the expression for the potential \cite{ihqcd},
\be\lab{potX0}
V(\l) = \frac{12}{\ell^2} (1-X_0^2) e^{\int_0^{\l} X_0(\tilde{\l})\frac{d\tilde{\l}}{\tilde{\l}}},
\ee
in (\ref{E2}) and see that it can be rewritten in the following form:
\be\lab{E3}
S_E^{TG} = -\frac{8}{3\ell} M \b'V_3' B_0(\l_0)^4 e^{-\frac43\int_0^{\l_0} X_0\frac{d\tilde{\l}}{\tilde{\l} } }
\int_{\l_0}^{\infty} \frac{d\l}{\l} \frac{1-X_0^2}{X_0} e^{\frac43 \int_{\l_0}^{\l} \frac{d\tilde{\l}}{\tilde{\l} } \frac{1-X_0^2}{X_0}}.
\ee
 To obtain this expression, we used (\ref{D0D}),  (\ref{B0B}) and (\ref{potX0}). We see that the integrand is a total derivative, hence the
 integral can be carried out exactly. One has,
\be\lab{E4}
S_E^{TG} = -\frac{2}{\ell} M^3 \b'V_3' B_0(\l_0)^4 e^{-\frac43\int_0^{\l_0} X_0\frac{d\tilde{\l}}{\tilde{\l}}}
e^{\frac43 \int_{\l_0}^{\l} \frac{d\tilde{\l}}{\tilde{\l}} \frac{1-X_0^2}{X_0}}\bigg|_{\l_0}^{\infty}.
\ee
  It is easy to see that the contribution from $\infty$ vanishes, as the expression above scales a $\l^{-2}$ for large $\l$
  for our confining IR asymptotics. Thus one only has the contribution at $\l_0$:
\be\lab{E5}
S_E^{TG} = \frac{2}{\ell} M^3 \b'V_3' B_0(\l_0)^4 e^{-\frac43\int_0^{\l_0} X_0\frac{d\tilde{\l}}{\tilde{\l} } } .
\ee
  This is our final expression for the Einstein term on the TG geometry.

  Let us now compute the analogous contribution on the BH geometry. From (\ref{E1}) we get,
 \be\lab{E6}
S_E^{BH} = \frac23 M^3 \b V_3 \int_{\l_0}^{\l_h} B(\l)^5 V(\l) D(\l)^{-1}.
\ee
Following the same steps as before, we substitute the expression for $D(\l)$, $B(\l)$  and $V(\l)$
from (\ref{D0D}), (\ref{B0B}) and (\ref{ftpot}), and obtain,
\be\lab{E7}
S_E^{BH} = -\frac{8}{3\ell} M^3 \b V_3 B(\l_0)^4 e^{-\frac43\int_0^{\l_0} (X-\frac{Y}{X})\frac{d\tilde{\l}}{\tilde{\l} } }
\int_{\l_0}^{\l_h} \frac{d\l}{\l} \frac{1-X^2+Y}{X} e^{\frac43 \int_{\l_0}^{\l} \frac{d\tilde{\l}}{\tilde{\l} } \frac{1-X^2+Y}{X}}.
\ee
Again, the integrand is a total derivative and one has,
\be\lab{E8}
S_E^{BH} = -\frac{2}{\ell} M^3 \b V_3 B(\l_0)^4 e^{-\frac43\int_0^{\l_0} X_0\frac{d\tilde{\l}}{\tilde{\l} } }
e^{\frac43 \int_{\l_0}^{\l} \frac{d\tilde{\l}}{\tilde{\l} } \frac{1-X^2+Y}{X}}\bigg|_{\l_0}^{\l_h}.
\ee
This can be simplified further: using (\ref{Yeq}), one realizes that the integrand in the exponent is a total derivative of
$\log Y(\l)$. Thus, one has,
\be\lab{E9}
S_E^{BH} = -\frac{2}{\ell} M^3 \b V_3 B(\l_0)^4 e^{-\frac43\int_0^{\l_0} X_0\frac{d\tilde{\l}}{\tilde{\l} } }
\le( \frac{Y(\l_0)}{Y(\l_h)}-1\ri).
\ee
But $Y(\l_h)=\infty$ by regularity condition at the horizon (see  section (\ref{reghor})), hence we have the final expression
for the Einstein contribution on the BH geometry:
\be\lab{E10}
S_E^{BH} = \frac{2}{\ell} M^3 \b V_3 B(\l_0)^4 e^{-\frac43\int_0^{\l_0} \le(X-\frac{Y}{X}\ri) \frac{d\tilde{\l}}{\tilde{\l} } } .
\ee

The Einstein contribution to the free energy follows from the difference of (\ref{E5}) and (\ref{E10}). In order to match the
two expressions we use the matching conditions (\ref{match1}).
Finally we also use (\ref{FL}) to obtain:
\be\lab{E11}
\delta S_E = S_{E}^{BH} - S_{E}^{TG} = \frac{2}{\ell} M^3 \b V_3 B(\l_0)^4\le(e^{-\frac43\int_0^{\l_0} \le(X-\frac{Y}{X}\ri) \frac{d\tilde{\l}}{\tilde{\l} } } - e^{-\frac43\int_0^{\l_0} (X_0-\frac{Y}{2X})\frac{d\tilde{\l}}{\tilde{\l} } } \ri).
\ee
As $\l_0$ is very small and the integrands in the expression above are very small in that region, one can expand the exponentials and obtain,
\be\lab{E12}
\delta S_E = \frac{2}{\ell} M^3 \b V_3 B(\l_0)^4 \frac23 \int_0^{\l_0}\le( \frac{Y}{X} - 2 \delta X \ri).
\ee
The functions $Y$ and $\delta X$ are given in section (\ref{soluv}). Using these expressions it is straightforward to carry out
the integrals. One obtains,
\be\lab{E13}
\delta S_E(\l_0) = -\frac{2}{\ell} M^3 \b V_3 C_0 B(\l_0)^4 e^{-\frac{4}{b_0\l_0}}(b_0\l_0)^{-4b}.
\ee
Finally we remove the cut-off by sending $\l_0$ to 0 and using the definition of the integration constant $\Lambda$:
\be\lab{defL}
\lim_{\l_0\to 0} B(\l_0) e^{-\frac{1}{b_0\l_0}}(b_0\l_0)^{-b} = \Lambda~\ell.
\ee
 Thus the final expression for the Einstein contribution to the free energy reads,
 \be\lab{freeE}
 {\cal F}_E = -\frac{2}{\ell} C_0 M^3 V_3  (\Lambda~\ell)^4.
 \ee

  \subsubsection{Gibbons-Hawking contribution}

We move on to the Gibbons-Hawking term that is given by the
frame-independent expression,  the second term in (\ref{a1}).

We first define two more functions of $\l$ in addition to (\ref{D0D}):
\be\lab{DA}
D_{A0} (\l)  = -\frac{1}{\ell} B_0(\l) e^{-\frac43\int_0^{\l} \frac{d\tilde{\l}}{\tilde{\l}} X_0(\tilde{\l})}, \qquad D_{A} (\l)  = -\frac{1}{\ell} B(\l) e^{-\frac43\int_0^{\l} \frac{d\tilde{\l}}{\tilde{\l}} X(\tilde{\l})}.
\ee
Just as in (\ref{D0D}) these are obtained from mapping the derivative of the scale factor from $r$ to $\l$ frame:
\be\lab{maprl}
D_{A0} (\l) = \frac{dA_0}{dr} \le(r_0(\l)\ri), \qquad D_{A} (\l) = \frac{dA}{dr} \le(r(\l)\ri)
\ee
where the functions $r(\l)$ and $r_0(\l)$ are given in (\ref{r0r}).

Computing the trace of the extrinsic curvature in the $\l$-frame on the TG solution (\ref{lambdaTG}), one obtains,
\be\lab{GH1}
S_{GH}^{TG} = 8M^3 \b' V_3' B_0(\l_0)^3 D_{A0}(\l_0).
\ee
Using the expressions above and (\ref{B0B}), one has,
\be\lab{GH2}
S_{GH}^{TG} = -\frac{8}{\ell}M^3 \b' V_3' B_0(\l_0)^4  e^{-\frac43\int_0^{\l_0} \frac{d\tilde{\l}}{\tilde{\l}} X_0(\tilde{\l})}
\ee
Similarly, the Gibbons-Hawking term evaluated on the BH solution reads,
\be\lab{GH3}
S_{GH}^{BH} = M^3 \b V_3 B(\l_0)^3 D_{A}(\l_0) \le(8 + 4Y(\l_0)\ri),
\ee
which gives,
\be\lab{GH4}
S_{GH}^{BH} = -\frac{8}{\ell} M^3 \b V_3 B(\l_0)^4  e^{-\frac43\int_0^{\l_0} \frac{d\tilde{\l}}{\tilde{\l}} X(\tilde{\l})} \le(1 + \half Y(\l_0)\ri),
\ee

Just as in the previous subsection, we compute the Gibbons-Hawking contribution to the free energy by taking the difference
of (\ref{GH2}) and (\ref{GH4}). Following the same steps as outlined above (\ref{E12}), one arrives at the following result:
\be\lab{GH5}
\delta S_{GH} =  S_{GH}^{BH} - S_{GH}^{TG}= -\frac{8}{\ell} M^3 \b V_3 B(\l_0)^4 \le( \frac23 \int_0^{\l_0}\le( \frac{Y}{X} - 2 \delta X \ri) +\half Y(\l_0)\ri).
\ee
Evaluating the integrals as before and using the small $\l$ asymptotics of $Y$ (see section \ref{soluv}) one arrives at the final expression
for the Gibbons-Hawking contribution to the free energy:
 \be\lab{freeGH}
 {\cal F}_E = \frac{8}{\ell} (C_0 -Y_0/2) M^3 V_3  (\Lambda~\ell)^4.
 \ee
Thus, the total free energy is obtained from combining (\ref{freeE}) and (\ref{freeGH}) as,
 \be\lab{freeGH2}
 {\cal F} = \frac{1}{\ell} (6C_0 -4Y_0) M^3 V_3  (\Lambda~\ell)^4.
 \ee

\subsection{Fluctuations of $\l$ and $A$ in the $\l$-frame:}

In order to compute the gluon condensate, we need to know $\delta\l(r) = \l(r)- \l_0(r)$, near the boundary.  This
quantity is defined in the r-frame. In the $\l$-frame the quantities $\l$ and $\l_0$ are the same, by definition.
However one still finds a non-zero value for $\delta\f(r)$ after a careful change of variables from r to $\l$ frame.

The quantity $\l_0(r)$ maps to $\l(r_0(\l))$ where $r_0$ is defined in (\ref{r0r}).
Thus, $\delta\l$ in the $\l$-frame is evaluated by:
\be\lab{ll1}
\delta\l \equiv  \l(r)- \l_0(r) = \l - \l(r_0(\l)) = \frac{d\l}{dr}\bigg|_{r_0(\l)} \delta r(\l) = D_0(\l) \delta r(\l),
\ee
where $D_0$ is defined in (\ref{D0D}) and $\delta r$ is given by, see (\ref{r0r}),
\be\lab{dr1}
\delta r(\l) = r(\l) - r_0(\l) = \int_0^{\l} d\tilde{\l} \le(D(\tilde{\l})^{-1} - D_0(\tilde{\l})^{-1}\ri).
\ee
It is straightforward to work out (\ref{dr1}) from (\ref{D0D}) and (\ref{B0B}). Let us define,
\be\lab{dAdef}
B(\l) = B_0(\l) \le(1 + \delta A(\l) \ri).
\ee
Then, one obtains,
\be\lab{dr2}
\delta r(\l) = \frac{\ell}{3} \int_0^{\l} \frac{d\tilde{\l}}{\tilde{\l}}\le[
\frac{\delta X(\tilde{\l})}{X_0(\tilde{\l})}  \delta A(\tilde{\l}) -\frac43\int_0^{\tilde{\l}} \delta X  \ri]
e^{\int_0^{\tilde{\l}}\frac{B_0(\tilde{\l})}{X_0(\tilde{\l})}}.
\ee
The last term is subleading in the limit $\l\to 0$, so we neglect it for the moment.
The second termcan be computed from (\ref{B0B}) and one finds,
\be\lab{dAres}
\delta A(\l) = -\frac94 G_0 e^{-\frac{4}{b_0\l}}(b_0\l)^{-4b-2} + \cdots
\ee
Finally the first term follows from (\ref{C0}). All in all, one finds,
\be\lab{dr3}
\delta r(\l) = \frac{9}{4\Lambda} g_0 e^{-\frac{5}{b_0\l}}(b_0\l)^{-5b-2} + \cdots
\ee
where we also used the definition of $\Lambda$ in (\ref{defL}). Now, $\delta\l$ follows
from (\ref{ll1}) as,
\be\lab{ll2}
\delta\l = D_0(\l) \delta r(\l) = \frac{9G_0}{4b_0}(\Lambda\ell)^4 e^{-\frac{4}{b_0\l}}(b_0\l)^{-4b},
\ee
which yields
\be\lab{df1}
\delta\f(r) = \frac{\delta \l}{\l_0} = \frac94 G_0 (\Lambda\ell)^4 (r/\ell)^4 \log(r\Lambda).
\ee
This is the correct coefficient that produces the mathching of the conformal anomaly in section \ref{holotrace}.

One can also compute $\delta b(r)$ in the r-frame, from the expressions that we obtained above
in the $\l$-frame. By definition,
\be\lab{db1}
\delta b(r) = b(r)- b_0(r) = b(r) - b_0(r_0)-\frac{db_0}{dr}\bigg|_{r_0}\delta r
= b_0(\l) \delta A(\l) - \frac{d B_0}{d\l} d_0(\l) \delta r(\l),
\ee
where $\delta A(\l)$ and $\delta r(\l)$ are given by (\ref{dr3}) and (\ref{dAres}) above.
We see that the leading terms cancel each other out, and one has to take into account
the subleading terms in (\ref{dr3}) and (\ref{dAres}). This is a straightforward but
lengthy computation and best carried out by a symbolical evaluation program. We shall
only present the result here:
\be\lab{db2}
\delta b(r) = \frac25 G_0 (\Lambda\ell)^4 (r/\ell)^4.
\ee
This is the result in the r-frame. In the u-frame, the coefficient becomes $\half G_0$.

\subsection{Higher order terms in the near horizon expansion}\lab{X1Y1App}

Demanding regularity at the horizon determines the higher terms in the expansion of (\ref{Yhor}) and (\ref{Xhor}).
One finds,
\be\lab{X1Y1}
X_1 = \frac{3}{16} \le(\frac{V''(\f_h)}{V(\f_h)}-\frac{V'(\f_h)^2}{V(\f_h)^2}\ri),
\qquad Y_1 =\frac{9}{64}\le(\frac{V''(\f_h)}{V(\f_h)}-2\frac{V'(\f_h)^2}{V(\f_h)^2}\ri)-1.
\ee

\subsection{Derivation of eq. (7.26)}\lab{detailT}

Here, we compute temperature in the scalar variables. By definition, $T$ is given by $4\pi T = |\dot{f}(r_h)| = |f \frac{d g}{d A}\frac{dA}{du}\frac{du}{dr}|$. Using the definition of $Y$ in (\ref{psv}),  eqs. (\ref{Ap}), (\ref{gp}), (\ref{gdet}), (\ref{Adet}) and  $du/dr = \exp(A)$ we obtain,
\be\lab{Tder1}
 T=  \frac{Y(\l_h)}{\pi\ell} e^{A_0-\int_{\l_0}^{\l_h}\frac{d\l}{\l} \frac{1}{X}} e^{\frac43 \int_{\l_0}^{\l_h} \frac{d\l}{\l} \frac{1+Y-X^2}{X}}.
 \ee
Now, using Y equation of motion (\ref{Yeq}) we see that the integrand in the last exponential is a total derivative and can easily be
integrated. One finds,
\be\lab{Tder2}
 T=  \frac{Y(\l_0)}{\pi\ell} e^{A_0-\int_{\l_0}^{\l_h}\frac{d\l}{\l} \frac{1}{X}}.
 \ee
Using the UV asymptotics of $Y$ in (\ref{Y0}) and (\ref{Adet}) with $b=\exp(A)$ one identifies the RHS of (\ref{Tder2}) as,
\be\lab{Tder3}
T = \frac{\Y0}{\pi b^3(\l_h)} [\ell^{-1}e^{4A_0-\frac{4}{b_0\l_0}}(b_0\l_0)^{-4b}].
\ee
The expression in the square brackets defines $\Lambda$, see (\ref{LQCD-2}). Thus, one finally obtains
eq. (\ref{T2u}).

\subsection{Integral representation for the free energy and the energy}\lab{intrep2}

Here, we provide further formulas regarding the integral representation of the free energy in section \ref{intrep}.
 For the big BH one has,
\be\lab{flh2}
{\cal F}_B(\l_h) = -4\pi M^3 V_3\int_{\infty}^{\l_h} b^3(\tilde\l_h) \frac{dT}{d\tilde\l_h} d\tilde\l_h , \qquad \l_h<\l_{min}.
\ee
Note that the two branches are combined in the integral, as $b(\l_h)$ is a single valued in the entire range
$\l_h\in\{\l_0,\infty\}$.
One can also put this equation in various useful forms. For example, one can write it in terms
of the $\Y0(\l_h)$ function. For this purpose, we make use of the relation between $T$ and $\Y0$ in
(\ref{T2u}).
Use this in (\ref{flh2}), carry out the integral and note that $\Y0(\infty) = 0$ to get:
 \be\lab{flh3}
{\cal F}_B(\l_h) = \Lambda^4 (M\ell)^3 V_3\le( 12 \int_{\infty}^{\l_h} \Y0(\tilde\l_h) \frac{dA}{d\tilde\l_h} d\tilde\l_h
-4\Y0(\l_h) \ri).
\ee
Comparison of this equation with (\ref{freeCY}) reveals an alternative expression for the energy density:
 \be\lab{en1}
\rho(\l_h) = \Lambda^4 (M\ell)^3 12 \int_{\infty}^{\l_h} \Y0(\tilde\l_h) \frac{dA}{d\tilde\l_h} d\tilde\l_h
\ee

\subsection{Derivation of eq. (7.38) }\lab{dersT3}

Reorganizing (\ref{ftpot}) we obtain, \be\lab{der1} V =
\frac{12}{\ell^2} (1-X^2+Y) e^{-\frac43\int_0^{\l}X
\frac{d\l}{\l}} \lim_{\epsilon\to 0} e^{\frac43
\int^{\l}_{\epsilon}\frac{d\l}{\l}(1-X^2-Y) -
\frac43\int^{\l}_{\epsilon} \frac{d\l}{\l X}}. \ee Now we use
(\ref{Yeq}) to perform the first integral in the exponent under
the limit sign and (\ref{Adet}) to obtain: \be\lab{der2} V =
\frac{12}{\ell^2} (1-X^2+Y) e^{-\frac43\int_0^{\l}X
\frac{d\l}{\l}} \lim_{\epsilon\to 0}
\frac{Y(\epsilon)}{Y(\l)}e^{4A(\epsilon)} e^{-4A(\l)}.\ee Now, we
can take the limit using the definition of $\Y0$ (\ref{Y0}):
$$\lim_{\epsilon\to 0} Y(\epsilon)e^{4A(\epsilon)} = \Y0 (\Lambda\ell)^4.$$
This yields a relation between enthalpy as a function of $\l_h$
and the dilaton potential: \be\lab{der3} V(\l) = \frac{12}{\ell^2}
(1-X^2+Y) \frac{\Y0 (\Lambda\ell)^4}{Y(\l)}e^{-\frac43\int_0^{\l}X
\frac{d\l}{\l}-4A(\l)}. \ee A remarkable simplification occurs if
we take the horizon limit $\l\to\l_h$. Using (\ref{Yhor}) and
(\ref{detyh}) one obtains, \be\lab{der4} e^{-\frac43\int_0^{\l}X
\frac{d\l}{\l}} = \frac{\ell^2 V(\l_h)}{12}\frac{e^{4A(\l_h)}}{\Y0
(\Lambda\ell)^4}. \ee Finally, using the Bekenstein-Hawking
formula for the entropy density $$s = 4\pi M_p^3 e^{3A(\l_h)},$$
and (\ref{ent2}), we obtain eq. (\ref{sT3}).

\section{High T  asymptotics}\lab{highT}

We can determine how the various quantities we described above approach
their ideal gas values for large-$T$,
by studying the next-to-leading order corrections to the AdS black-hole.
To leading order $\Y0$ in (\ref{G0}) is determined by the AdS black-hole:
\be\lab{adsBH}
\Y0 = \pi \frac{T}{\La}\frac{1}{(r_h \La)^3},\qquad T = \frac{1}{\pi r_h}.
\ee
The subleading corrections to the AdS scale factor, close to the boundary, is presented in section (\ref{UVasymps}).
From (\ref{Y0}) and (\ref{T}), one arrives at the following expression:
\be\lab{Y0T}
\Y0(T) = {\tT}^4\le[1-\frac43 \frac{1}{\log(\tT)} -\frac{16}{9} ~b~ \frac{\log(\log(\tT))}{\log^2(\tT)}+\cdots
 \ri], \qquad \tT \equiv \pi \frac{T}{\La},
\ee
Let us first compute how $(\rho-3P)/T^4$ approaches to 0 at high-$T$. This determines the high-$T$ asymptotics
of the gluon condensate. It follows from (\ref{Y0T}) and integration by parts in (\ref{G0}) that,
\be\lab{univ1}
\frac{\rho-3p}{T^4} \to \frac{4\pi^2}{135}\le[\frac{1}{\log^2\le(\frac{T}{T_c}\ri)}
+8~b~ \frac{\log\le(\log\le(\frac{T}{T_c}\ri)\ri)}{\log^3\le(\frac{T}{T_c}\ri)}+\cdots\ri]
\ee
This computation can easily be extended to the large-$T$ asymptotics of $p$, $\rho$ and $s$.
We find:
\bea\lab{fhT2}
\frac{p}{T^4} &=& \frac{\pi^2}{45} - \frac{4\pi^2}{135}\frac{1}{\log\le(\frac{T}{T_c}\ri)} -
\frac{16b}{135} \frac{\log\le(\log\le(\frac{T}{T_c}\ri)\ri)}{\log^2\le(\frac{T}{T_c}\ri)}+\cdots\\
\lab{shT2}
\frac{s}{T^3} &=& \frac{4\pi^2}{45} - \frac{16\pi^2}{135} \frac{1}{\log\le(\frac{T}{T_c}\ri)} -
\frac{64\pi^2 b}{135} \frac{\log\le(\log\le(\frac{T}{T_c}\ri)\ri)}{\log^2\le(\frac{T}{T_c}\ri)}+\cdots\\
\lab{ehT}
\frac{\rho}{T^4} &=& \frac{3\pi^2}{45} - \frac{4\pi^2}{45} \frac{1}{\log\le(\frac{T}{T_c}\ri)} -
\frac{16\pi^2 b}{45} \frac{\log\le(\log\le(\frac{T}{T_c}\ri)\ri)}{\log^2\le(\frac{T}{T_c}\ri)}+\cdots
\eea
We note that the pressure, entropy density and the energy density approach their ideal gas limits
{\em from below} as they should.

It is also useful to derive the high-$T$ asymptotics of the speed of sound.
It is obtained from eqs. (\ref{Y0T}) and (\ref{soundspeed}):
\be\lab{cshT}
\frac{1}{c_s^2}-3 = \frac{4}{3}\frac{1}{\log^2\le(\frac{T}{T_c}\ri)}
+\frac{32b}{9} \frac{\log\le(\log\le(\frac{T}{T_c}\ri)\ri)}{\log^3\le(\frac{T}{T_c}\ri)}+\cdots
\ee
We note that the speed of sound approaches to its ideal gas value $1/3$ also {\em from below} as
it should.

We remark that, all of these expressions are completely independent of the parameters of our theory {\em and}
any modification to the dilaton potential.
It follows directly from demanding an asymptotically AdS solution dual to the UV of the field theory.
Moreover, the coefficients of the first terms are even independent of the parameters of the gauge theory,
i.e. the $\beta$-function coefficients. Thus, we expect this form hold universally for any large-$N_c$ gauge
theory that exhibits logarithmic running in the UV.

\section{Analytic solutions}\label{AnalyticSolutions}

One can easily obtain analytic solutions to the system, by
restricting to the fixed points of (\ref{Xeq}) and (\ref{Yeq}).
One obvious fixed point of (\ref{Yeq}) is $Y=0$. This takes us back
to the zero-$T$ analysis which was studied in \cite{ihqcd}. In the
following we always consider the case $Y\ne 0$. We present our solutions
in the domain-wall coordinate system, see section \ref{DWframe}. They can easily be converted
into the conformal frame using (\ref{dudr}).

\subsection{Analytic solutions: zero potential}
\lab{ana1}
 Another obvious fixed point of the system is $X=const.$,
$Y=const$ and $1+Y-X^2=0$. From (\ref{ftpot}), we see that this
corresponds to vanishing dilaton potential, thus it is not very
interesting for our purposes regarding holography. Nevertheless, it produces the
following interesting analytic solution.

It is straightforward to solve (\ref{Ap}), (\ref{Fp}) and (\ref{gp}). One finds,
for the dilaton:
\begin{equation}\label{dilsol}
    \l = e^{\f} = \le(C_1 - 4X^2\frac{u}{\ell}\ri)^{\frac{3}{4X}}.
\end{equation}
Then the metric can be found as,
\begin{equation}\label{metsol1}
    ds^2 = \le(C_1-4X^2\frac{u}{\ell}\ri)^{\frac{1}{2X^2}}
    \le(dx_idx^i - \le(C_1-4X^2\frac{u}{\ell}\ri)^{\frac{X^2-1}{X^2}}dt^2\ri)
    +\le(C_1-4X^2\frac{u}{\ell}\ri)^{\frac{1-X^2}{X^2}}du^2.
\end{equation}
There is a curvature singularity at,
\begin{equation}\label{cursini}
    \frac{u_0}{\ell}  = \frac{C_1}{4X^2},
\end{equation}
where the dilaton blows up and the metric shrinks to a point (for
$X^2<1$) or a line (for $X^2>1$). Note that $X<0$ in all of our
solutions.

In order to understand the physics of this solution, one has to
distinguish these two cases. For $X^2>1$ the same point $u_0$
coincides with the event horizon. Therefore there is a curvature
singularity at the event horizon $u_h=u_0$ where the geometry
shrinks to a point. We note also that $Y>0$ in this case. Therefore,
from (\ref{gp}) we see that $g$ (or $f$) is monotonically
decreasing. It decreases from 1 at the boundary to 0 at the horizon.

In the other case, $X^2<1$, there is no event horizon. There is a
curvature singularity at $u_0$, where the geometry shrinks to a
line. Also in this case $g$ is monotonically increasing.

\subsection{Analytic solutions: exponential potential}

A less obvious fixed point of (\ref{Xeq}) is when $X=const.$ and
the dilaton potential is exponential:
\begin{equation}\label{exppot}
    V  = V_0 (1-X^2) \l^{-\frac83 X}.
\end{equation}
The proportionality constant will become clear below.

Also in this case one can find the most general analytic solution to
the system. This case is of more interest because of the following
reasons. We find below that in this case $Y$ does not need to be
constant or zero and it can be a function of $\f$. However, as
(\ref{exppot}) does not depend on $Y$, we find that it is a moduli
of the exponential potential. This fact will allow us to obtain both
thermal gas solutions and the black-hole solutions to the same
potential. Moreover, we note from \cite{ihqcd} that the leading IR
behavior of the dilaton potential, in most of the confining theories
is an exponential. Since the confinement-deconfinement phase
transition is expected to take place in the IR of the theory,
(\ref{exppot}) can be taken a first approximation to understand the
finite temperature dynamics of the interesting confining theories.

One first solves (\ref{Yeq}) to obtain $Y$ as,
\begin{equation}\label{Ysol}
    Y = \frac{C_2(1-X^2)}{\l^{\a}-C_2},
\end{equation}
where $C_2\ge 0$ and we defined,
\begin{equation}\label{adef}
    \a = \frac{4(1-X^2)}{3X}.
\end{equation}
Note that for a monotonically decreasing $g$ we need $Y<0$. As we
also require $Y\to 0$ near the boundary, where $\l\to 0$, one should
take $X^2<1$, hence $\a<0$. We stress that, this case covers the
rest of the physically interesting, constant $X$ solutions, as the
physical solution in the previous subsection covered the range
$X^2>1$.

As we demonstrate below, the solution (\ref{Ysol}) describes a
black-hole for $C_2>0$ and a thermal gas for $C_2=0$. To check that
indeed $Y$ is a moduli of (\ref{exppot}), one inserts (\ref{Ysol})
in (\ref{ftpot}) and finds (\ref{exppot}) after nice cancelations.

Both for the black-hole and the thermal gas, one finds the same $\l$
behavior for the dilaton:
\begin{equation}\label{dilsoll}
    \l = e^{\f} = \le(C_1 - 4X^2\frac{u}{\ell}\ri)^{\frac{3}{4X}},
\end{equation}
and the scale factor,
\begin{equation}\label{Asol}
    e^A = e^{A_0} \l^{\frac{1}{3X}}.
\end{equation}
We note that these are the same as in the previous subsection, as
they follow directly from the fact that $X$ is constant.

One finds the location of the horizon by solving for $f$ from
(\ref{gdet}):
\begin{equation}\label{fsol}
    f = e^g = 1- C_2\l^{-\a}.
\end{equation}
We find that indeed $f\to 1$ on the boundary, ($\l\to 0$) as $\a<0$.
There is an event horizon located at (using (\ref{dilsoll})),
\begin{equation}\label{ehor}
    \l_h = C_2^{\frac{1}{\a}}\qquad i.e. \qquad \frac{u_h}{\ell} =
    \frac{C_1}{4X^2}- \frac{C_2^{\frac{X^2}{1-X^2}}}{4X^2} .
\end{equation}
The curvature singularity is located at $\l=\infty$ \ie,
\begin{equation}\label{cursin}
    \frac{u_0}{\ell}  = \frac{C_1}{4X^2}.
\end{equation}
We note that $u_h<u_0$ and indeed there is a well-behaved black-hole
solution to the system. The metric of the black-hole is given by,
\bea
    ds^2 &=& e^{2A_0}\le(C_1-4X^2\frac{u}{\ell}\ri)^{\frac{1}{2X^2}}
    \le\{dx_idx^i -
    \le(1-C_2(C_1-4X^2\frac{u}{\ell})^{\frac{1-X^2}{X^2}}\ri)dt^2\ri\}\nonumber\\
   {}&& + \le(1-C_2(C_1-4X^2\frac{u}{\ell})^{\frac{1-X^2}{X^2}}\ri)^{-1}du^2\label{BHmet}.
\eea

The temperature of the black-hole is determined by requiring
regularity of the Euclidean continuation at $u_h$:
\begin{equation}\label{tempgen}
    \b=\frac{1}{T} = \frac{4\pi}{|f'(u_h)|e^{A(u_h)}}.
\end{equation}
One finds,
\begin{equation}\label{temp}
    \b = \pi \ell
    \frac{e^{-A_0}C_2^{\frac{\frac14-X^2}{1-X^2}}}{1-X^2}.
\end{equation}
The physically most interesting case corresponds to the value
$X=-1/2$, see \cite{ihqcd}. Very interestingly, in this case the
temperature is only given by the integration constant $A_0$:
\begin{equation}\label{tempsp}
    \b = \frac1T = \frac{4\pi \ell}{3 e^{A_0}}.
\end{equation}
Otherwise the temperature is determined by the combination of $A_0$
and $C_2$, namely the string tension and the location of the event
horizon.

The thermal gas solution is found by setting $C_2=0$ in
(\ref{Ysol}), hence $f=1$. The dilaton is given again by
(\ref{dilsoll}) and the metric is,
\begin{equation}\label{TGmet}
    ds^2 = e^{2A_0}\le(C_1-4X^2\frac{u}{\ell}\ri)^{\frac{1}{2X^2}}
    \le\{dx_idx^i + dt^2\ri\} + du^2.
\end{equation}
Here we required the same integration constant for $A$ as the
black-hole solution (\ref{BHmet}). This is because they should have
the same asymptotics at the boundary. Euclidean time is compactified
with circumference, $\bar{\b}$. We note that there is a curvature
singularity ar $u_0$ that is given by (\ref{cursin}). It is the same
locus as the curvature singularity of the black-hole solution--that
is cloaked behind the event horizon-- resides.

\vspace{0.5cm} {\bf Computation of the energy of the
solutions}\vspace{0.5cm}

Here we prove that the analytic solutions describe above do not demonstrate
a Hawking-Page transition. Hence they are not interesting for holographic purposes.
The action is given by (\ref{a1}). One finds that the trace of the
intrinsic curvature is given by,
\begin{equation}\label{intcuru}
    K = \frac{\sqrt{f}}{2}(8A'+g')
\end{equation}
in the domain-wall coordinate system. Thus, the boundary
contribution to the action becomes,
\begin{equation}\label{sboun}
    S_{bnd} = -M^3 V_3 \b \le\{ e^{g+4A}(8A'+g')\ri\}_{u_b},
\end{equation}
where $u_b$ denotes the regulated boundary of the geometry
infinitesimally close to $-\infty$.

 The bulk contribution to the action, evaluated on the
solution can be simplified as,
\begin{eqnarray}
  S_{bulk} &=& 2 M^3 V_3 \b \int_{u_b}^{u_s} du \frac{d}{du}\le(f e^{4A} A'\ri) \nonumber \\
  {} &=& 2 M^3 V_3 \b \le\{f(u_s)e^{4A(u_s)}A'(u_s) -
  f(u_b)e^{4A(u_b)}A'(u_b)\ri\}\label{sbulk}.
\end{eqnarray}
Here $u_s$ denotes $u_0$ or $u_h$ depending on which appears first.
Thus, for the black-hole solution $u_s=u_h$, whereas for the thermal
gas $u_s = u_0$.

The first term in (\ref{sbulk}) deserves attention. Clearly it
vanishes for the black-hole, as $f(u_h)=0$ by definition. However,
it is not a priori clear that it also vanishes for the thermal gas.
A straightforward computation using (\ref{Asol}),(\ref{dilsoll})
and,
\begin{equation}\label{Apsol}
    A' = -\frac{1}{\ell}\l^{-\frac{4X}{3}}
\end{equation}
shows that it indeed vanishes for our physically interesting case
$X^2<1$. Therefore, one obtains the following total expression for
the action from (\ref{sboun}) and (\ref{sbulk}) by dropping the
first term in (\ref{sbulk}):
\begin{equation}\label{stotl}
    S = -2M^3V_3\b e^{g(u_b)+4A(u_b)}\le(5A'(u_b)+\half g'(u_b)\ri).
\end{equation}

In order to compare the energies of the black-hole and the thermal
gas geometries, we fix the UV asymptotics of the thermal gas
geometry by requiring the same circumference for the Euclidean time
at $u_b$:
\begin{equation}\label{tfix}
    \bar{\b} = \b \sqrt{f(u_b)}.
\end{equation}

Now, it is straightforward to compute the energy of the geometries.
For the black-hole (\ref{BHmet}), one finds:
\begin{equation}\label{sBH}
    S_{BH} = - 2M^3 V_3\le(\frac{\b}{\ell}\ri) e^{4A_0} \le(C_2(3+2X^2)-5\l_b^{\a}\ri).
\end{equation}
Here $\l_b$ is the value of the dilaton on the regulated boundary
$u_b$. As $\a<0$ and $\l\to 0$ near the boundary, it is a divergent
piece that should be regulated.

For the thermal gas one finds, using (\ref{tfix}),
\begin{equation}\label{sTG}
    S_{TG} = - 10 M^3 V_3\le(\frac{\b}{\ell}\ri) e^{4A_0} \le(\frac{C_2}{2} - \l_b^{\a}\ri).
\end{equation}
We note that the divergent terms in (\ref{sBH}) and (\ref{sTG})
cancels in the difference and one finds,
\begin{equation}\label{sdif}
    S_{BH} - S_{TG} = - 2 M^3 V_3\le(\frac{\b}{\ell}\ri) e^{4A_0} C_2\le(3X^2+\half\ri).
\end{equation}

We note from (\ref{temp}) that the temperature is given by,
\begin{equation}\label{tempfin}
    e^{A_0} = \frac{\pi T \ell}{1-X^2}C_2^{\frac{\frac14-X^2}{1-X^2}}.
\end{equation}
As (\ref{sdif}) is always negative, we observe that the BH solution
always minimizes the action, hence if it exist it is the dominant
solution. Therefore, there is no phase transition in this geometry.

\end{document}